\documentclass[11pt]{article}
\usepackage[margin=1in]{geometry}
\usepackage{latexsym}
\usepackage{graphicx}
\usepackage{amsfonts}
\usepackage[colorlinks,allcolors=blue]{hyperref}
\usepackage{amsmath}
\usepackage{amsthm}
\usepackage{amssymb}
\usepackage{mathtools}
\usepackage{comment}
\usepackage{authblk}
\usepackage{setspace}
\usepackage[parfill]{parskip}
\RequirePackage{natbib}
\usepackage{bm}
\usepackage{threeparttable}
\usepackage{booktabs}
\usepackage{enumerate}
\usepackage{bbm}
\usepackage{floatpag}
\usepackage[font=small,labelfont=bf]{caption}
\usepackage{siunitx}
\usepackage{color}
\usepackage{subcaption}

\usepackage{algorithm}
\usepackage{algpseudocode}

\usepackage{tikz}
\usetikzlibrary{shapes.geometric, arrows.meta, positioning}

\definecolor{outline}{RGB}{8,24,96}
\definecolor{colT0}{RGB}{232,152,152}
\definecolor{colPD}{RGB}{200,16,16}
\definecolor{colSW}{RGB}{192,32,232}

\newtheorem{lemma*}{Lemma}

\newtheorem{theorem*}{Theorem}

\newcommand{\R}{\mathbb{R}}

\usepackage{tikz}
\usetikzlibrary{fit,positioning,arrows,automata,calc}
\tikzset{
  main/.style={circle, minimum size = 5mm, thick, draw =black!80, node distance = 10mm},
  connect/.style={-latex, thick},
  box/.style={rectangle, draw=black!100}
}

\usepackage{xcolor}

\newcommand{\calEi}{\mathcal{E}_i}

\begin{document}

\pagestyle{plain}

\newcommand{\blind}{0}

\newcommand{\tit}{Leveraging External Controls for Treatment Switching in Randomized Controlled Trials: A Weighted Causal Inference Framework for Overall Survival}

\if0\blind

{\title{\vspace{-1em}\tit\thanks{Includes work completed while A.S. was an intern and R.L. was employed at Genentech/Roche.}}

\author[1,2]{Andy A. Shen\thanks{Email: \texttt{shen.andy@gene.com}}}
\author[1]{Chenqi Fu\thanks{Email: \texttt{fu.chenqi@gene.com}}}
\author[1]{Ray Lin\thanks{Email: \texttt{raylin.stanford@gmail.com}}}

\affil[1]{Product Development Data Sciences, Genentech Inc, South San Francisco, CA}
\affil[2]{Department of Statistics, University of California, Berkeley}
\date{\today}
\maketitle
}\fi

\if1\blind

{\title{\bf \tit}
\author{}
\date{}
\maketitle
}\fi

\begin{abstract}
In many oncology clinical trials where overall survival is a key endpoint, patients are permitted to switch from the control arm to the experimental treatment arm or other suitable therapies. Switching can occur for various reasons, including disease progression. This violates the causal guarantees of randomized treatment assignment, resulting in biased treatment effect estimates. Existing methods often require strong assumptions, complicated model specifications, or both. In this paper, we propose a general framework that incorporates external controls to account for treatment switching in randomized controlled trials. Leveraging the synthetic control method and balancing weights from observational causal inference, we propose several estimators that use multiple imputation and time-varying weights to adjust for treatment switching. We also discuss approaches to selecting the risk set of external controls to impute from. Through extensive simulation studies, we show that our proposed methods lead to meaningful statistical improvements relative to standard adjustment methods that utilize external controls in naive ways or those that do not utilize external controls at all. We then demonstrate the utility of our external control-based approaches with two phase III oncology trials. 
\end{abstract}

\noindent \emph{Keywords}: treatment switching, clinical trials, causal inference, weighting, inverse probability weighting, balancing weights

\clearpage
\onehalfspacing

\section{Introduction} \label{sec:introduction}
Randomized controlled trials (RCTs) are the gold standard for causal inference in clinical research because random assignment tends to create comparable groups that differ only in treatment assignment. As a result, differences in outcomes between randomized arms are typically interpreted under the intention-to-treat (ITT) principle (also referred to as a treatment-policy strategy) \citep{guideline2017addendum,manitz2022estimands}. 
\emph{Treatment switching} occurs when participants deviate from their assigned regimen. A common instance in oncology is \emph{one-way switching}, where control-arm patients cross over to the experimental therapy or initiate non-protocol therapy (NPT) after experiencing disease progression (PD) \citep{manitz2022estimands,latimer2014adjusting}.
This paper focuses specifically on the one-way case from control to experimental treatment or NPT. 
While our empirical example focuses on crossover to experimental therapy (a specific instance of switching), our framework can be applied to any deviation of the control arm therapy, including switching to a NPT.
When switching occurs, the ITT estimand tends to underestimate the true treatment benefit, as ``switchers'' generally have a better baseline prognosis than non-switchers.

Various statistical adjustment methods, such as the rank-preserving structural failure time model (RPSFTM) and inverse probability of censoring weighting (IPCW), exist to account for switching bias (see Appendix \ref{sec:existing-methods} for a detailed review) \citep{latimer2014adjusting,latimer2019two,zhao2024multiple,robins2000correcting,watkins2025further}.
However, these traditional methods share a common limitation: they rely on internal non-switchers to infer counterfactual survival times for switchers. This introduces selection bias if the two groups differ systematically in baseline characteristics and prognosis \citep{latimer2016treatment}. As a result, recent discussion has highlighted the potential use of external data (i.e.,  ``external controls (ECs)'') to adjust for treatment switching following the publication of documents from the National Institute for Health and Care Excellence (NICE) Decision Support Unit (TSD 24) \citep{latimer2014nice,gorrod2024nice} and many studies have followed \citep{ishak2011adjusting,kuehne2022estimating,campbell2025augmented}. Broadly speaking, EC data can include data from previous clinical trials or real-world data collected from medical records or registries. While promising, incorporating ECs requires strong assumptions of no unmeasured confounding and careful statistical adjustment \citep{gorrod2024nice}.

We propose a causal inference framework incorporating ECs to address treatment switching bias in time-to-event settings. In particular, we review commonly used weighting methods from observational causal inference and show how inverse propensity score weighting (IPW) and the synthetic control method can be used to align covariate distributions of ECs with those of switchers. We then introduce approaches to estimate the treatment effect using multiple imputation and time-varying weights. Our simulation results demonstrate that our EC-based methods reduce bias and improve Type I error and power, relative to existing methods. We then provide practical guidance on the application of our framework by augmenting the Phase III IMpower130 trial \citep{west2019atezolizumab} with the OAK trial \citep{rittmeyer2017atezolizumab,mazieres2021atezolizumab} serving as EC data.

Our paper proceeds as follows: Section \ref{sec:background} details the setup and a review of weighting approaches. Section \ref{sec:methods} introduces our EC-based methodology. Section \ref{sec:simulation} evaluates performance via simulation, Section \ref{sec:impower130} presents the empirical application, and Section \ref{sec:discussion} concludes.

\section{Background} \label{sec:background}

\subsection{Notation, assumptions, and estimands}
Consider a randomized trial $\mathcal{R}$ of $n_r$ subjects, and a set of \textit{external controls} (ECs) $\mathcal{E}$ of size $n_e$, totaling $n = n_r + n_e$ patients. For each patient $i = 1, \dots, n$, let $R_i= \mathbbm{1}\left\lbrace i \in \mathcal{R} \right \rbrace$ denote trial participation. Let $Z_i \in \left\lbrace 0,1 \right\rbrace$ indicate assignment to the experimental or control arm, respectively, with $n_1$ and $n_0$ denoting the randomized arm sizes ($n_r = n_1 + n_0$).

We assume \textit{one-way switching}, where control-arm patients may initiate the experimental treatment or NPT strictly after experiencing disease progression (PD). Let $D_i\in\{0,1\}$ indicate observed PD with (possibly censored) time $T_{\mathrm{PD},i}$. Let $S_i \in \{0,1\}$ denote whether patient $i$ switches, resulting in $n_s = \sum_{i: Z_i=0} S_i$ total switchers.
Each patient has baseline covariates $X_i \in \mathcal{X} \subset \R^{d}$. For time-to-event (TTE) outcomes, we observe $Y_i = (T_{\mathrm{OS},i}, \delta_i)$, where $T_{\mathrm{OS},i} = \min(T_i, C_i)$ is the censored time-to-death, $\delta_i = \mathbbm{1}\left\lbrace T_i < C_i\right \rbrace$ is the event indicator, and $T_i$ and $C_i$ are the underlying death and censoring times. We drop the subscript $i$ where convenient.

Under the estimand framework, treatment switching is considered an intercurrent event \citep{manitz2022estimands}. We focus on the hypothetical strategy for handling these intercurrent events by estimating the treatment effect assuming treatment switching had not occurred. In particular, we consider the \emph{hypothetical hazard ratio}, the ratio of hazard rates between treatment arms in the scenario where all patients remained on their assigned arms and no switching occurred:
\begin{equation}
\theta(t) = \frac{\lambda^{(1)}(t)}{\lambda^{(0)}(t)},
\end{equation}
where $\lambda^{(z)}(t)$ represents the hazard function for the potential time-to-event outcome $T(z)$ under treatment assignment $z \in \{0,1\}$ had no treatment switching occurred. In our paper, we follow the standard proportional hazards assumption, so $\theta(t)$ reduces to a constant $\theta$, which is the target of the Cox models we fit in Sections \ref{sec:methods}-\ref{sec:impower130}.

For patients who adhere to their assigned therapy throughout the trial, the observed event time $T_{\mathrm{OS},i}$ equals the potential outcome $T(z)$ (absent non-switching censoring), so these patients contribute directly to estimation of $\theta$. The challenge lies with switchers: after switching, their observed outcomes no longer reflect the assigned regimen, and their counterfactual survival under continued control must be estimated. This is the focus of our paper.

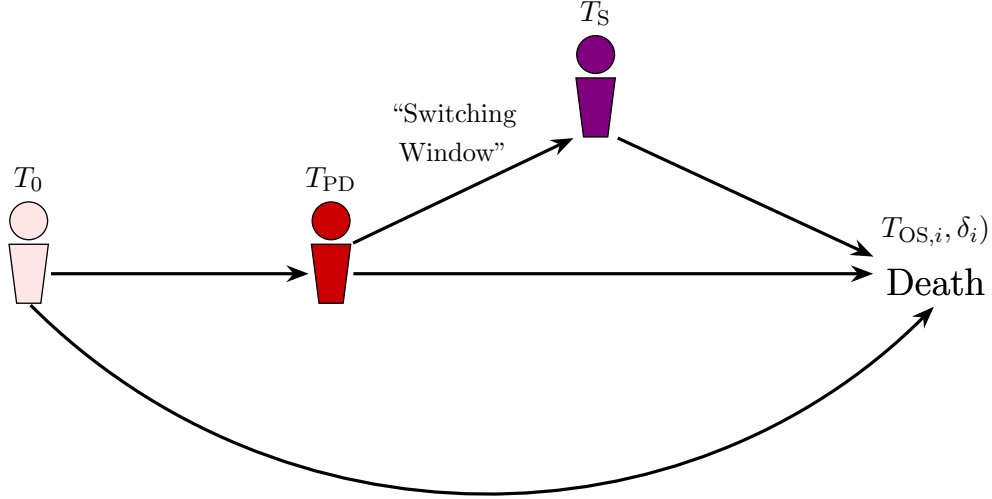
\begin{figure}[ht]
\centering

\newcommand{\DrawBody}[2]{%
  \filldraw[fill=#2, draw=black, line width=0.5pt]
    (#1.north west) -- (#1.north east) --
    ([xshift=-0.10cm]#1.south east) -- ([xshift=0.10cm]#1.south west) -- cycle;
}

\begin{tikzpicture}[
    person/.style={circle, minimum size=0.5cm, draw=black, line width=0.5pt},
    bodybox/.style={minimum width=0.52cm, minimum height=0.78cm, inner sep=0pt, outer sep=0pt, draw=none},
    arrow/.style={-Stealth, line width=1.2pt, black, shorten <=1pt, shorten >=1pt}
]

\node[person, fill=pink!40] (head0) at (0, 1.2) {};
\node[bodybox] (body0) at (0, 0.5) {};
\DrawBody{body0}{pink!40}
\node at ($(head0)+(0,0.55)$) {$T_0$};

\node[person, fill=red!80!black] (headPD) at (4, 1.2) {};
\node[bodybox] (bodyPD) at (4, 0.5) {};
\DrawBody{bodyPD}{red!80!black}
\node at ($(headPD)+(0,0.55)$) {$T_{\mathrm{PD}}$};

\node[person, fill=violet] (headS) at (7.5, 3.4) {};
\node[bodybox] (bodyS) at (7.5, 2.7) {};
\DrawBody{bodyS}{violet}
\node at ($(headS)+(0,0.55)$) {$T_{\mathrm{S}}$};

\node[font=\Large] (death) at (12, 0.4) {Death};

\node[font=\Large] (death) at (12, 0.4) {Death};
\node[above=0.6pt] at (death.north) {$T_{\mathrm{OS},i}, \delta_i)$};

\draw[arrow] (body0.east) -- (bodyPD.west);

\coordinate (deathPD) at (death.west |- bodyPD.east);
\draw[arrow] (bodyPD.east) -- (deathPD);

\draw[arrow] (bodyPD.north east) --
  node[pos=0.55, above, align=center, font=\small, yshift=5pt,xshift=-8pt]{``Switching\\Window''}
  (bodyS.south west);

\draw[arrow] (bodyS.south east) --  (death.north west);

\draw[arrow] (body0.south) to[out=-45, in=-135] (death.south);

\end{tikzpicture}
\caption{Treatment switching mechanism for control arm patients. Patients who experience disease progression (PD) are provided the option to switch/crossover within a certain window. For all patients, we observe $T_{\mathrm{OS}}$, the censored time-to-death.}
\label{fig:switching-window}
\end{figure}

\paragraph{Switching mechanism.} As mentioned previously, we assume one-way switching, meaning that only patients assigned to the control arm are eligible to switch. Moreover, we assume control arm patients are only permitted to switch if they experience disease progression (PD), a commonly established policy for oncology trials where overall survival or progression-free survival are desired endpoints. Therefore, after PD at time $T_{\mathrm{PD},i}$, a control-arm patient may switch at time $T_{\mathrm{S},i}$ within a certain time window, and overall survival (or censoring) is observed as $Y_i=(T_{\mathrm{OS},i},\delta_i)$. Figure \ref{fig:switching-window} provides a visual of this timeline for control-arm patients.

\newcommand{\TOS}{T_{\mathrm{OS}}}

\subsection{External controls and population-level weighting} \label{sec:leveraging-ec}
As mentioned in Section \ref{sec:introduction}, external controls (ECs) are a common source of patient data that can strengthen the results of a clinical trial by acting as a secondary comparison arm. As the name suggests, EC data comes from outside of the trial and can include control arm data from previous clinical trials or real-world patient data from other registries, such as electronic health records. In particular, EC-based causal inference frameworks have been used in trials with longitudinal outcomes \citep{zhou2025causal}, hybrid trials \citep{valancius2024causal,liao2025prognostic}, and to account for treatment switching with generic continuous outcomes as well as repeated outcome measurements \citep{zhou2024estimating,shi2025rdborrow}. In this paper, we use entire external control cohorts in which no patient switched therapies. This section also introduces population-level reweighting of ECs, a conventional weighting technique derived from observational causal inference. Our main methodological contribution of switcher-specific reweighting is presented in Section \ref{sec:methods}.

While the use of EC data can significantly augment trial findings, its success depends on whether the endpoints and patient characteristics align with the study at hand or can be adequately handled through statistical adjustment. This adjustment achieves \textit{covariate balance}, which ensures that the EC population and its corresponding target population are comparable on observed baseline characteristics. Weighting methods achieve covariate balance by re-weighting the EC cohort to have a comparable covariate distribution to its target population (i.e., the entire control arm or only switchers), typically via inverse propensity score weighting (IPW). The \textit{propensity score} $e(x)$ is generally defined as the probability of receiving treatment, conditional on observed covariates \citep{rosenbaum1983central}, which we denote as: $e(x)= \mathbb{P}(Z = 1 \mid X = x)$. 

To connect ECs with propensity score weighting, consider the case in which the trial’s internal control arm is replaced entirely with the EC cohort. To illustrate the standard single-arm trial analogy, let $G \in \{0,1\}$ denote group membership, with $G=1$ for trial participants assigned to the experimental arm and $G=0$ for EC participants. Under the standard overlap assumption of non-deterministic treatment assignment $(e(x)<1)$, weighting EC participants by $e(x)/(1-e(x))$ balances baseline covariates in the sense that
\begin{align}
\label{eqn:covbal-ps}
\mathbb{E}[X \mid G=1]
=
\mathbb{E}\!\left[\underbrace{\frac{e(X)}{1-e(X)}}_{\text{IPW}}\,X \,\middle|\, G=0\right].
\end{align}
In other words, ${e(x)}/({1 - e(x)})$ are the inverse propensity (IP) weights that balance covariate distributions by reweighting the EC group so its weighted covariate distribution aligns with that of the treatment arm of the trial \citep{ben2021balancing,ding2024first}. 

Since our paper is concerned about control arm patients in the trial who switch therapies, we can generalize the propensity score to target populations beyond the treatment group, namely to the population of switchers. Let $A_i \in \left\lbrace 0,1 \right\rbrace$ be an indicator variable denoting switcher vs EC status: 
\begin{align}
    A_i = \begin{cases}
        1 \quad \text{if } i \in \left\lbrace \mathcal{R} : S_i = 1 \right\rbrace\\
        0 \quad \text{if } i \in \left\lbrace \mathcal{E} : D_i = 1 \right\rbrace, 
    \end{cases}
\end{align}
where we assume that disease-progression status $D_i$ and progression time $T_{\mathrm{PD},i}$ are observed for EC patients as well.
Note that, in this reweighting scheme, we restrict the EC cohort only to those with disease progression ($D_i = 1$) as it is a necessary requirement to switch.
Here we consider the switcher propensity score $e_a(x) = \mathbb{P}(A = 1 \mid X = x)$, which measures the probability of being a switcher vs an EC, conditional on observed covariates. Note that the propensity score here is different from the probability of switching, which refers to a specific action as opposed to belonging to a certain group. Under a similar overlap assumption $(e_a(x) < 1)$ We then have that
\begin{align}
\label{eqn:covbal-ps-switch}
   \mathbb{E}\left[X \mid A = 1\right] = \mathbb{E}\left[{\frac{e_a(X)}{1 - e_a(X)}} X \mid A = 0\right],
\end{align}
which states that ${e_a(x)}/({1 - e_a(x)})$ are the IP weights that balance covariate distributions by reweighting the EC group so its weighted covariate distribution aligns with that of the switchers. 

Traditionally, IP weights are computed by first estimating the propensity score with logistic regression and then plugging the estimated values $\hat{e}(x)$ or $\hat{e}_a(x)$ into the analytic form of the weights. This approach can lead to unstable weights due to model misspecification or extreme values of the propensity score, causing the resulting estimator to ``blow up'' \citep[see][]{ben2021balancing}. Instead, we focus on estimating the weights directly using the \emph{balancing weights} framework, which we introduce in Section \ref{sec:synth-balwgt}. In contrast to the plug-in approach, balancing weight methods solve an optimization problem to find weights that directly minimize covariate imbalance, providing a stable alternative to traditional IPW estimation \citep{ben2021balancing,benmichael2021_lor,keele2025balancing}.

Given these weights, we can estimate a treatment effect by fitting a weighted Cox proportional hazards model using the inverse propensity weights in the partial likelihood.
In what follows, we treat these two population-level weighting approaches (reweighting towards the treatment arm or only the switchers) as conventional comparators. They appear alongside traditional methods (ITT, censoring all switchers, RPSFTM, IPCW) in both our simulations and applied example, serving as baselines against which we evaluate the switcher-specific methods introduced in Section \ref{sec:methods}.

\section{Switcher-Specific Methodology} \label{sec:methods}
Our main contribution addresses a limitation of population-level EC reweighting: because switchers progress and cross over at different times, no single population-level weight vector can align the EC cohort with each and every switcher's disease trajectory at their unique switch time. We resolve this by constructing weights at the individual switcher level, leveraging methods from observational causal inference. We first present a weighting framework based on the synthetic control method \citep{abadie2010synthetic,abadie2015comparative,abadie2021penalized} and balancing weights \citep{ben2021balancing,benmichael2021_lor,keele2025balancing} before discussing our estimation approaches. For each switcher, we build a nearest-neighbor risk set from the EC cohort and then estimate weights that balance that risk set to the switcher’s observed baseline and time-varying covariates. We discuss both procedures in detail in the subsequent sections.
\textbf{Importantly, the risk set for a switcher $i$ must only consist of external controls that a) have survived past the time of switcher $i$'s switch time and b) have experienced disease progression.}

In principle, one may further restrict this set to the $K$ nearest-neighbor ECs to reduce computation. In practice, results are typically insensitive to $K$ once it is moderately large, since the weighting algorithm assigns negligible (often zero) weight to poorly matched ECs. For this reason, a simple default is to take the risk set as the full set of ECs at risk at the switch time, which is what we adopt hereafter.
We then use these switcher-specific weights in three estimators of the treatment effect: two imputation-based estimators and one time-varying weights estimator.
These three frameworks build on the multiple imputation methods of \citet{zhao2024multiple} which consider time-to-event studies but not ECs, as well as the EC weighting-based frameworks of \citet{zhou2024estimating,zhou2025causal} and \citet{valancius2024causal}.

\subsection{Accounting for PD Time} \label{sec:pd-adjust}
As mentioned earlier, having disease progression is a necessary but not sufficient condition for switching: all switchers have $D_i = 1$, but not all who progressed will eventually switch. Therefore, our imputing EC risk set only includes those who experience disease progression. Within this specific group, we want to impute ECs who have a similar disease status as the switcher by accounting for a patient's time of PD ($T_{\mathrm{PD}}$) because PD is highly predictive of OS. Because PD status and its timing are post-baseline variables, they cannot be included as covariates in the balancing-weight optimization without risking bias from conditioning on a post-treatment event. However, progression is still a prerequisite for switching, so we need a mechanism to restrict the EC risk set to patients at a comparable stage of disease. We achieve this through a progression-aligned selection strategy described below. 

\begin{figure}[ht]
    \centering
    \includegraphics[width=0.95\linewidth]{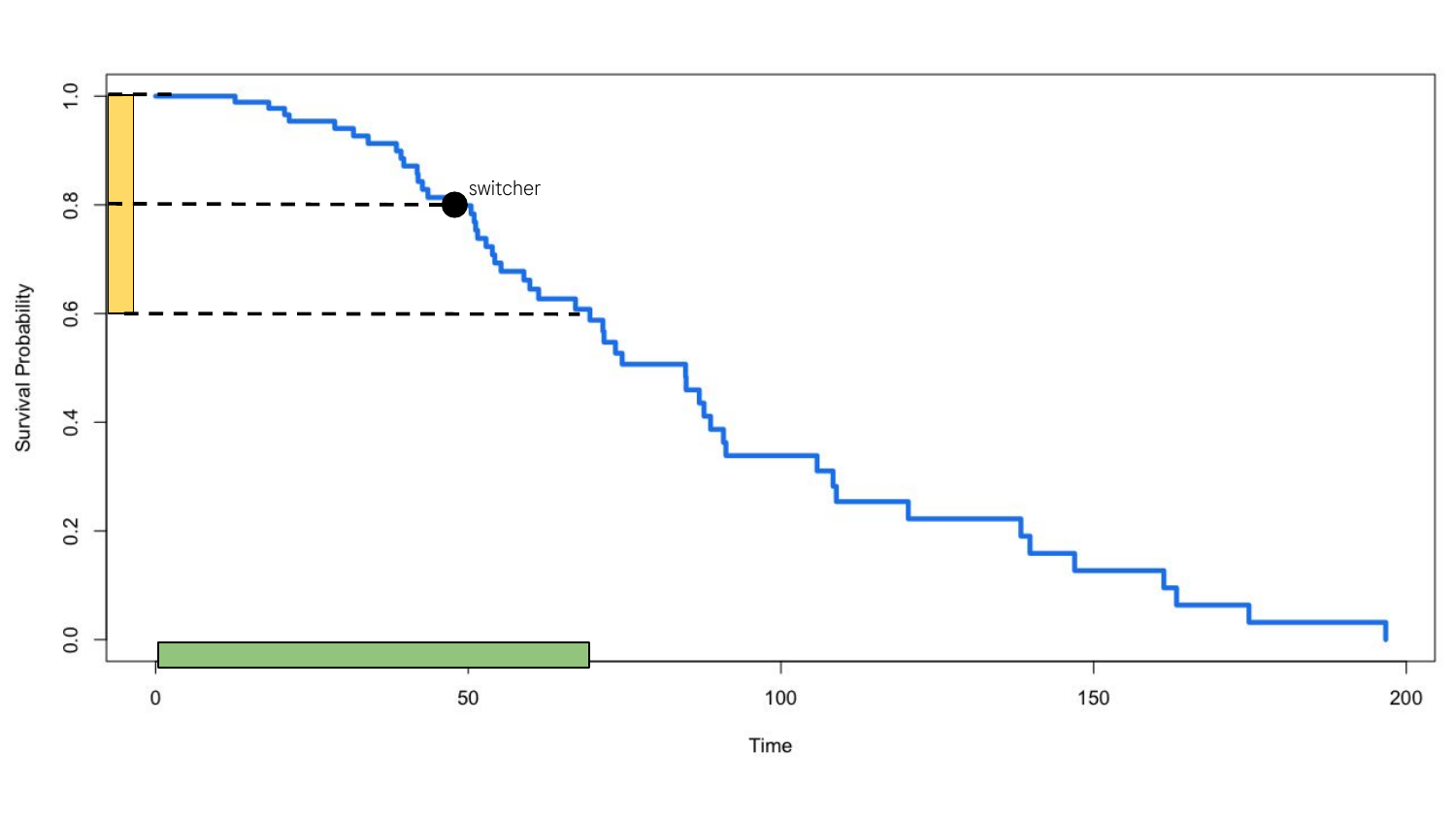}
    \caption{PD-aligned risk set selection method. A switcher’s progression time is mapped to its corresponding survival probability quantile on a Kaplan-Meier curve derived from EC progression events. A symmetric probability interval (yellow bar) is constructed around the switcher's position to identify ECs with comparable disease trajectories (green bar).}
    \label{fig:pd-window}
\end{figure}

To address this, we employ a progression-aligned EC selection strategy. We first construct a Kaplan-Meier (KM) curve using only EC patients who experienced disease progression, where the event itself is PD. 
For a given switcher with progression time $T_{\mathrm{PD}}$, we let $P = \hat{S}(T_{\mathrm{PD}})$ denote their corresponding survival probability  on this curve.
To identify comparable ECs, we construct a probability window around $P$ that is as wide as possible while remaining within $[0,1]$. Specifically, we set $d = \min(P, 1-P)$ and retain in the risk set $\mathcal{E}_i$ only those ECs whose progression times correspond to survival probabilities in $[P - d, P + d]$. This symmetric construction ensures that the window is centered on the switcher's quantile and avoids edge effects near $P=0$ or $P=1$.  For instance, a switcher at the 80th percentile ($P=0.8$) is aligned with ECs in the $[0.6,~1.0]$ range (shown in Figure \ref{fig:pd-window}), while a switcher at the 20th percentile ($P=0.2$) is aligned with the $[0.0,~0.4]$ range. This approach ensures that the switchers are compared to external counterparts at a similar stage of their disease trajectory and can be visualized in Figure \ref{fig:pd-window}. 
We note that the choice of a symmetric interval is a pragmatic default; alternative window rules (e.g., fixed-width quantile bands) could be substituted without changing the downstream weighting procedure.
Following this selection step, we apply balancing weights to the filtered EC pool to adjust for baseline covariates as described in the following section.

\subsection{Synthetic controls and balancing weights} \label{sec:synth-balwgt}
Synthetic control methods \citep{abadie2010synthetic,abadie2015comparative,abadie2021penalized} were developed for settings with one treated unit and a donor pool of untreated units. In our setting, the ``treated'' unit is a single switcher $i$ and the donor pool is the risk set of ECs selected using the steps described in Section \ref{sec:pd-adjust}, denoted as $\mathcal{E}_i$. 
The method forms a weighted combination of donors that matches the treated unit on pre-treatment characteristics \citep{abadie2010synthetic}. Importantly, we use the synthetic control method solely as a weight-construction device: we do not use (or require) the usual outcome-modeling structure (e.g., a linear factor model or a long pre-intervention outcome history). Instead, the weights feed into survival-based estimators described in Sections \ref{sec:ec-kmi-rsi} and \ref{sec:tvw}.

The construction of a weighted combination of ECs (the ``synthetic control'') to match the covariate profile of a single switcher is a special case of the \emph{balancing weights} framework in traditional causal inference where there exists only one ``treated'' unit. As mentioned in Section \ref{sec:leveraging-ec}, balancing weight methods solve an optimization problem to find weights that directly target in-sample covariate balance. 
Let $\mathcal{E}_i$ denote the unique risk set of ECs for switcher $i \in \left\lbrace \mathcal{R} : S_i = 1\right\rbrace$.
Following the developments in \citet{benmichael2021_lor,benmichael2021_drp,keele2023hospital,keele2025balancing,shen2025forest},
we consider the following balancing weights objective function for a single switcher $i$:
\begin{align}
\label{eqn:bal-wgt}
\min_{w_i} \underbrace{\left\lVert \sum_{j \in \mathcal{E}_i}  w_{ij} \mathbf{X}_j - \mathbf{X}_i \right\rVert^2_2}_{\text{imbalance}} + \lambda \underbrace{\sum_{j \in \mathcal{E}_i} w_{ij}^2}_{\text{dispersion}},
\end{align}
where $w_{ij}$ represents the weight of EC $j$ for switcher $i$. We pre-process each covariate $X_k$ to have mean 0 and variance 1 which assigns equal importance to each covariate, though this can be adjusted based on user preference \citep{shen2025forest}. This weight estimation procedure is then repeated for each switcher.
This objective represents a ``bias-variance tradeoff'' where the imbalance term measures the squared Euclidean covariate imbalance between switcher $i$ and a weighted combination of the risk set ECs, while the penalty term $\lambda$ controls the dispersion of the weights \citep{benmichael2021_lor,keele2025balancing,Hirshberg2019_amle,ben2023using}. Following \citet{benmichael2021_lor}, we set this hyperparameter by testing a grid of $\lambda$ values and selecting the one that results in the most reduction in bias (imbalance). 
See \citet{keele2025balancing} for further discussion on configuring the optimization problem.

\subsection{Multiple imputation methods} \label{sec:ec-kmi-rsi}
Once the weights are computed for each switcher's corresponding risk set $\mathcal{E}_i$, we can estimate the treatment effect under the hypothetical strategy (assuming no treatment switching) using imputation. The imputation step replaces each switcher’s unobserved counterfactual outcome with multiple draws from $\mathcal{E}_i$. We consider two multiple-imputation approaches: weighted risk set imputation (wRSI) and weighted Kaplan–Meier imputation (wKMI). For each imputed dataset, we estimate the treatment effect by fitting a Cox proportional hazards model with adjustment or stratification for background covariates as needed. We combine results across multiple imputed datasets using Rubin’s rules \citep{little2019statistical}.

\subsubsection{Weighted risk set imputation.}
Risk-set imputation (RSI) samples an external control at random from the risk set and replaces the switcher’s observed outcome using the sampled EC's event time and event indicator. This method was first proposed in general censored survival data settings \citep{taylor2002survival,hsu2006survival}. RSI samples uniformly from the risk set, which treats all at-risk ECs as equally plausible matches for the switcher. However, when baseline risk or covariate profiles vary within the risk set, this assumption fails and uniform RSI can be biased.
Our weighted proposal for RSI \texttt{(wRSI)} samples ECs with probabilities proportional to the normalized balancing weights estimated from Equation \eqref{eqn:bal-wgt} so that ECs with more similar covariate profiles to the switcher are selected more often. Repeating this procedure yields multiple imputations, which we carry forward to estimation and pooling as described above. A step-by-step procedure is outlined below:

\begin{algorithm}[H]
\caption{Weighted risk set imputation (RSI)}\label{alg:rsi}
\begin{algorithmic}[1]
\For{each switcher $i = 1 \dots n_s$}
    \State \parbox[t]{\dimexpr\linewidth-\algorithmicindent\relax}{%
        Define a risk set $\mathcal{E}_i$ which consists of the $K$ nearest ECs $k = 1 \dots K$ with observed time ${T}_{\mathrm{OS},k} \geq {T}_{\mathrm{OS},i}$. In general, we set $K = n_e$ so that the entire pool of ECs is used.\strut}
    \State \parbox[t]{\dimexpr\linewidth-\algorithmicindent\relax}{%
        Using Equation \eqref{eqn:bal-wgt}, estimate the balancing weights for each EC in $\mathcal{E}_i$.\strut}
    \State \parbox[t]{\dimexpr\linewidth-\algorithmicindent\relax}{%
        Draw an EC $j$ from $\mathcal{E}_i$ at random, where the probability of drawing EC $j$ is $w_{ij} / \sum_{k \in \mathcal{E}_i} w_{ik}$.\strut} 
    \State \parbox[t]{\dimexpr\linewidth-\algorithmicindent\relax}{%
        Replace switcher $i$'s time and event status pair with that of EC $j$.\strut}
\EndFor
\State \parbox[t]{\linewidth}{%
    Estimate the treatment effect for the imputed dataset via a Cox proportional hazards model.\strut}
\State \parbox[t]{\linewidth}{%
    Repeat the entire imputation and estimation process multiple times and combine the results using Rubin's rules.\strut}
\end{algorithmic}
\end{algorithm}

\begin{figure}[ht]
    \centering
    \hspace*{-1cm} 
    \includegraphics[width=1.1\linewidth]{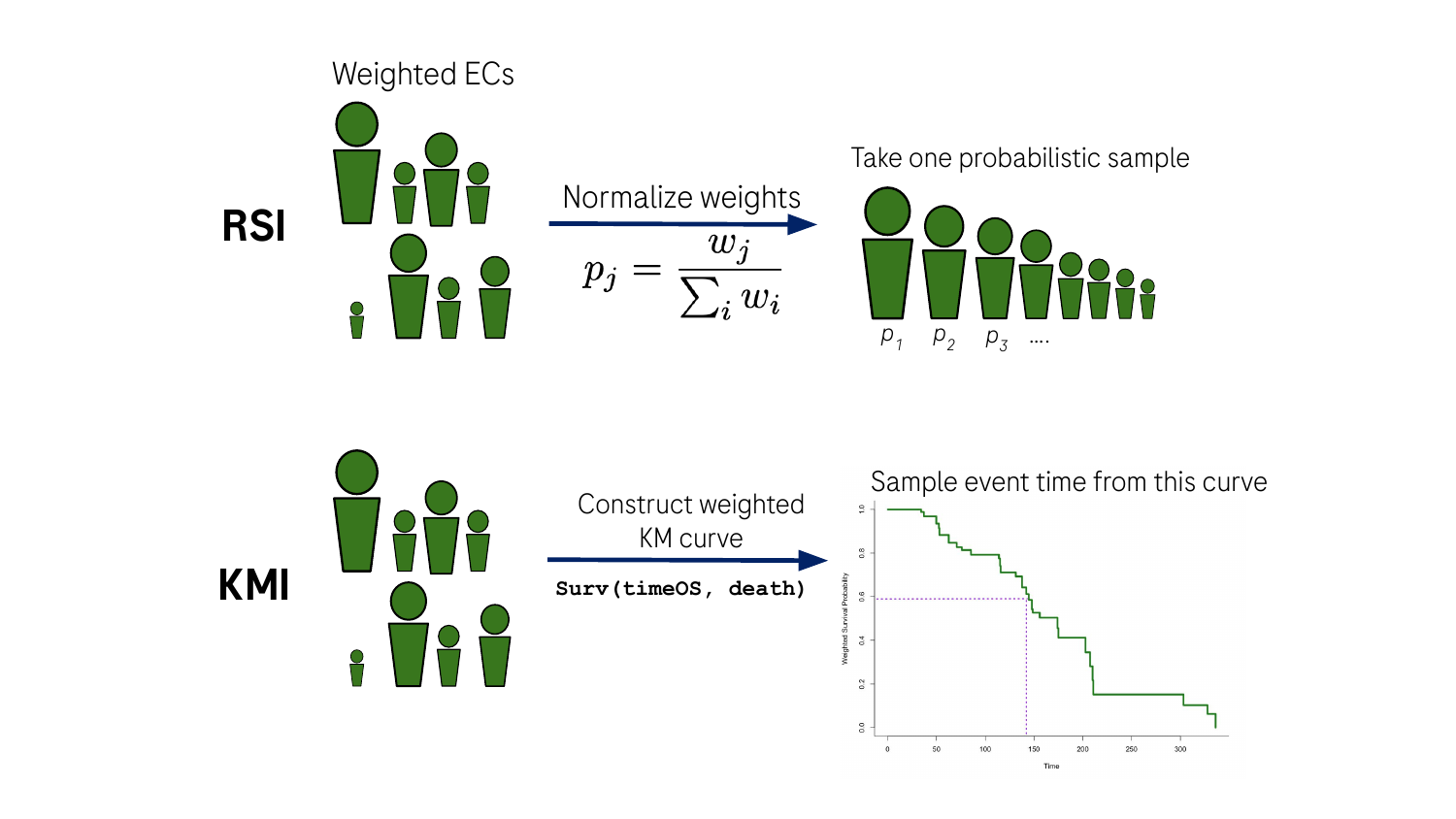}
    \caption{Risk-set imputation method (RSI, top) and Kaplan-Meier imputation method (KMI, bottom). In RSI, ECs are randomly sampled based on their weight (ECs with larger weight are more likely to be selected). In KMI, a weighted Kaplan-Meier curve is drawn using the ECs, and an event time is sampled from the CDF implied by the curve.}
    \label{fig:rsi-kmi}
\end{figure}

\subsubsection{Weighted Kaplan-Meier imputation.}
Kaplan-Meier imputation mirrors RSI by utilizing the EC risk set as a donor pool, but samples from an estimated weighted survival curve instead of individual ECs. For each switcher $i$, we first estimate a Kaplan-Meier curve using $\mathcal{E}_i$. Our weighted version \texttt{(wKMI)} fits a weighted Kaplan–Meier estimator using the balancing weights estimated from Equation \eqref{eqn:bal-wgt}. We draw $u \sim \text{Uniform}(0,1)$ and set the imputed event time to $\hat{S}_i^{-1}(u) = \inf\{t : \hat{S}_i(t) \leq u\}$, where $\hat{S}_i$ is the weighted Kaplan–Meier survival function estimated from $\mathcal{E}_i$. In general, Kaplan-Meier imputation will impute event times. However, if the Kaplan-Meier curve ends above 0 because the final observed time(s) are censored, then values of $u$ resulting in draws at or beyond the last event time are imputed as censored. We repeat this procedure for each switcher to form one imputed dataset and repeat the entire procedure to generate multiple imputations. We then fit the Cox model described above and pool estimates using Rubin’s rules. A step-by-step procedure is provided below. For theory on Kaplan-Meier and weighted Kaplan-Meier imputation, see \citet{murray1996nonparametric,taylor2002survival,hsu2006survival}.

\begin{algorithm}[H]
\caption{Weighted Kaplan--Meier Imputation (wKMI)}\label{alg:wkmi}
\begin{algorithmic}[1]
\For{each switcher $i = 1,\dots,n_s$}
    \State \parbox[t]{\dimexpr\linewidth-\algorithmicindent\relax}{%
        Define a risk set $\mathcal{E}_i$ which consists of the $K$ nearest ECs $k = 1,\dots,K$ with observed time ${T}_{\mathrm{OS},k} \geq {T}_{\mathrm{OS},i}$. In general, we set $K = n_e$ so that the entire pool of ECs is used.\strut}
    \State Using Equation \eqref{eqn:bal-wgt}, estimate the balancing weights for each EC in $\mathcal{E}_i$.
    \State \parbox[t]{\dimexpr\linewidth-\algorithmicindent\relax}{%
        Construct a weighted Kaplan--Meier curve using the ECs in $\mathcal{E}_i$, with weights $\{w_k : k \in \mathcal{E}_i\}$.\strut} 
    \State \parbox[t]{\dimexpr\linewidth-\algorithmicindent\relax}{%
        Draw $u \sim \mathrm{Unif}(0,1)$ and sample an imputed time from the implied CDF (equivalently, the corresponding quantile of the weighted Kaplan--Meier curve).\strut} 
    \State \parbox[t]{\dimexpr\linewidth-\algorithmicindent\relax}{%
        If the draw falls beyond the last event time (i.e., the curve ends above zero because the final time is censored), impute a censored outcome at the last observed time in $\mathcal{E}_i$.\strut}
    \State Replace switcher $i$'s time and event pair with the imputed pair.
\EndFor
\State \parbox[t]{\linewidth}{%
    Estimate the treatment effect for the imputed dataset via a Cox proportional hazards model.\strut}
\State \parbox[t]{\linewidth}{%
    Repeat the entire imputation and estimation process multiple times and combine the results using Rubin's rules.\strut}
\end{algorithmic}
\end{algorithm}

\newcommand{\switchtimeordered}{t_{(i)}^{S}}
\newcommand{\calEiorder}{\mathcal{E}_{(i)}}

\subsection{Time-varying balancing weights} \label{sec:tvw}
\begin{figure}[ht]
    \centering
    \includegraphics[width=0.95\linewidth]{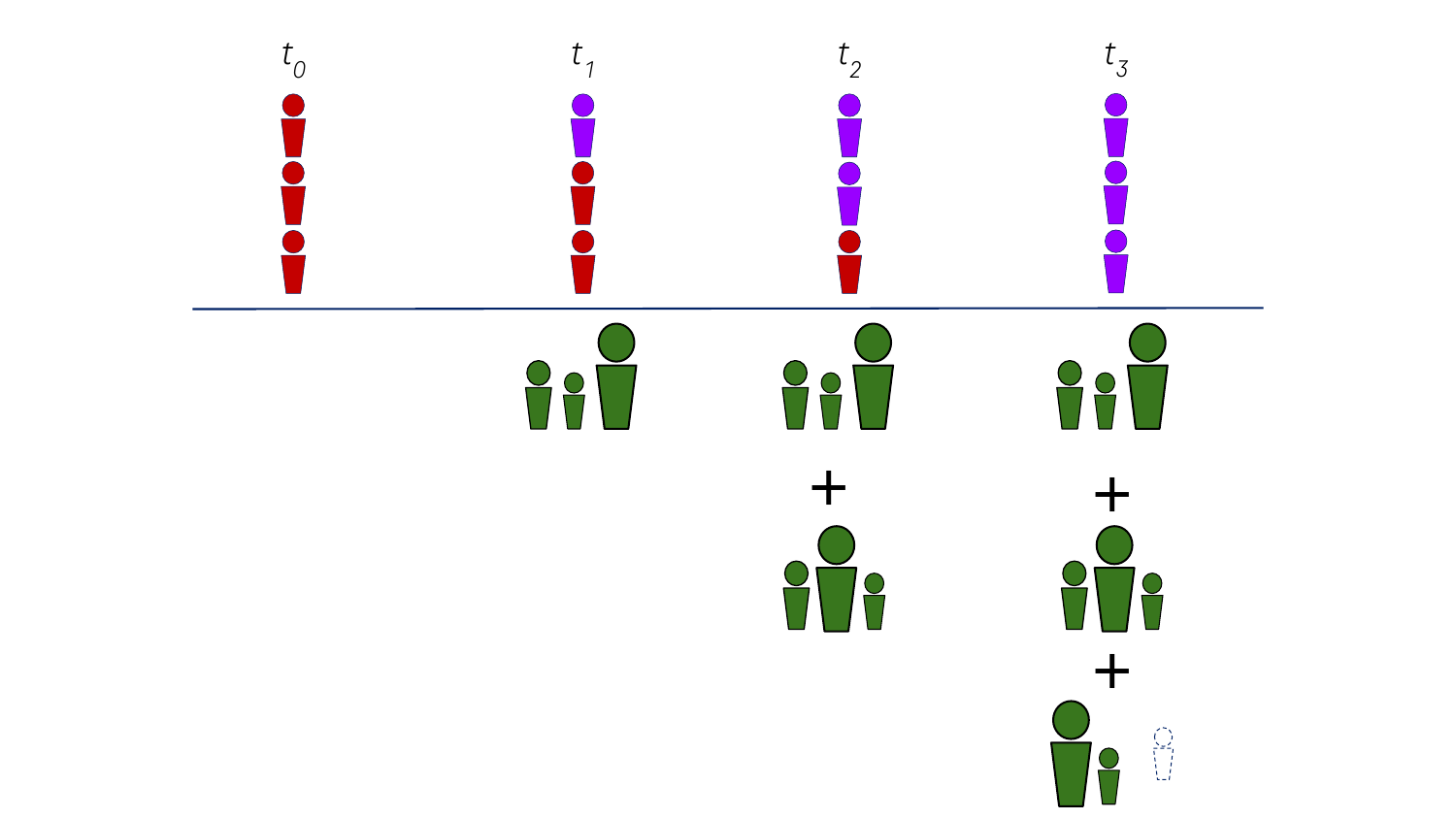}
    \caption{Time-varying balancing weights method. Internal switchers (top row) contribute to the control arm while in the original therapy state (red) and are substituted by weighted external controls (green) upon switching (purple). At each sequential switch time ($t_1, t_2, t_3$), EC weights are aggregated, increasing the contribution of each EC to compensate for the loss of information from switching. If an EC unit experiences death or is censored in its source study (hollow figure at $t_3$), it exits the risk set and its weight becomes zero.}
    \label{fig:tvw-fig}
\end{figure}

Our last method incorporates EC information \emph{after switching occurs}; in particular, switchers contribute internal trial data prior to their switch time and are replaced by weighted EC counterparts thereafter. 
Specifically, we allow for EC contribution to increase over time as more patients switch by aggregating the estimated weights across switch times. This strategy avoids the loss of prognostic information caused by simple censoring while ensuring that the external data only compensates for the information gaps introduced by switching. We refer to this as the ``time-varying balancing weights'' method (TVW).

Let $t_i^{S}$ denote the switch time for all switchers $i \in \left\lbrace \mathcal{R}: S_i = 1 \right\rbrace$ and let $\switchtimeordered$ denote the $i$th order statistic of the switching time; in other words, $\switchtimeordered$ is the $i$th smallest switching time. To simplify notation, we relabel switchers by their switch times and write
$t_{(1)}^S \le \cdots \le t_{(n_s)}^S$ for the ordered switch times.
Let $\calEiorder$ represent the imputing EC risk set for the $i$th smallest switching time. 
Similar to the procedures in wRSI and wKMI, we estimate a vector of weights with respect to each switcher's imputing risk set $\mathcal{E}_i$, such that $w_{(i) j}$ corresponds to the weight for EC $j \in \calEiorder$. Instead of imputing from this weighted risk set or drawing a Kaplan-Meier curve, we aggregate these weights based on switch times so that the EC contribution grows as more switchers are censored due to switching.

We define the \emph{time-varying weight} of EC $j$ at time $t$ as:
\begin{align}
    w_{j}(t) = \begin{cases}
        0 \quad \text{if } t < t_{(1)}^S \\
        0 \quad \text{if } t \geq {T}_{\mathrm{OS},j} \\
        \sum\limits_{k=1}^i w_{(k)j} \quad \text{if } t_{(i)}^S \leq t < t_{(i+1)}^S \forall i = 1 \dots n_s.        
    \end{cases}
\end{align}
Because this weight aggregates across all switchers, we drop the switcher subscript $i$ and write $w_{j}(t)$.
This can be equivalently expressed as
\begin{align}
  w_{j}(t)=\mathbbm{1}\{t \geq t^{S}_{(1)}\}\,\mathbbm{1}\{t < {T}_{\mathrm{OS},j}\}\sum_{i=1}^{n_{s}}\mathbbm{1}\{j\in\mathcal{E}_{(i)}\}\,\mathbbm{1}\{t \geq t^{S}_{(i)}\}\,w_{(i)j}.
\end{align}

Concretely, $w_{j}(t)$ accounts for the fact that each EC must contribute greater weight as the number of switchers increases, to compensate for post-switch information loss.
Prior to the first switching event ($t < t_{(1)}^S$), all switchers remain in the analysis and ECs contribute zero weight to the analysis. 
 Once the first switcher crosses over, EC $j$ contributes weight $w_{(1) j}(t)$. If an EC unit $j$ experiences death or is censored in its source study ($t > \Tilde{T}_j$), it exits the risk set and carries a weight of 0 thereafter.

Once these time-varying weights are computed, we estimate the treatment effect using a weighted Cox proportional hazards model in the counting process format. Substituting weighted EC information for switchers post-switch provides a dynamic estimate of the control arm's survival trajectory while maintaining trial-specific data until switching occurs. 

\begin{algorithm}[H]
\caption{Time-varying balancing weights (TVW)}\label{alg:tvw}
\begin{algorithmic}[1]
\For{each switcher $i = 1,\dots,n_s$ with switch time $t_i^S$}
    \State \parbox[t]{\dimexpr\linewidth-\algorithmicindent\relax}{%
        Define a risk set $\mathcal{E}_i$ which consists of the $K$ nearest ECs $k = 1,\dots,K$ with observed time $\tilde{T}_k \ge t_i^S$. In general, we set $K = n_e$ so that the entire pool of ECs is used.\strut}
    \State \parbox[t]{\dimexpr\linewidth-\algorithmicindent\relax}{%
        Using Equation \eqref{eqn:bal-wgt}, estimate the balancing weights for each EC in $\mathcal{E}_i$, and denote these weights by $\{w_{(i)j} : j \in \mathcal{E}_i\}$.\strut}
\EndFor
\State \parbox[t]{\linewidth}{%
    Order the switch times as $t_{(1)}^S \le \cdots \le t_{(n_s)}^S$ and, for ties, treat all switchers with the same switch time as switching simultaneously.\strut}
\For{each EC $j$}
    \State \parbox[t]{\dimexpr\linewidth-\algorithmicindent\relax}{%
        Construct a time-varying weight function $w_j(t)$ by aggregating switcher-specific weights across switch times. For $t < t_{(1)}^S$ set $w_j(t)=0$; for $t \ge \tilde{T}_j$ set $w_j(t)=0$; and for $t_{(i)}^S \le t < t_{(i+1)}^S$ set $w_j(t)=\sum_{k=1}^{i} 1\{j \in \mathcal{E}_{(k)}\} w_{(k)j}$.\strut}
\EndFor
\State \parbox[t]{\linewidth}{%
    Fit a weighted Cox proportional hazards model (using the counting process data format), including trial experimental-arm participants and internal controls with weight 1, censoring control-arm switchers at their switch times, and including ECs as controls with time-varying weights $w_j(t)$.\strut}
\end{algorithmic}
\end{algorithm}

\subsection{Summary of methods} \label{sec:method-summary}
Table \ref{tab:methods_summary} summarizes the statistical methods discussed so far. For clarity, we categorize these approaches into three groups: traditional methods, population-level EC reweighting, and switcher-specific EC reweighting. We investigate the performance of these methods in Section \ref{sec:simulation} (simulation) and Section \ref{sec:impower130} (empirical application).

\begin{table}[ht]
\centering
\resizebox{\textwidth}{!}{%
\begin{tabular}{ccp{7cm}}
\toprule
\textbf{Category} & \textbf{Method} & \textbf{Description} \\
\midrule
\textbf{Traditional} & ITT & Treatment policy strategy \\
 & Censoring Switchers & Censors switchers at switch time \\
 & RPSFTM & Rank-preserving structural failure time model \\
 & IPCW & Inverse probability of censoring weighting \\
\midrule
\textbf{Population-Level EC} & EC-Treatment Weighting & Reweights all ECs to match the treatment arm (baseline covariates) \\
 & EC-Switcher Weighting & Reweights ECs with PD ($D_i=1$) to match switchers (baseline covariates) \\
\midrule
\textbf{Switcher-Specific EC} & Weighted Risk-Set Imputation (wRSI) & Randomly select EC from $\calEi$ \\
 & Weighted Kaplan-Meier Imputation (wKMI) & Create KM curve with $\calEi$ and impute from implied CDF \\
 & Time-Varying Balancing Weights (TVW) & Substitutes switchers post-switch with dynamically weighted ECs \\
\bottomrule
\end{tabular}%
}
\caption{Summary of methods compared in this paper. Traditional methods use only internal trial data. Population-level EC methods reweight the entire EC cohort to a target population. Switcher-specific EC methods (our contribution) construct per-switcher weights from the EC risk set.}
\label{tab:methods_summary}
\end{table}




\section{Simulation Study} \label{sec:simulation}
We conduct a simulation study to compare the methods described in Section \ref{sec:method-summary}. The design varies the strength of unmeasured confounding and the correlation between progression and overall survival to test robustness. We implement the IPCW and RPSFTM methods using the \texttt{trtswitch} package in \texttt{R} \citep{kaifengtrtswitch}. We use the \texttt{balancer} package \citep{benmichael2021_lor} in \texttt{R} to estimate balancing weights for the population-level and switcher-specific EC reweighting methods. 

\subsection{Simulation setup} \label{sec:sim-setup}
Our simulation is adapted from the atezolizumab metastatic lung cancer trials \citep{west2019atezolizumab,rittmeyer2017atezolizumab,mazieres2021atezolizumab}; we refer to this manuscript for further details. 
In particular, we adapt the parameters of the data-generating process (DGP) to reflect the results of this trial, which we investigate in Section \ref{sec:impower130}. See Table \ref{tab:simulation_covariates} for a summary of this DGP. See Appendix Section \ref{sec:existing-methods} for a description of the traditional methods (ITT, censoring switchers, RPSFTM, IPCW).

In this simulation, our internal trial arm consists of 600 patients with 1:1 randomization into treatment ($Z=1$) and control ($Z=0$) groups. The treatment effect, defined as hazard ratios, for PD and OS are 0.64 and 0.72, respectively. The median PD and OS times in the control group are 6 and 14 months, respectively. 

We generate four baseline covariates $X_1 \dots X_4$ as follows:
\begin{align*}
    &X_1 \sim \mathrm{Bern}(0.5)\\
    &X_2 \sim \mathrm{Bern}(0.5)\\
    &X_3 \sim \mathrm{Cat}(1/3,~ 1/3,~1/3)\\
    &X_4 \sim \text{LogNormal}(\log(80), ~0.66).
\end{align*}
In this simulation $X_1$ and $X_2$ represent age (1 for 65+, 0 otherwise) and sex (1 for female, 0 for male), respectively. The respective conditional hazard ratios for $X_1$ and $X_2$ are 0.9 and 0.8.
$X_3$ represents a generic biomarker with three expression levels: low (reference), median, and high. The hazard ratio of the ``median'' group is 0.8 and 0.7 for the ``high'' group. Finally, $X_4$ represents any generic continuous variable with hazard ratio 1.005, such as the ``sum of longest diameters of target tumor lesions (SLD)'' \citep{RECISTv1.1}. 
The hazard ratios for all covariates are equivalent for both PD and OS. In addition to the observed covariates, we also include an unmeasured confounder $U \sim \mathrm{Bern}(0.5)$ with varying strength, discussed further in Section \ref{sec:factors-vary-sim}.

\begin{table}[ht]
\centering
\begin{tabular}{lllc}
\toprule
\textbf{Variable} & \textbf{Description} & \textbf{Distribution} & \textbf{Hazard Ratio (PD \& OS)} \\
\midrule
$X_1$ & Age ($\geq65$ vs. $< 65$) & $\text{Bern}(0.5)$ & 0.90 \\
$X_2$ & Sex (Female vs. Male) & $\text{Bern}(0.5)$ & 0.80 \\
$X_3$ & Biomarker (Median vs. Low) & $\text{Cat}(1/3, 1/3, 1/3)$ & 0.80 \\
      & Biomarker (High vs. Low) &  & 0.70 \\
$X_4$ & Tumor Burden (SLD) & $\text{LogNormal}(\ln(80), 0.66)$ & 1.005 \\
$U$   & Unmeasured Confounder & $\text{Bern}(0.5)$ & Varying \\
\midrule
$Z$ & Binary Treatment & 1:1 Randomization & 0.72 (OS) \\
          &                          &                   & 0.64 (PD) \\
\bottomrule
\end{tabular}
\caption{Summary of baseline covariate distributions and conditional hazard ratios used in the simulation study.}
\label{tab:simulation_covariates}
\end{table}

PD and OS are usually positively correlated as it represents the aggressive nature of the disease --- those who experience shorter PD times are more likely to experience shorter death times. As a result, we generate PD and OS survival times using a bivariate Gaussian copula to represent such correlation \citep{othus2010gaussian,weber2019quantifying}. Specifically, we generate latent correlated uniform variables $(u_{\mathrm{PD}}, u_{\mathrm{OS}})$ from a bivariate Gaussian copula with correlation parameter $\rho$. These variables are mapped to survival times using exponential marginal distributions, where the rate parameters $\lambda_{\mathrm{PD}}$ and $\lambda_{\mathrm{OS}}$ are defined as the exponential of the linear predictor: for $g \in \left\lbrace \text{PD}, \text{OS} \right\rbrace$, we have $\lambda_{g} = \exp(\eta_{g})$, where
\begin{equation}
\label{eqn:linear-copula}
    \eta_{g} = \beta_{0g} + \beta_g ^{\top} X + \beta_{ug} U + \beta_{zg} Z.
\end{equation}
This specification directly incorporates the conditional hazard ratios for treatment (e.g., $\exp(\beta_{\mathrm{OS}}) = 0.72$ for OS) and baseline covariates into the scale of the survival times. The strength of this association is controlled by $\rho$, allowing us to simulate varying degrees of simultaneity bias and residual correlation unaccounted for by covariates (see Section \ref{sec:factors-vary-sim}).
Finally, we enforce a logical constraint where progression cannot occur after death by censoring PD events if the simulated OS time precedes the PD time. In Equation \eqref{eqn:linear-copula}, all coefficients except $\beta_{0k}$ are fixed; we calibrate the baseline hazard intercepts ($\beta_{0k}$) such that the true median PD time in the control group is around 6 months and the true median OS time in the control group is around 14 months, as mentioned above. 
We generate independent dropout times and apply administrative censoring at clinical data cut, defined as the time when 420 OS events (70\% of the 600 patients) were observed in the simulated trial. Regarding treatment switching, we assume a switching probability of 0.5 for each control arm patient experiencing PD. For switchers, we generate their switch time as a random uniform number between 0 and 2 to represent that switching only occurs within two months after PD.  

The EC cohort consists of 1000 patients in order to reflect a wider patient pool than the internal trial arm. We vary several aspects of the EC arm in our simulation (see Section \ref{sec:factors-vary-sim}), which primarily concern the covariate distributions and confounding structures. Namely, however, we assume the copula for generating OS and PD survival distributions are the same in both the internal trial and external control arms, leaving deviations of this assumption to future work. 

We consider a variety of treatment switching methods, including those with internal trial data only and those involving ECs; see Section \ref{sec:method-summary} for a complete list of methods we assess.
For each method, we estimate the treatment effect using a Cox proportional hazards model, stratifying on all binary and categorical covariates (all covariates except $X_4$). We use the robust sandwich variance estimator for inference \citep{ding2024linear}.  We run 5000 Monte Carlo replications for each method. In addition to measuring the bias between the true and estimated hazard ratios, we also measure the Type I error and power of each method.

\subsection{Factors to vary in simulations} \label{sec:factors-vary-sim}
We vary several factors of the data-generating process to assess robustness --- 
a critical factor impacting the validity of EC-based analyses is the compatibility of the EC and trial populations. These factors control the baseline covariate distributions between EC and internal trial groups, the correlation between PD and OS as generated by the copula model, and the strength of unmeasured confounding, expressed as omitted variable bias.

\paragraph{EC baseline covariate prevalence.} We vary the prevalence of certain groups of covariates in the EC population by changing the parameters of the covariate DGP introduced in the previous section to create ``healthier'' and ``sicker'' EC populations. For instance, we can adjust the Bernoulli parameter of $X_1$ from 0.5 to 0.25 to resemble a ``healthier'' cohort and to 0.75 to resemble a sicker cohort as a result of the hazard ratios being less than 1. See Section \ref{sec:app-ec-dgp} in the appendix for the full covariate DGP for these EC populations. 

\paragraph{PD/OS correlation (simultaneity).} We also measure the (conditional) correlation $\rho$ between PD and OS in the Copula model. This controls the residual dependence between PD and OS after conditioning on $X$ and $U$: larger values of $\rho$ imply that earlier progression is more predictive of earlier death. In econometrics, this phrase is commonly referred to as ``simultaneity bias'' \citep{zaefarian2017endogeneity}. We consider three correlation values: $\rho = 0$ (no dependence), $\rho=0.5$ (moderate dependence), $\rho=0.9$ (strong dependence). As a result of this correlation structure, certain estimators, including censoring of switchers and IPCW, will be biased by informative censoring. 

\paragraph{Unmeasured confounding (omitted variable bias).} In addition to the observed covariates, we also introduce a binary baseline covariate $U$ that is unseen to the analyst but still carries linear prognostic value $\beta_u$ for predicting PD and OS. In particular, we vary $U$ such that its associated hazard ratios are $\exp(\beta_u) = 1$ (no unmeasured confounding), $\exp(\beta_u) = 0.6$ (moderate unmeasured confounding), and $\exp(\beta_u) = 0.2$ (high unmeasured confounding). This is typically referred to in the causal inference and econometrics literatures as omitted variable bias (OVB) \citep{zaefarian2017endogeneity,cinelli2020making}.

The biases resulting from omitted variables and simultaneity are common sources of \emph{endogeneity}, which broadly refers to any situation where the cause-and-effect relationship between two variables is disentangled due to unaccounted errors in the modeling process, resulting biased treatment effect estimates \citep{zaefarian2017endogeneity}. 
Our simulation considers these sources of endogeneity to assess the extent to which methods are biased by these sources.  

\subsection{Simulation results} \label{sec:sim-results}
\begin{figure}[ht]
    \centering
    \begin{subfigure}[b]{0.49\linewidth}
        \centering
        \includegraphics[width=\linewidth]{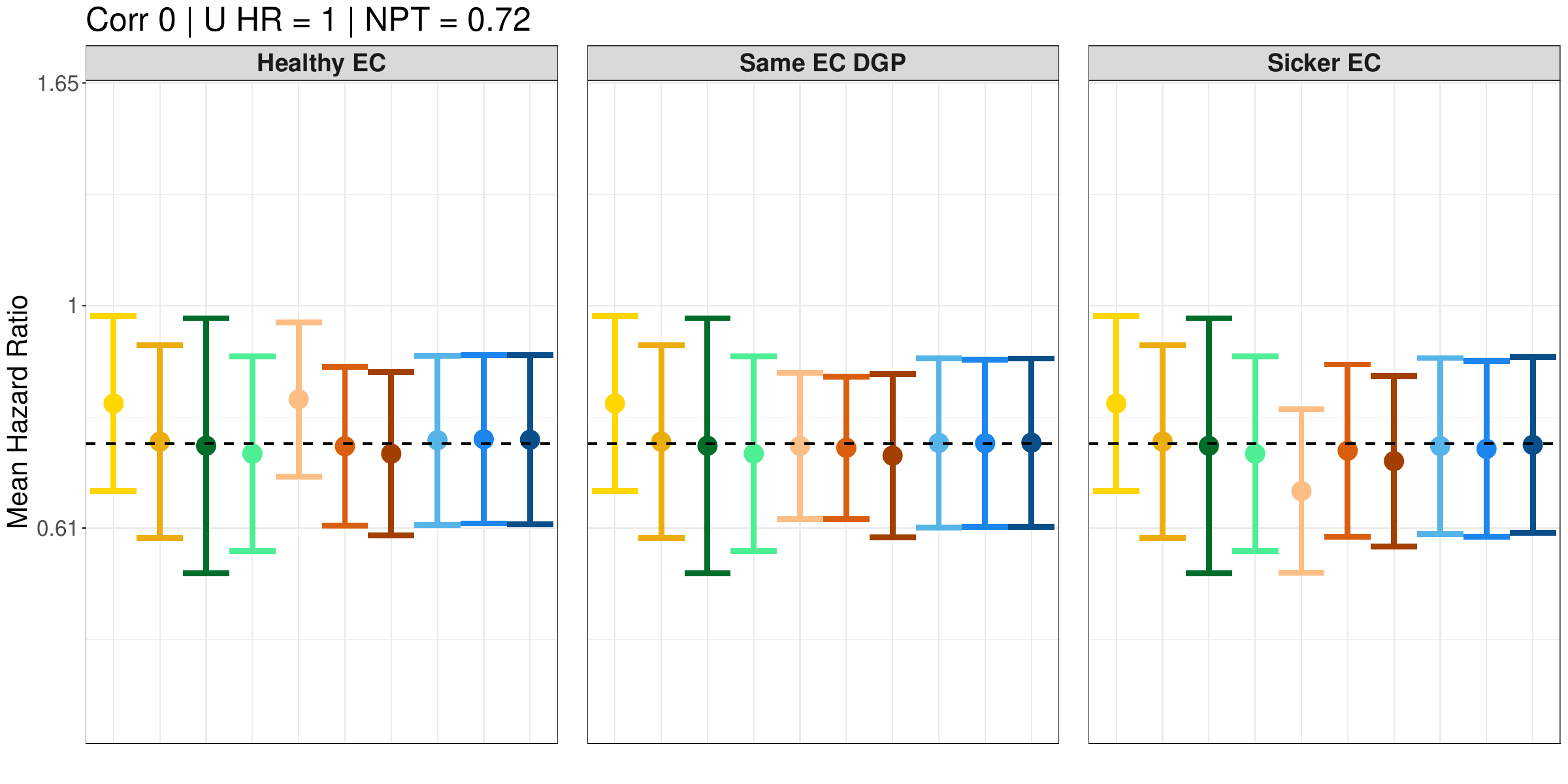}
        \caption{Hazard Ratio estimates}
        \label{fig:corr0-u1-npt72-a}
    \end{subfigure}
    \hfill
    \begin{subfigure}[b]{0.49\linewidth}
        \centering
        \includegraphics[width=\linewidth]{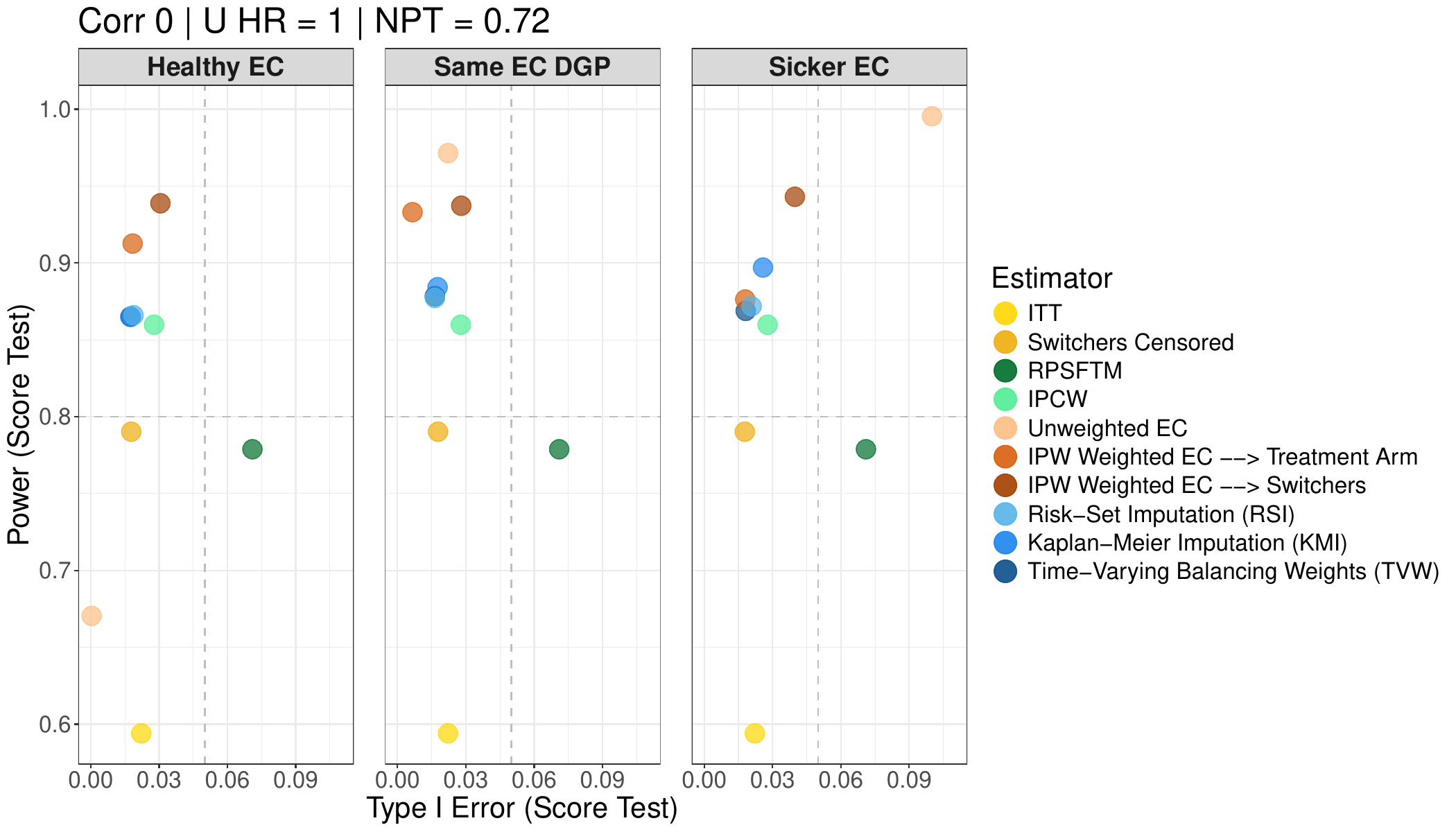}
        \caption{Power and Type I error estimates}
        \label{fig:corr0-u1-npt72-b}
    \end{subfigure}
    
    \caption{Simulation results for $\rho = 0$, $\exp(\beta_u) = 1$ and constant NPT. Individual panels represent covariate distribution of ECs relative to internal controls (e.g., healthier/sicker/same DGP). The first four estimates in each panel are identical since they do not use EC data. Refer to Section \ref{sec:method-summary} for a description of all methods used. Precise Type I error rates for estimators exceeding the visual truncation threshold of 0.1 include: Unweighted EC in the Sicker EC panel (0.212).}
    \label{fig:corr0-u1-npt72}
\end{figure}

\begin{figure}[ht]
    \centering
    \begin{subfigure}[b]{0.49\linewidth}
        \centering
        \includegraphics[width=\linewidth]{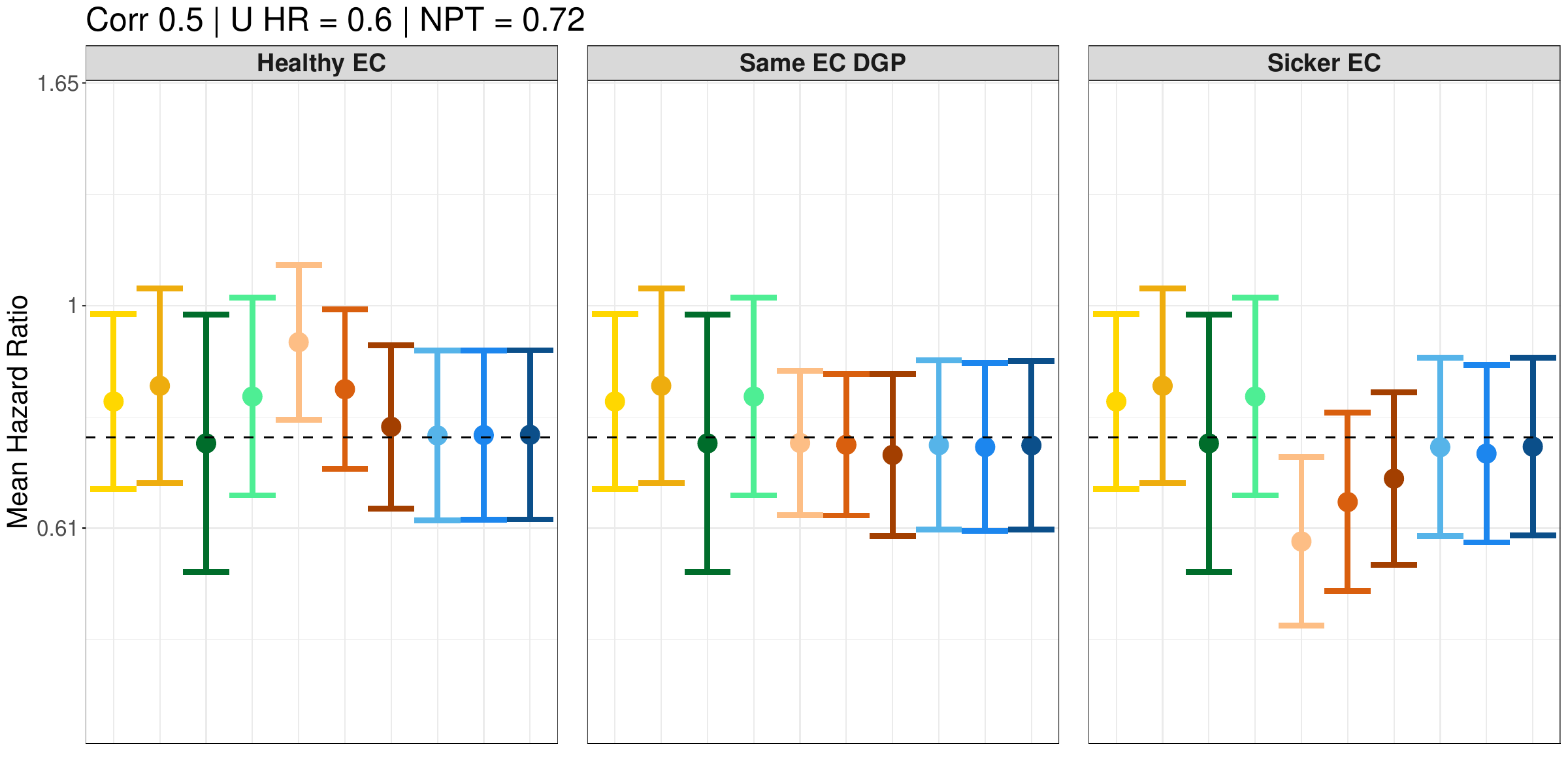}
        \caption{Hazard Ratio estimates}
        \label{fig:corr0.5-u0.6-npt72-a}
    \end{subfigure}
    \hfill
    \begin{subfigure}[b]{0.49\linewidth}
        \centering
        \includegraphics[width=\linewidth]{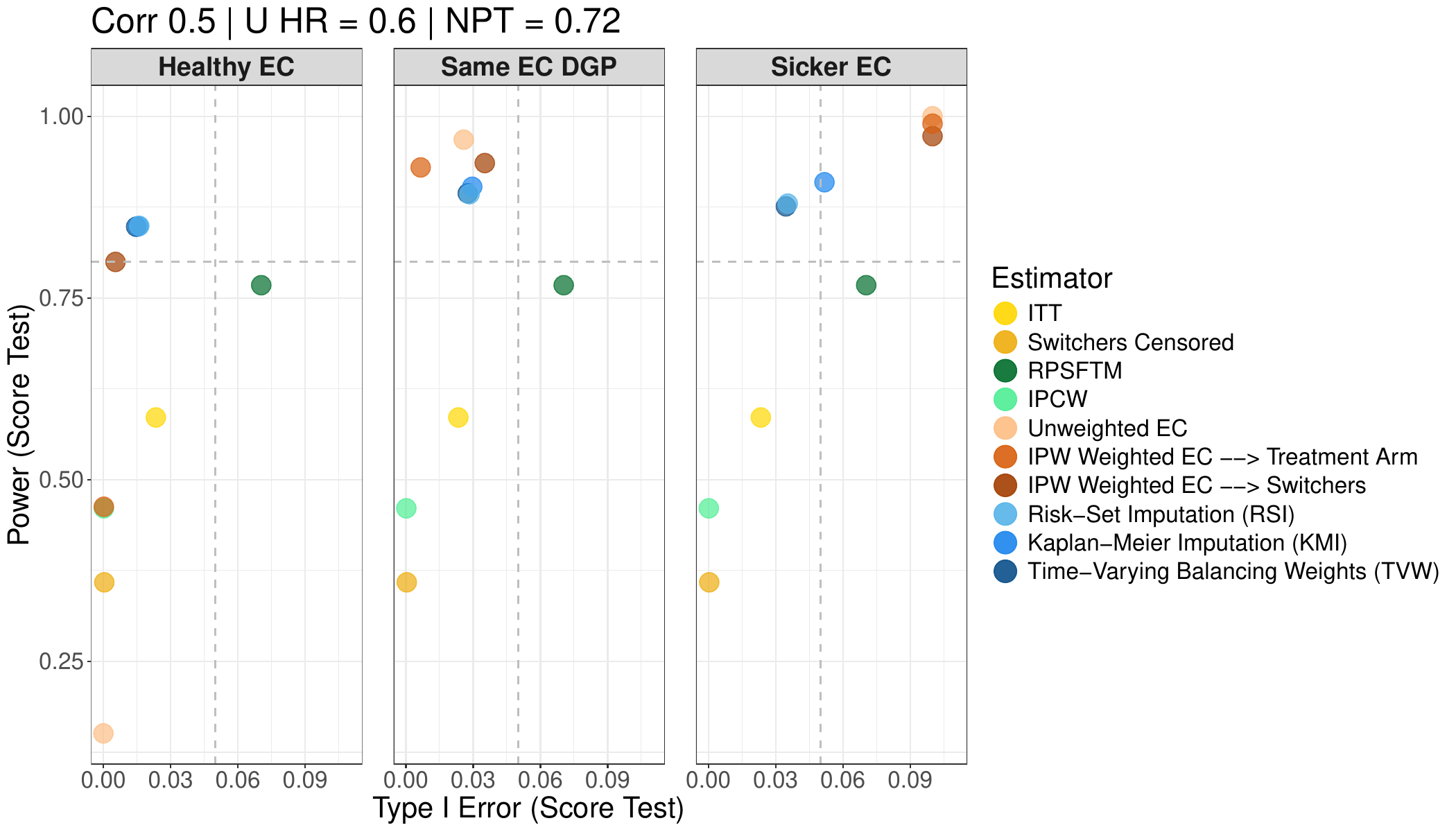}
        \caption{Power and Type I error estimates}
        \label{fig:corr0.5-u0.6-npt72-b}
    \end{subfigure}
    
    \caption{Simulation results for $\rho = 0.5$, $\exp(\beta_u) = 0.6$ and constant NPT. Individual panels represent covariate distribution of ECs relative to internal controls (e.g., healthier/sicker/same DGP). The first four estimates in each panel are identical since they do not use EC data. Refer to Section \ref{sec:method-summary} for a description of all methods used. Precise Type I error rates for estimators exceeding the visual truncation threshold of 0.1 include: Unweighted EC in the Sicker EC panel (0.692), Weighted EC (treatment arm) in the Sicker EC panel (0.205) and Weighted EC (switchers) in the Sicker EC panel (0.11).}
    \label{fig:corr0.5-u0.6-npt72}
\end{figure}

\begin{figure}[ht]
    \centering
    \begin{subfigure}[b]{0.49\linewidth}
        \centering
        \includegraphics[width=\linewidth]{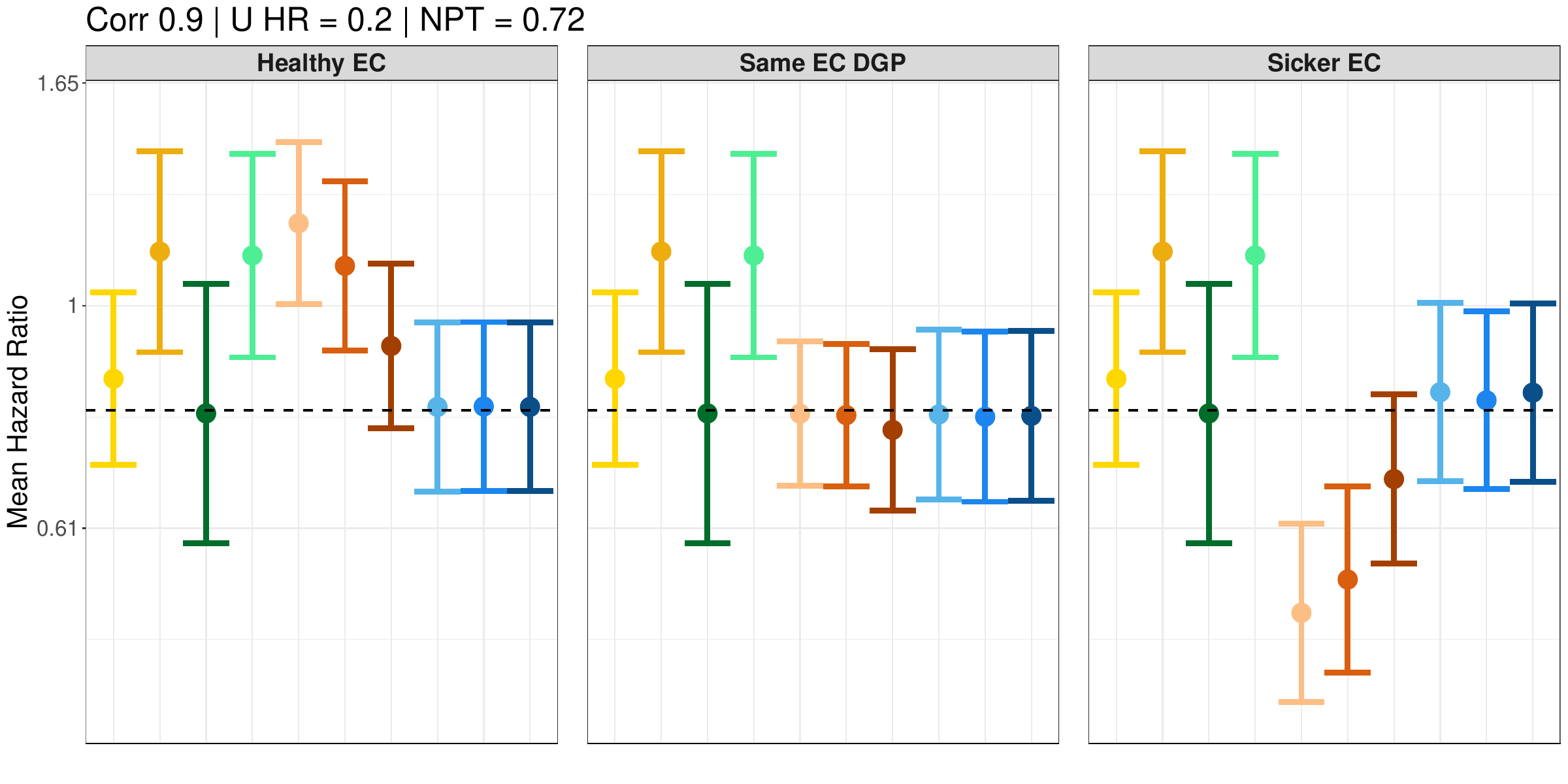}
        \caption{Hazard Ratio estimates}
        \label{fig:corr0.9-u0.2-npt72-a}
    \end{subfigure}
    \hfill
    \begin{subfigure}[b]{0.49\linewidth}
        \centering
        \includegraphics[width=\linewidth]{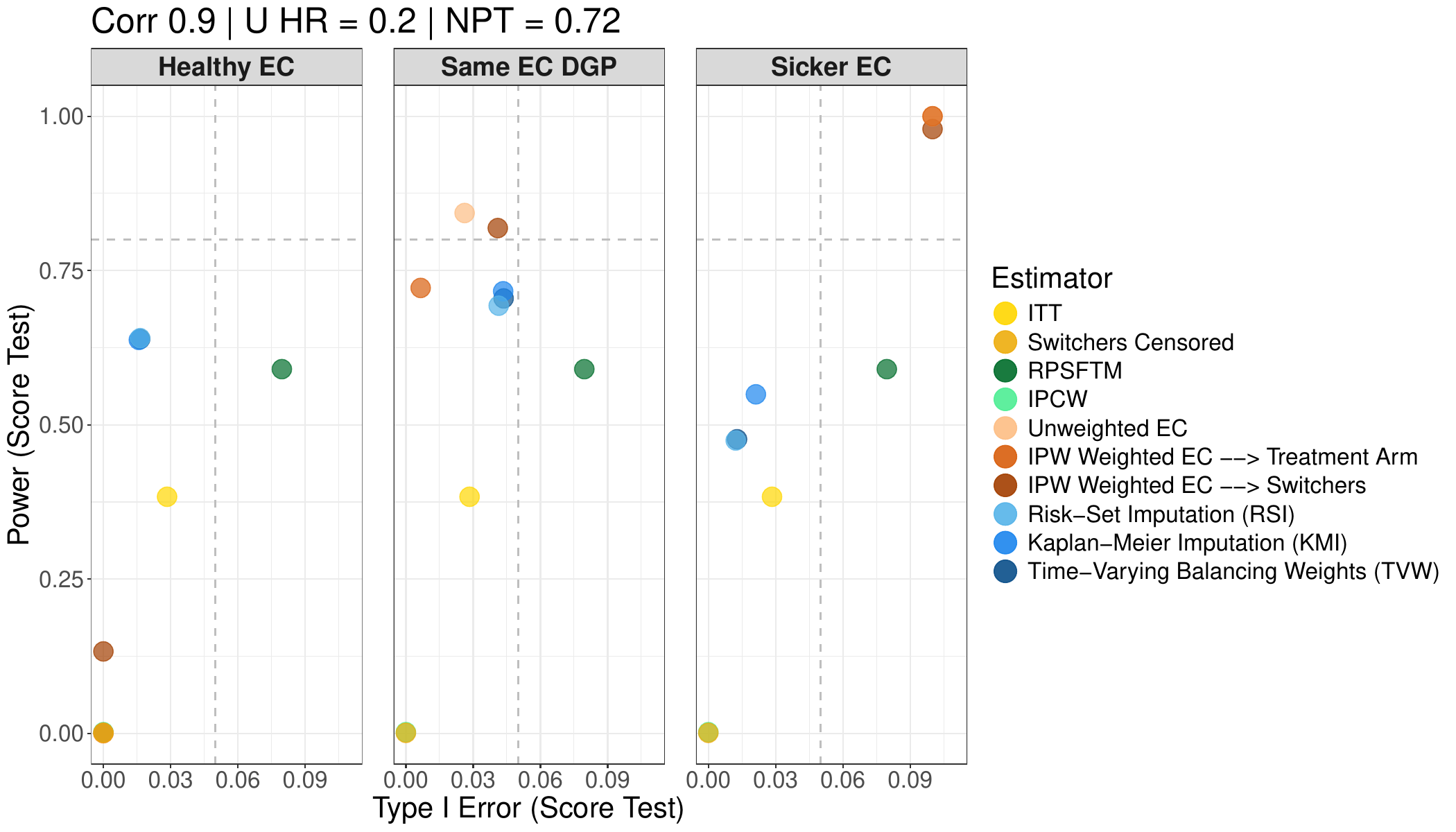}
        \caption{Power and Type I error estimates}
        \label{fig:corr0.9-u0.2-npt72-b}
    \end{subfigure}
    
    \caption{Simulation results for $\rho = 0.9$, $\exp(\beta_u) = 0.2$ and constant NPT. Individual panels represent covariate distribution of ECs relative to internal controls (e.g., healthier/sicker/same DGP). The first four estimates in each panel are identical since they do not use EC data. Refer to Section \ref{sec:method-summary} for a description of all methods used. Precise Type I error rates for estimators exceeding the visual truncation threshold of 0.1 include: Unweighted EC in the Sicker EC panel (0.997), Weighted EC (treatment arm) in the Sicker EC panel (0.907) and Weighted EC (switchers) in the Sicker EC panel (0.283).}
    \label{fig:corr0.9-u0.2-npt72}
\end{figure}

We investigate the performance of each method across varying degrees of simultaneity bias ($\rho$) and unmeasured confounding ($\beta_u$). These trends are displayed in Figures \ref{fig:corr0-u1-npt72} through \ref{fig:corr0.9-u0.2-npt72}, which highlight how different methods respond to increasingly complex DGPs; a comprehensive breakdown of all simulation scenarios is provided in Section \ref{sec:addl-simulation} of the Appendix.
Subfigure (a) in each figure displays point estimates for each method to characterize bias and variance. Points represent the mean log hazard ratio across 5000 simulation runs, while error bars depict the 95\% empirical interval ($\pm 1.96 \times$ SD of the log hazard ratios). To assist with interpretability, the y-axis is scaled to the corresponding hazard ratio while maintaining the underlying log-symmetry.
Subfigure (b) evaluates the trade-off between Type I error ($\alpha = 0.05$, $x$-axis) and statistical power ($y$-axis) using a score test. To maintain a meaningful visual scale, the Type I error rate is truncated at a maximum of 0.1. For estimators where the actual Type I error rate exceeds this threshold, the precise values are documented in the respective figure captions.

Figure \ref{fig:corr0-u1-npt72} presents the base case scenario where $\rho = 0$ and $\exp(\beta_u) = 1$. In this setting, PD and OS are completely independent conditioned on the covariates $X$ and no omitted variable bias exists. This scenario serves as a reference point to evaluate estimator performance under ideal conditions before introducing bias. 
Several patterns emerge from Figure 5. First, the four traditional methods (ITT, censoring of switchers, RPSFTM, IPCW) are identical across the three panels because they do not use EC data. Among them, only ITT shows meaningful bias (a diluted treatment effect due to switching); the others are nearly unbiased in this idealized setting (no informative censoring or omitted-variable bias), though IPCW performs best in terms of the tradeoff between power and Type I error. We also note that the ITT estimator struggles to achieve even 60\% power, the lowest across all methods.

Second, among EC-based approaches, all weighted methods, both population-level and switcher-specific, perform well when neither confounding nor PD/OS correlation is present. Third, replacing the internal control arm with unweighted ECs introduces bias proportional to the covariate shift: sicker ECs overstate the treatment effect, and healthier ECs understate it. The weighting-based methods tend to achieve power above 80\% while all methods generally maintain nominal Type I error rates (with the exception of the unweighted EC estimator when the covariate distributions vary). 


In Figure \ref{fig:corr0.5-u0.6-npt72}, we introduce both moderate correlation ($\rho=0.5$) and omitted variable bias ($\exp(\beta_u) = 0.6$). Most traditional methods, including censoring switchers and IPCW, underestimate the marginal treatment effect as a result of collapsibility from the unmeasured variable being unaccounted for in the weights and the final Cox model.
While RPSFTM remains unbiased due to the common treatment effect assumption being maintained, it has the largest variance across all methods. In the Appendix \ref{sec:non-constant-npt}, we present results with a different switcher treatment effect and demonstrate how RPSFTM fails to recover the true treatment effect in this setting while our switcher-specific EC methods maintain strong performance. 
We also observe greater bias in the population-level weighting methods as $U$ is unaccounted for in both the weighting and estimation schemes. Moreover, the population-level weighting methods exhibit extreme variability in power and Type I error: when ECs are healthier, the false positive rate is low at the expense of power due to underestimated treatment effect. When ECs are sicker, we observe the opposite where the false positive rate is unreasonably high. Figures \ref{fig:corr0.5-u0.6-npt72-b} and \ref{fig:corr0.9-u0.2-npt72-b} display Type I error rates at around 0.1 solely for the sake of visual clarity and in these rates are much higher in reality (see the figure captions for the precise values).
In contrast, all three switcher-specific strategies (RSI, KMI, and TVW) remain robust and nearly unbiased while maintaining a reasonable power and Type I error tradeoff. This performance likely stems from the progression-aligned selection process, which ensures that switchers are matched with external counterparts at a comparable stage of their disease trajectory, thereby mitigating the bias introduced by post-randomization confounding. 

Finally, Figure \ref{fig:corr0.9-u0.2-npt72} shows the most extreme scenario, where correlation and omitted variable bias are at their most extreme. As expected, the bias observed in Figure \ref{fig:corr0.5-u0.6-npt72} is significantly amplified for traditional and population-level weighting methods, in addition to the tradeoff between Type I error and power. In contrast, the switcher-specific methods (RSI, KMI, and TVW) remain remarkably stable and robust. This strong performance is driven by the progression-aligned selection process which facilitates a precise match of disease status at the moment of switching without weighting/balancing on PD time or status. Because each patient progresses at a different time, the specific PD time of an individual switcher provides the necessary temporal anchor required to identify a comparable external risk set. However, this procedure is logically impossible for the population-level weighting estimators. The varying PD times for each switcher means there is no single temporal anchor from which to construct a unified risk set, which results in the risk set consisting of all eligible ECs\footnote{When reweighting ECs to the internal treatment arm, the risk set is all ECs; when reweighting to the switchers, the risk set is all ECs that experience PD.}. Moreover, attempting to balance on PD time at the population level would require conditioning on a post-randomization variable, which violates standard causal identification assumptions \citep{ding2024first}.

\section{Empirical Application: IMpower130 and OAK Trials} \label{sec:impower130}
We also assess our treatment switching methods in the practical setting using an oncology trial conducted by Roche/Genentech and an external dataset from a comparable trial from Roche/Genentech. 
We discuss the studies in Section \ref{sec:impower-background} and present the results in Section \ref{sec:impower-results}.

\subsection{Background} \label{sec:impower-background}
The IMpower130 study \citep{west2019atezolizumab} was a Phase 3 oncology study conducted by Roche/Genentech from 2015 to 2021 to evaluate the efficacy and safety of Atezolizumab in combination with chemotherapy carboplatin and nab-paclitaxel as a first-line treatment for metastatic non-squamous non-small-cell lung cancer (NSCLC). In total, 723 patients were randomized in a 2:1 ratio: Arm A (treatment arm) received atezolizumab plus chemotherapy (carboplatin and nab-paclitaxel), and Arm B (control arm) received chemotherapy alone. The initial study protocol allowed patients in Arm B to cross over to receive atezolizumab monotherapy upon documented disease progression. Atezolizumab monotherapy reflects the treatment arm of our EC dataset (the OAK study), which is described in the following paragraph. 
The study demonstrated improved overall survival (OS) and progression-free survival (PFS) in the primary analysis population (EGRF and ALK wildtype) \citep{west2019atezolizumab}. 
While a strong PFS benefit was observed with HR 0.64 (95\% CI $0.54–0.77$; $p<0.0001$), the crossover-unadjusted observed OS benefit of HR 0.79 (95\% CI $0.64–0.98$; $p=0.033$) might be underestimated due to the high proportion of switching observed in Arm B patients (40\%, $n_s = 93$) at the time of primary analysis \citep{west2019atezolizumab}. The adjusted OS results assuming no switching (ie., hypothetical strategy) demonstrated a stronger effect of HR 0.71 using the RPSFTM method \citep{EMA2019Tecentriq}.

The OAK study \citep{rittmeyer2017atezolizumab,mazieres2021atezolizumab} was a Phase 3 study conducted from 2014 to 2019 to compare atezolizumab monotherapy to docetaxel in NSCLC patients who had progressed after prior platinum-based chemotherapy regimens (Mazieres et al., 2021). In total, 1225 patients were enrolled and 612 were randomized to receive docetaxel (external control arm), which is the second-line therapy for IMpower130 control patients (Arm B). In other words, docxetaxel is standard of care that non-switcher IMpower130 Arm B patients would have received after experiencing PD, making it a suitable EC arm. Excluding patients with known EGFR and ALK mutations, 529 patients were identified as external controls for our analysis.

As mentioned previously, the treatment arm of the OAK study is identical to the therapy provided to switchers in IMpower130 (atezolizumab monotherapy). In addition, the control arm therapy of OAK is identical to the therapy that the IMpower130 switchers would have received if they had not switched (docetaxel). Therefore, the OS treatment effect estimated in the OAK study serves as a valid benchmark for the OS treatment effect for the IMpower130 switchers and allows us to assess the common treatment effect assumption of RPSFTM, which is discussed further in Section \ref{sec:impower-results}. Table \ref{tab:baseline_covariates} in the Appendix presents demographics tables for these two studies.

\subsection{Data analysis results} \label{sec:impower-results}
The results from the IMpower130 analysis are presented in Table \ref{tab:impower-results-tbl}.\footnote{Consistent with the IMpower130 study, our analysis is stratified by sex and PD-L1 expression on tumor cells (TC) and tumor infiltrating cells (IC).} The pattern of results closely mirrors the simulation scenario with moderate confounding and PD/OS correlation ($\rho=0.5$, $\exp(\beta_u)=0.6$; see the ``Sicker EC'' panel of Figure \ref{fig:corr0.5-u0.6-npt72}).
In particular, we observe that censoring switchers at their switch time weakens the estimated treatment effect. This results in informative censoring, reflecting the bias that results from omitting the potentially sicker subgroup of switchers from the study. We also observe a similar direction and magnitude of bias when using IPCW, which may reflect the presence of strong unmeasured confounding and/or informative censoring.

\begin{table}[ht]
    \centering
    \begin{threeparttable}

        \footnotesize
        \setlength{\tabcolsep}{3pt} 
        
        \begin{tabular}{lcccc}
            \toprule
            \textbf{Estimator} & \textbf{Hazard Ratio} & \textbf{Log Hazard Ratio} & \textbf{SE (log HR)} & \textbf{\begin{tabular}{@{}c@{}}p-value\\(score test)\end{tabular}} \\
            \midrule
            ITT & 0.791 & -0.23 & 0.11 & 0.033 \\
            Censor Switchers & 0.847 & -0.17 & 0.13 & 0.2 \\
            RPSFTM & 0.721\tnote{*} & -0.33 & 0.12 & 0.005 \\
            IPCW & 0.850 & -0.16 & 0.13 & 0.21 \\
            Unweighted EC & 0.562 & -0.58 & 0.08 & $< 0.0001$ \\
            IPW Weighted EC $\to$ Treatment Group & 0.577 & -0.55 & 0.08 & $< 0.0001$ \\
            IPW Weighted EC $\to$ Switchers & 0.574 & -0.55 & 0.09 & $< 0.0001$ \\
            RSI & 0.738 & -0.3 & 0.12 & 0.004 \\
            KMI & 0.735 & -0.31 & 0.12 & 0.0036 \\
            Time-varying balancing weights & 0.725 & -0.32 & 0.09 & 0.001 \\
            \bottomrule
        \end{tabular}
        
        \begin{tablenotes}
            \scriptsize
            \item[*] A minor difference compared to the RPSFTM results reported in EPAR due to rounding.
        \end{tablenotes}
    \end{threeparttable}
    \caption{Estimated hazard ratios for the IMpower130 trial ($n_r = 723$; Arm B switchers $n_s = 93$) using OAK docetaxel-arm patients as external controls ($n_e = 529$). All Cox models are stratified by sex and PD-L1 expression (TC and IC categories). SE denotes the robust standard error of the log hazard ratio. The $p$-value is from the score test.}
        \label{tab:impower-results-tbl}
\end{table}

However, we observe that RPSFTM yields a hazard ratio estimate similar to those obtained using our switcher-specific EC-based approaches. While RPSFTM does not utilize baseline covariates for adjustment, its strong performance in this application suggests that the underlying constant treatment effect assumption is satisfied. To understand this, we consider the differing control arms and lines of therapy. In IMpower130, the RPSFTM assumes that the survival benefit of crossing over to atezolizumab monotherapy in the second-line setting (relative to the standard-of-care docetaxel) is equivalent to the primary trial effect of atezolizumab plus chemotherapy in the first-line setting. As established in Section \ref{sec:impower-background}, the OAK study directly estimates this exact second-line crossover effect (atezolizumab monotherapy versus docetaxel) and yielded a hazard ratio of 0.73 \citep{rittmeyer2017atezolizumab,mazieres2021atezolizumab,EMA2019Tecentriq}. The OAK second-line effect aligns with the RPSFTM-adjusted first-line overall hazard ratio of 0.71 for IMpower130 \citep{west2019atezolizumab}. Because the empirical treatment effect for the crossover therapy (benchmarked by OAK) mirrors the overall trial effect, the constant treatment effect assumption required by RPSFTM is highly plausible. Consequently, the strong alignment between the RPSFTM and our switcher-specific EC methods not only validates our proposed framework, but also underscores the robustness of the identified treatment effect across entirely different modeling choices.


With regards to the EC-based methods, we again observe a nearly identical trend to our simulation results. Using the raw external controls without weighting adjustment results in a stronger treatment effect, suggesting that the OAK control population is sicker. This is a plausible assumption since all patients in this study experienced disease progression (i.e., including the second line patients as well as later lines of patients), whereas the switchers in IMpower130 were strictly second line patients \citep{rittmeyer2017atezolizumab,mazieres2021atezolizumab}. We also observe that both population-based IPW approaches result in similar treatment effect estimates, underscoring a significant degree of unmeasured confounding. 

Most importantly, our switcher-specific approaches (KMI, RSI, time-varying weights), which exhibited the strongest performance in simulation, yield results that closely align with RPSFTM, suggesting that these methods appropriately adjust for the bias introduced by switching to better estimate the true underlying treatment effect, even in the face of the population drift from the OAK trial.  The consistency of these distinct estimators, one relying on internal trial assumptions and others leveraging external data, provides stronger empirical evidence that our results are robust and not merely an artifact of model choice.

\section{Discussion} \label{sec:discussion}
Treatment switching is a common occurrence in oncology trials and accounting for its bias is crucial to identifying relevant and accurate treatment effects. A standard ITT analysis may produce biased results that under-report the true treatment effect. Standard adjustment methods such as RPSFTM and IPCW rely on strong assumptions: a common treatment effect for RPSFTM and no unmeasured confounding for IPCW. Results from these methods can be substantially biased when these assumptions are violated via informative censoring or omitted-variable bias. 
As such, the use of external controls is an emerging space where other real-world data sources can be used to gain further insight the true effect of a therapy. In this paper, we present a weighting-based causal inference framework to incorporate EC data with internal trial data with one-way treatment switching. In particular, we discuss how inverse propensity score weighting can be used to equalize baseline covariate distributions between the EC group and the treatment group or the group of control patients who switch. We then show how the synthetic control method can be used to reweight ECs to match the switcher in terms of baseline covariates and discuss how these weights can be used in meaningful ways to assist with imputation or upweight switchers with later switch times. Our simulation results suggest that our EC-based switcher-specific methods show the strongest performance in the presence of population drift and unmeasured confounding. In the empirical application, the switcher-specific EC approaches yielded treatment effect estimates similar to those obtained from RPSFTM and to previously reported analyses, illustrating how the proposed framework can be implemented in a realistic oncology setting and providing additional values for estimating the treatment effect under the hypothetical strategy for treatment switching.

There are many promising areas for future work, one of which is the development of a sensitivity analysis framework (or ``tipping-point analysis'') to better handle unmeasured confounding. These frameworks posit the strength of an unmeasured confounder necessary to overturn an observed treatment effect. There exists a rich body of sensitivity analyses for weighted estimators in traditional causal inference \citep{zhao2019sensitivity,soriano2023interpretable,huang2025variance,shen2025calibrated}, so extending these frameworks to our weighting framework offers a promising line of research. Regarding our setting, future research should explore the effects of differing confounding strength between the internal and EC population. In addition, as our time-varying weights method is largely empirical, future work could involve developing asymptotic and/or other theoretical arguments to bolster its statistical underpinning. Another future direction is to investigate the utility of weighted causal inference frameworks in two-way switching settings, where patients in both arms are eligible to switch \citep{xu2022bias}.

\if0\blind{
\section*{Acknowledgments}
The authors would like to thank the following individuals for helpful comments and suggestions: Yulia Deng, Peng Ding, Avi Feller, Jessie Hsu, Kaifeng Lu, Vivian Ng, Herb Pang, Sam Pimentel, Na Xu, Godwin Yung, Wei Zhang, and Jiawen Zhu.
Andy Shen's graduate studies are partially supported by the National Science Foundation Graduate Research Fellowship under Grant No. 2146752. Any opinion, findings, and conclusions or recommendations expressed in this material are those of the authors(s) and do not necessarily reflect the views of the NSF. Andy Shen and Chenqi Fu are employees of Genentech and stockholders of Roche.
}\fi

\singlespacing
\bibliographystyle{apalike}
\bibliography{paper_refs}  

\clearpage

\onehalfspacing
\setcounter{page}{1}
\begin{center}
    \singlespacing
    \Large
    \textbf{Supplementary Materials:} \\Leveraging External Controls for Treatment Switching in Randomized Controlled Trials: A Weighted Causal Inference Framework for Overall Survival
\end{center}

\appendix

\counterwithin{figure}{section} 

\counterwithin{table}{section} 

\section{Existing methods for one-way switching} \label{sec:existing-methods}
There are several approaches for handling post-randomization treatment changes that rely only on data collected within the randomized trial. Throughout this paper, ``switching'' refers to any post-randomization deviation from the assigned control-arm therapy. This includes switching to the experimental treatment and initiation of a non-protocol therapy (NPT). We note below where specific methods handle only one of these forms. In this section, we review three commonly used methods: the rank-preserving structural failure time model (RPSFTM), inverse probability of censoring weighting (IPCW), and multiple imputation (MI). We refer the reader to \citet{latimer2016treatment,watkins2013adjusting,watkins2025further} for in-depth reviews of these methods.

\paragraph{Censoring of switchers.} Because treatment switching disrupts the exchangeability provided by initial randomization, a naive approach is to censor each switcher at their switch time. Specifically, we replace a switcher's observed overall survival time $\TOS$ with their switch time $T_\mathrm{S}$ and set the event indicator to $\delta = 0$. While this prevents the direct contamination of the control arm with post-switch survival data, it induces \emph{informative censoring}. Since switching typically occurs following disease progression, a strong predictor of mortality, the censoring time is no longer independent of the underlying survival time. This violation of the independent censoring assumption leads to biased treatment effect estimates, as the censored patients are generally ``sicker'' than those remaining in the study.

\paragraph{RPSFTM.} \label{sec:rpsftm} 
The rank-preserving structural failure time model (RPSFTM) \citep{robins1993information,white2006estimating} parameterizes treatment exposure using time on the experimental treatment. This setup matches a one-way switching setting where patients switch only from control to the experimental treatment (no other post-randomization or non-protocol therapies). Let $\tilde{T}_i$ denote the observed event or censoring time for subject $i$ such that $\tilde{T}_i = T_i^{C} + T_i^{E}$, where $T_i^{C}$ is the total time spent on control and $T_i^{E}$ is the total time spent receiving the experimental treatment. For non-switchers, if subject $i$ is assigned to control, then $T_i^{E}=0$ and $\tilde{T}_i=T_i^{C}$; if subject $i$ is assigned to the experimental arm, then $T_i^{C}=0$ and $\tilde{T}_i=T_i^{E}$. 

The RPSFTM posits a counterfactual time of the form
\begin{equation*}
T_i^{(0)} = T_i^{C} + \exp(\psi) T_i^{E},
\end{equation*}
where $T_i^{(0)}$ denotes the counterfactual event time under no experimental treatment, and $\exp(\psi)$ is an acceleration factor that rescales time spent on the experimental treatment.

To estimate $\psi$, one can construct counterfactual treatment and control survival curves with varying values of $\psi$ and select the value of $\psi$ that yields a test statistic as close to 0 as possible. This estimation strategy exploits the standard causal assumption that the counterfactual survival time $T_i^{(0)}$ is independent of treatment assignment.
Thus, RPSFTM also assumes a common acceleration factor across subjects and over time, and it implies rank preservation across counterfactual outcomes. A key limitation of this method is that when switching includes NPT initiation, the single-parameter RPSFTM no longer matches the data-generating process unless additional structure is introduced (e.g., a second exposure parameter for NPT).


\paragraph{IPCW.} \label{sec:ipcw}
Inverse probability of censoring weighting (IPCW) treats treatment
switching as a censoring event and reweights individuals who remain on their
assigned treatment to represent those who switch and become censored \citep{robins2000correcting}. The weights are constructed from an outcome model (either Cox or logit) for the switching process conditional on observed covariate and outcome history, so causal estimand identification relies on having observed all baseline time-varying factors that jointly predict switching and the endpoint (no unmeasured confounding/informative censoring).

\paragraph{Multiple imputation.} \label{sec:rsi-kmi}
Recent literature has also explored multiple imputation (MI) methods to correct for treatment switching. In particular, \citet{zhao2024multiple} propose a \emph{Kaplan-Meier-based bootstrap imputation approach}. Rather than estimating inverse censoring weights, each switcher is matched to a nearest-neighbor risk set of non-switchers who are still at risk at the switching time and have similar observed history. This risk set is used to estimate a Kaplan–Meier (KM) curve for the post-switch survival distribution, and event times for each switcher are imputed by sampling from the implied cumulative distribution function (CDF) from this Kaplan-Meier curve.
A related approach directly samples a non-switching control patient from the risk set and imputes the donor’s remaining time and event indicator for the switcher, which is known as \textit{risk-set imputation (RSI)}.
Both imputation procedures can be repeated over multiple bootstrap samples and the imputations are combined using Rubin's rules \citep{little2019statistical}. As with IPCW, these methods rely on the assumption that the observed history used to form the risk set captures the factors that drive both switching and the endpoint, and they work best when each switcher has adequate overlap with available non-switchers.

\section{Additional Simulation Results} \label{sec:addl-simulation}
This Section presents additional simulation results not included in the main text. For clarity, Table \ref{tab:sim-scenarios-all} contains all simulation scenarios and their location in the manuscript.

\begin{table}[ht]
    \centering
    \begin{tabular}{cccl}
        \toprule
        \textbf{Switcher Effect (NPT)} & \textbf{Correlation ($\rho$)} & \textbf{OVB ($\exp(\beta_u)$)} & \textbf{Location} \\
        \midrule
        0.72 (Constant) & 0.0 & 1.0 & \textbf{Main Text (Figure \ref{fig:corr0-u1-npt72})} \\
                        & 0.0 & 0.6 & Appendix \ref{sec:remain-sim} \\
                        & 0.0 & 0.2 & Appendix \ref{sec:remain-sim} \\
        \addlinespace
                        & 0.5 & 1.0 & Appendix \ref{sec:remain-sim} \\
                        & 0.5 & 0.6 & \textbf{Main Text (Figure \ref{fig:corr0.5-u0.6-npt72})} \\
                        & 0.5 & 0.2 & Appendix \ref{sec:remain-sim} \\
        \addlinespace
                        & 0.9 & 1.0 & Appendix \ref{sec:remain-sim} \\
                        & 0.9 & 0.6 & Appendix \ref{sec:remain-sim} \\
                        & 0.9 & 0.2 & \textbf{Main Text (Figure \ref{fig:corr0.9-u0.2-npt72})} \\
        \midrule
        0.52 (Non-constant) & 0.0 & 1.0 & Appendix \ref{sec:non-constant-npt} \\
                        & 0.0 & 0.6 & Appendix \ref{sec:non-constant-npt} \\
                        & 0.0 & 0.2 & Appendix \ref{sec:non-constant-npt} \\
        \addlinespace
                        & 0.5 & 1.0 & Appendix \ref{sec:non-constant-npt} \\
                        & 0.5 & 0.6 & Appendix \ref{sec:non-constant-npt} \\
                        & 0.5 & 0.2 & Appendix \ref{sec:non-constant-npt} \\
        \addlinespace
                        & 0.9 & 1.0 & Appendix \ref{sec:non-constant-npt} \\
                        & 0.9 & 0.6 & Appendix \ref{sec:non-constant-npt} \\
                        & 0.9 & 0.2 & Appendix \ref{sec:non-constant-npt} \\
        \bottomrule
    \end{tabular}
    \caption{Summary of all simulation scenarios, varying the switcher treatment effect (NPT HR), PD/OS correlation ($\rho$), and unmeasured confounding ($\exp(\beta_u)$). Each individual scenario (row) includes three sub-panels representing the EC covariate distribution (Healthier EC, Same EC DGP, Sicker EC). Scenarios explicitly discussed in the main text are bolded.}
    \label{tab:sim-scenarios-all}
\end{table}

\subsection{Remaining simulation settings} \label{sec:remain-sim}
\begin{figure}[ht]
    \centering
    \begin{subfigure}[b]{0.49\linewidth}
        \centering
        \includegraphics[width=\linewidth]{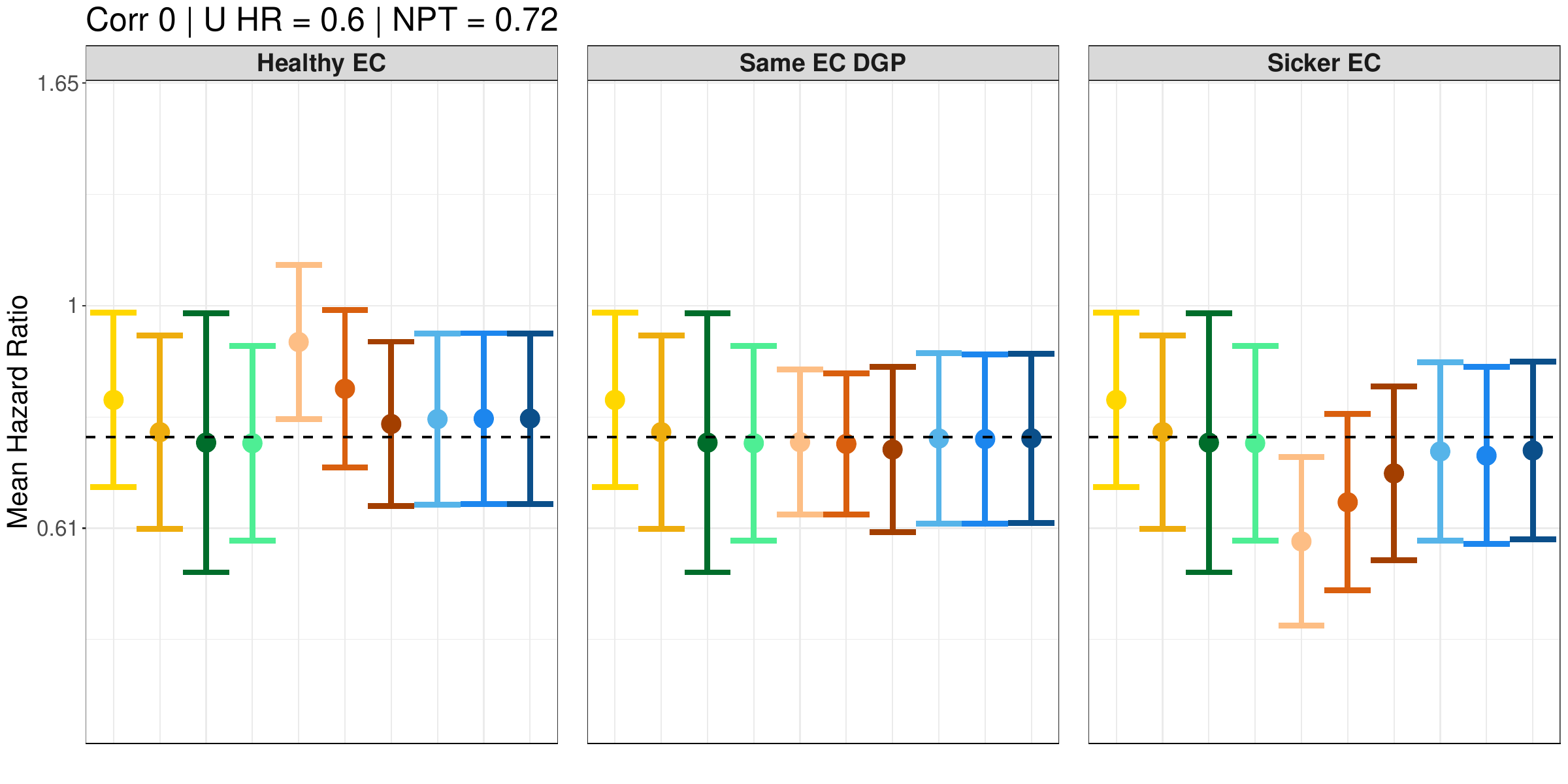}
        \caption{Hazard Ratio estimates}
        \label{fig:corr0-u0.6-npt72-a}
    \end{subfigure}
    \hfill
    \begin{subfigure}[b]{0.49\linewidth}
        \centering
        \includegraphics[width=\linewidth]{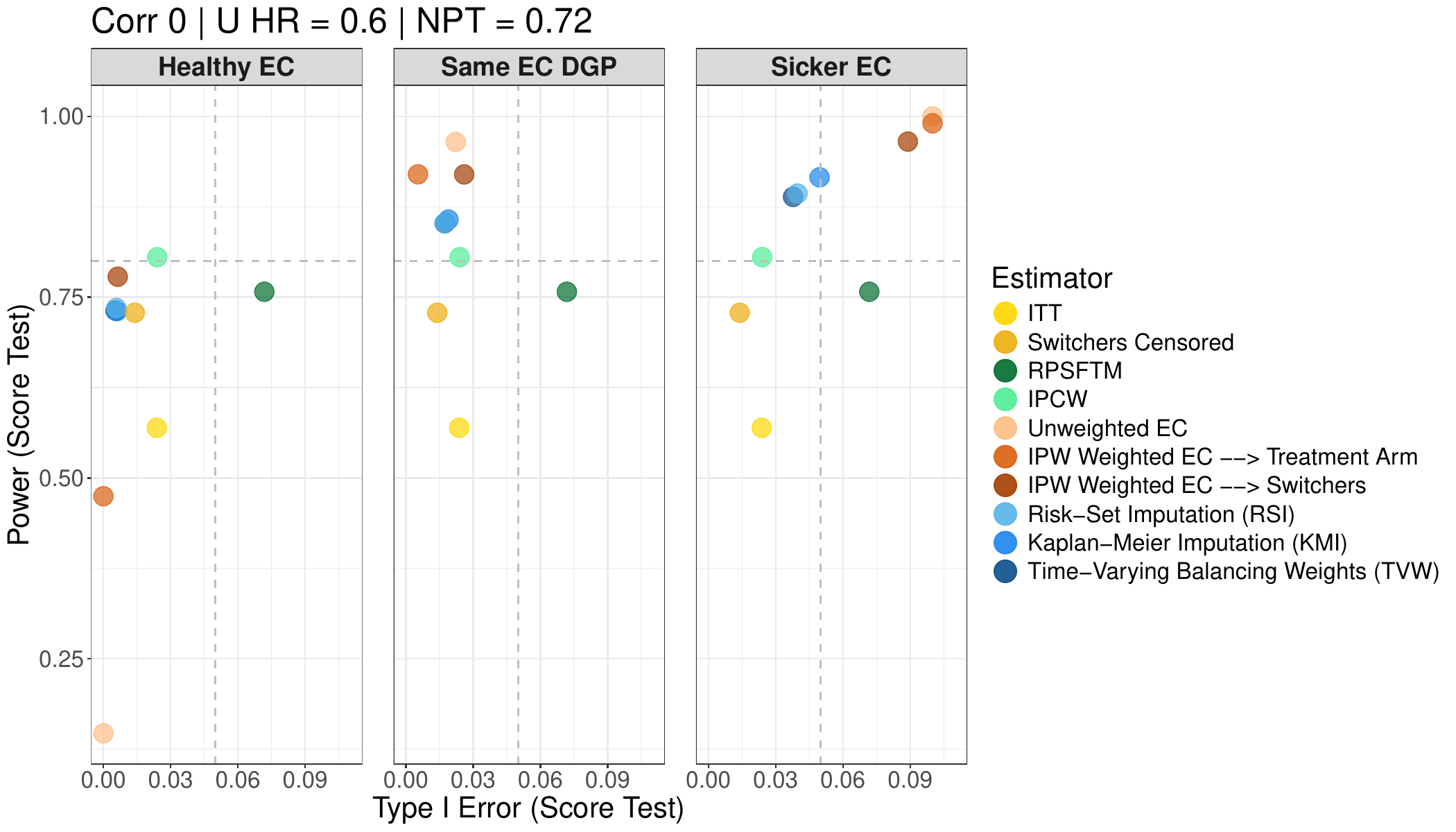}
        \caption{Power and Type I error estimates}
        \label{fig:corr0-u0.6-npt72-b}
    \end{subfigure}
    
    \caption{Simulation results for $\rho = 0$, $\exp(\beta_u) = 0.6$ and constant NPT. Individual panels represent covariate distribution of ECs relative to internal controls (e.g., healthier/sicker/same DGP). The first four estimates in each panel are identical since they do not use EC data. Refer to Section \ref{sec:method-summary} for a description of all methods used. Precise Type I error rates for estimators exceeding the visual truncation threshold of 0.1 include: Unweighted EC in the Sicker EC panel (0.698) and Weighted EC (treatment arm) in the Sicker EC panel (0.194).}
    \label{fig:corr0-u0.6-npt72}
\end{figure}

\begin{figure}[ht]
    \centering
    \begin{subfigure}[b]{0.49\linewidth}
        \centering
        \includegraphics[width=\linewidth]{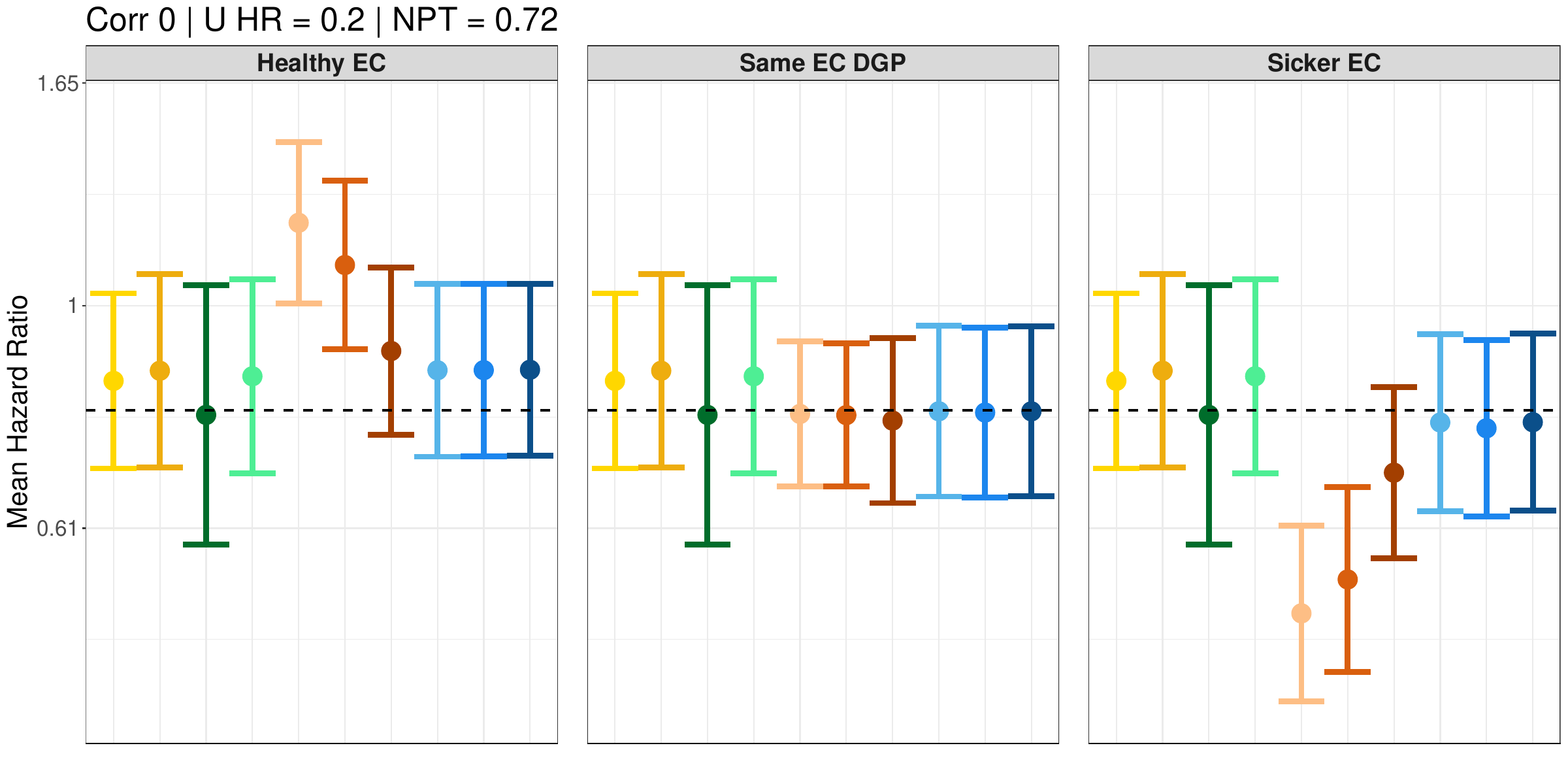}
        \caption{Hazard Ratio estimates}
        \label{fig:corr0-u0.2-npt72-a}
    \end{subfigure}
    \hfill
    \begin{subfigure}[b]{0.49\linewidth}
        \centering
        \includegraphics[width=\linewidth]{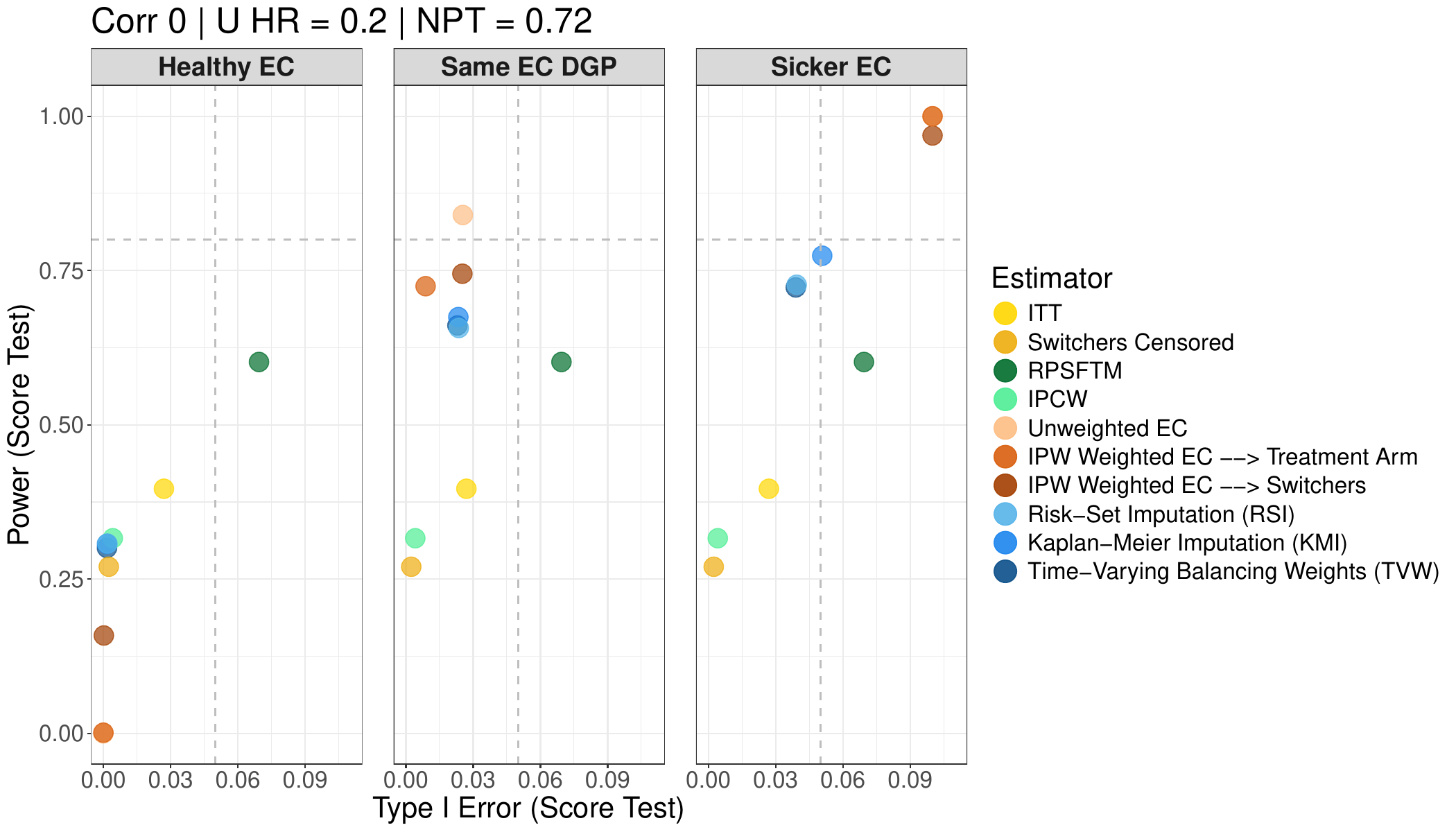}
        \caption{Power and Type I error estimates}
        \label{fig:corr0-u0.2-npt72-b}
    \end{subfigure}
    
    \caption{Simulation results for $\rho = 0$, $\exp(\beta_u) = 0.2$ and constant NPT. Individual panels represent covariate distribution of ECs relative to internal controls (e.g., healthier/sicker/same DGP). The first four estimates in each panel are identical since they do not use EC data. Refer to Section \ref{sec:method-summary} for a description of all methods used. Precise Type I error rates for estimators exceeding the visual truncation threshold of 0.1 include: Unweighted EC in the Sicker EC panel (0.997), Weighted EC (treatment arm) in the Sicker EC panel (0.918) and Weighted EC (switchers) in the Sicker EC panel (0.233).}
    \label{fig:corr0-u0.2-npt72}
\end{figure}

\begin{figure}[ht]
    \centering
    \begin{subfigure}[b]{0.49\linewidth}
        \centering
        \includegraphics[width=\linewidth]{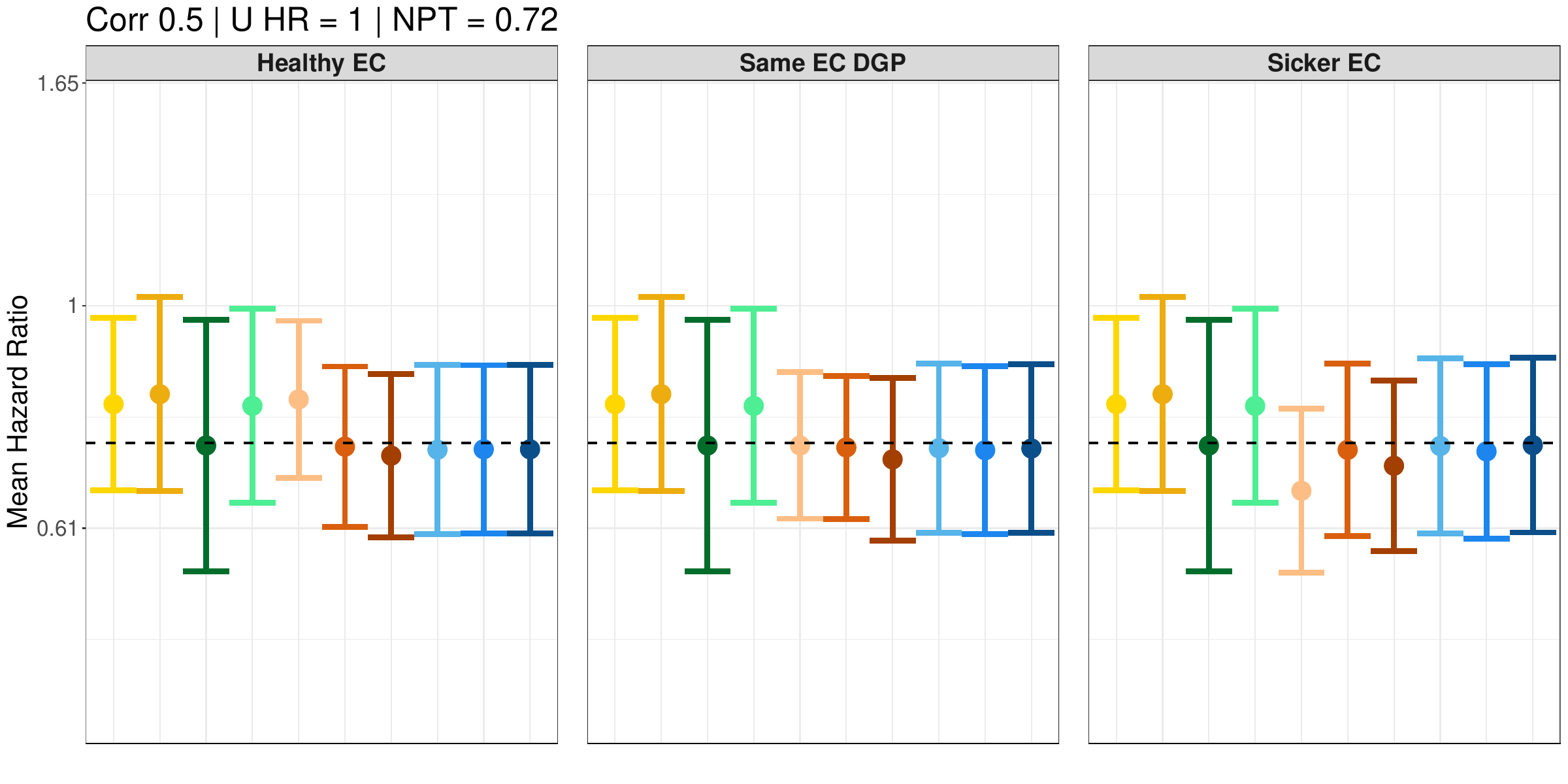}
        \caption{Hazard Ratio estimates}
        \label{fig:corr0.5-u1-npt72-a}
    \end{subfigure}
    \hfill
    \begin{subfigure}[b]{0.49\linewidth}
        \centering
        \includegraphics[width=\linewidth]{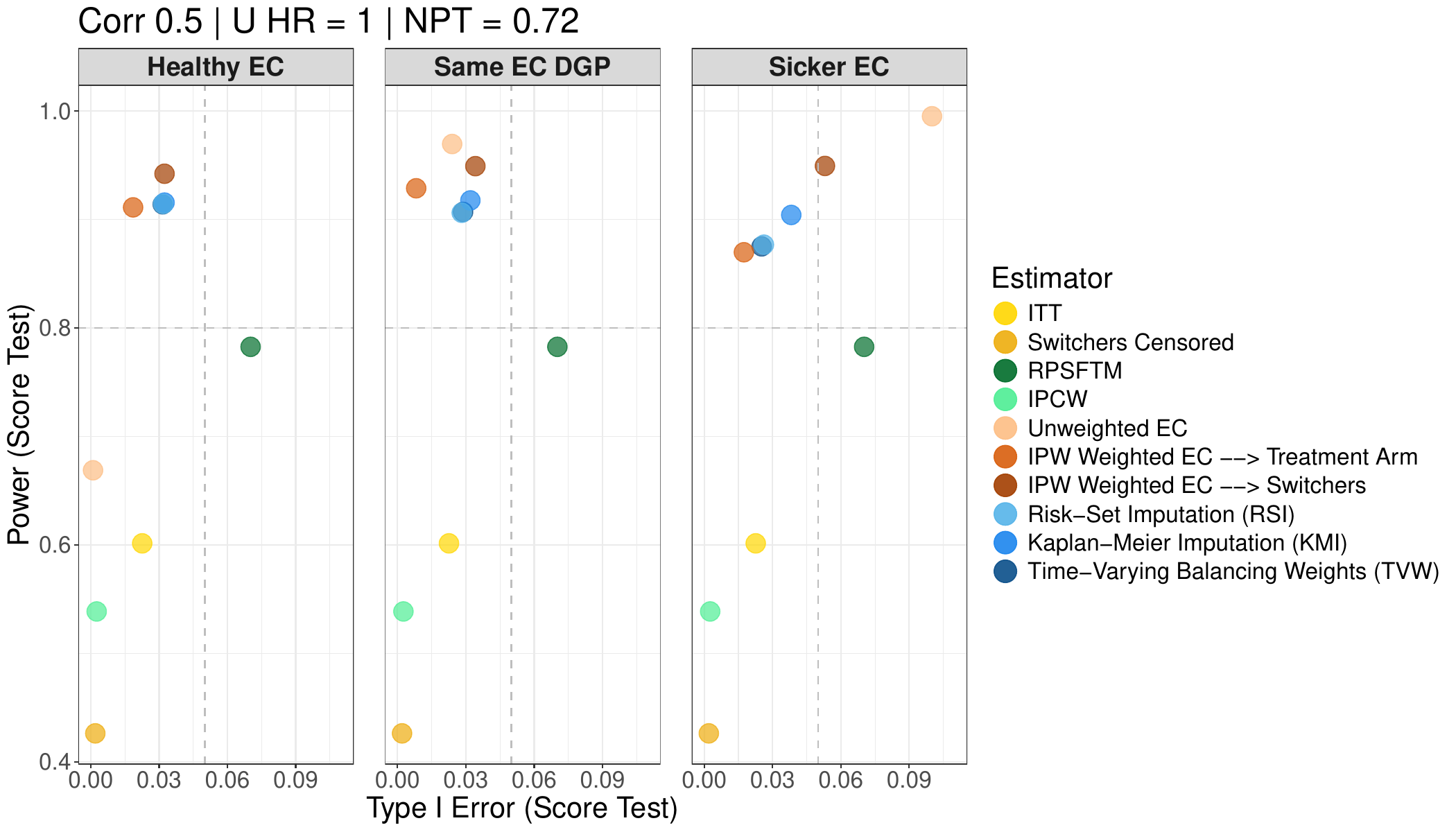}
        \caption{Power and Type I error estimates}
        \label{fig:corr0.5-u1-npt72-b}
    \end{subfigure}
    
    \caption{Simulation results for $\rho = 0.5$, $\exp(\beta_u) = 1$ and constant NPT. Individual panels represent covariate distribution of ECs relative to internal controls (e.g., healthier/sicker/same DGP). The first four estimates in each panel are identical since they do not use EC data. Refer to Section \ref{sec:method-summary} for a description of all methods used. Precise Type I error rates for estimators exceeding the visual truncation threshold of 0.1 include: Unweighted EC in the Sicker EC panel (0.216).}
    \label{fig:corr0.5-u1-npt72}
\end{figure}

\begin{figure}[ht]
    \centering
    \begin{subfigure}[b]{0.49\linewidth}
        \centering
        \includegraphics[width=\linewidth]{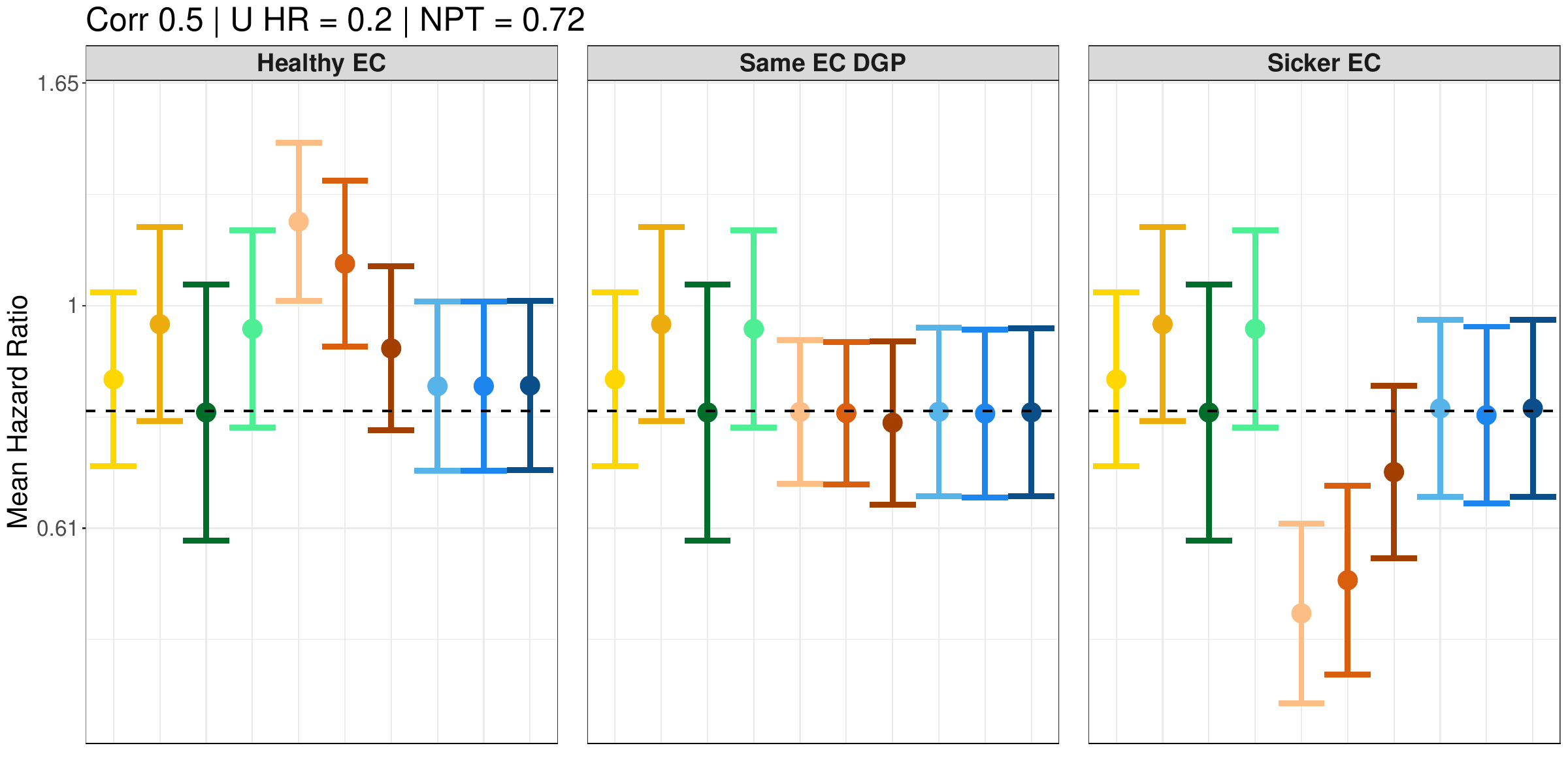}
        \caption{Hazard Ratio estimates}
        \label{fig:corr0.5-u0.2-npt72-a}
    \end{subfigure}
    \hfill
    \begin{subfigure}[b]{0.49\linewidth}
        \centering
        \includegraphics[width=\linewidth]{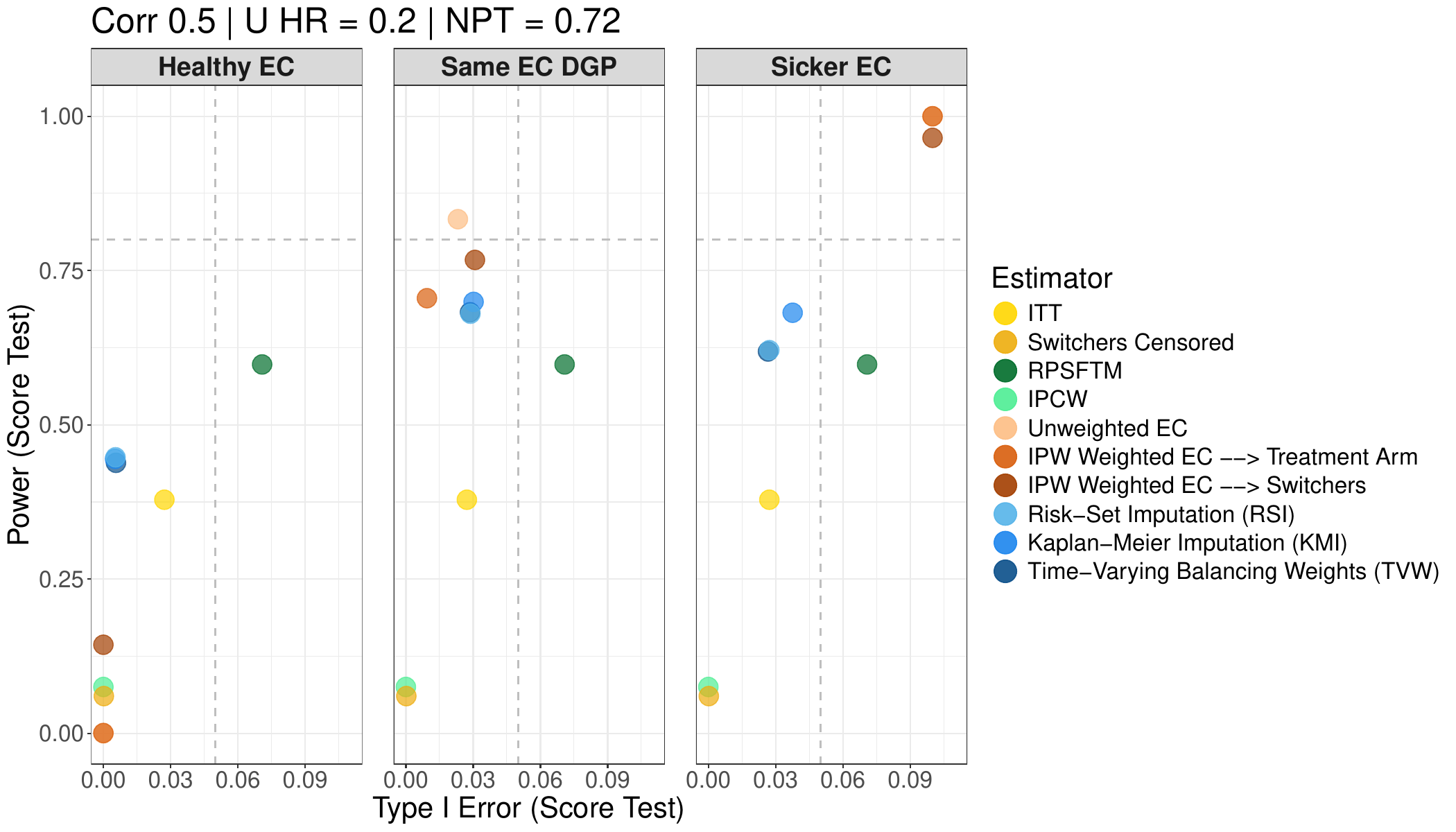}
        \caption{Power and Type I error estimates}
        \label{fig:corr0.5-u0.2-npt72-b}
    \end{subfigure}
    
    \caption{Simulation results for $\rho = 0.5$, $\exp(\beta_u) = 0.2$ and constant NPT. Individual panels represent covariate distribution of ECs relative to internal controls (e.g., healthier/sicker/same DGP). The first four estimates in each panel are identical since they do not use EC data. Refer to Section \ref{sec:method-summary} for a description of all methods used. Precise Type I error rates for estimators exceeding the visual truncation threshold of 0.1 include: Unweighted EC in the Sicker EC panel (0.995), Weighted EC (treatment arm) in the Sicker EC panel (0.911) and Weighted EC (switchers) in the Sicker EC panel (0.24).}
    \label{fig:corr0.5-u0.2-npt72}
\end{figure}

\begin{figure}[ht]
    \centering
    \begin{subfigure}[b]{0.49\linewidth}
        \centering
        \includegraphics[width=\linewidth]{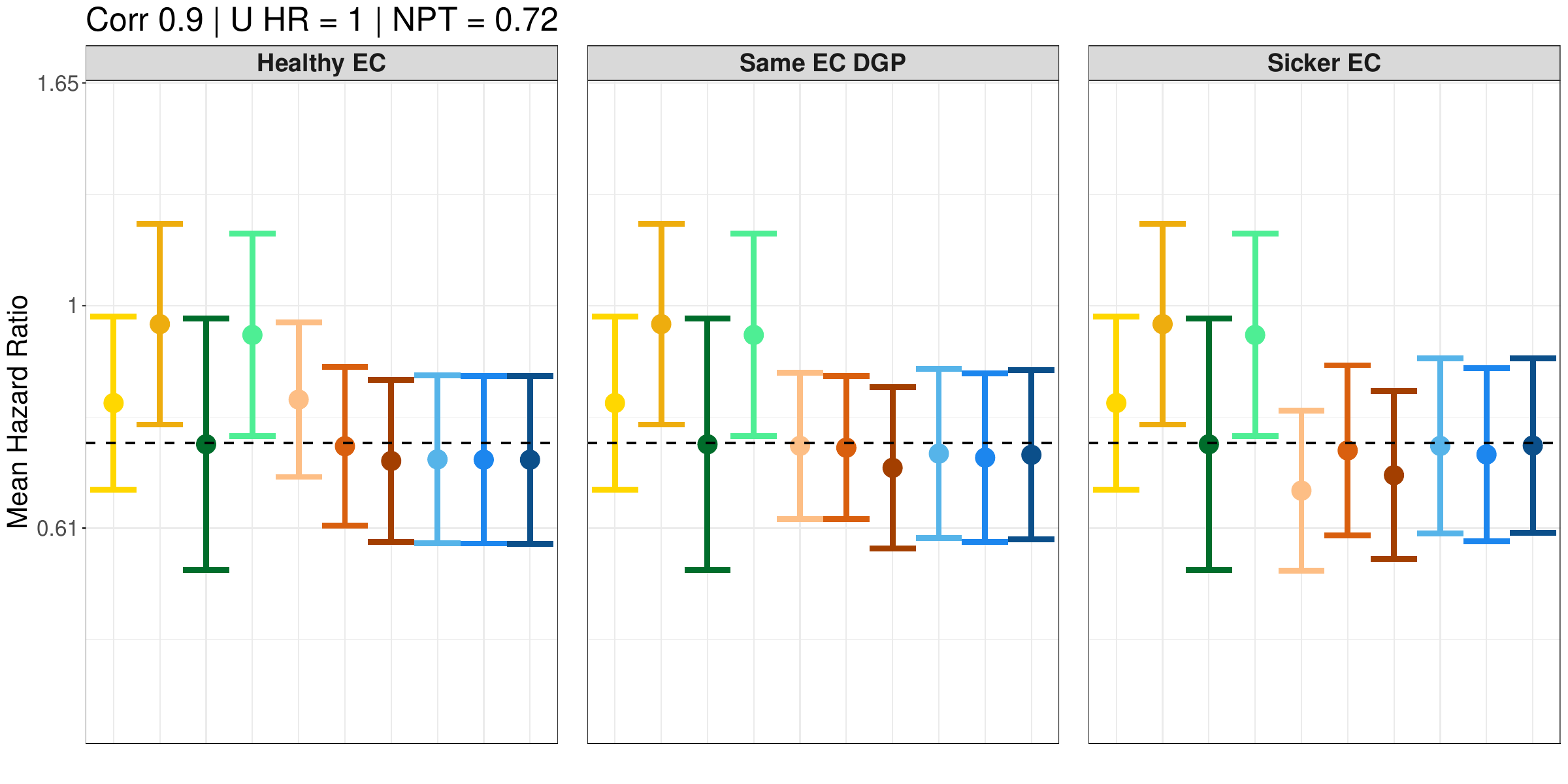}
        \caption{Hazard Ratio estimates}
        \label{fig:corr0.9-u1-npt72-a}
    \end{subfigure}
    \hfill
    \begin{subfigure}[b]{0.49\linewidth}
        \centering
        \includegraphics[width=\linewidth]{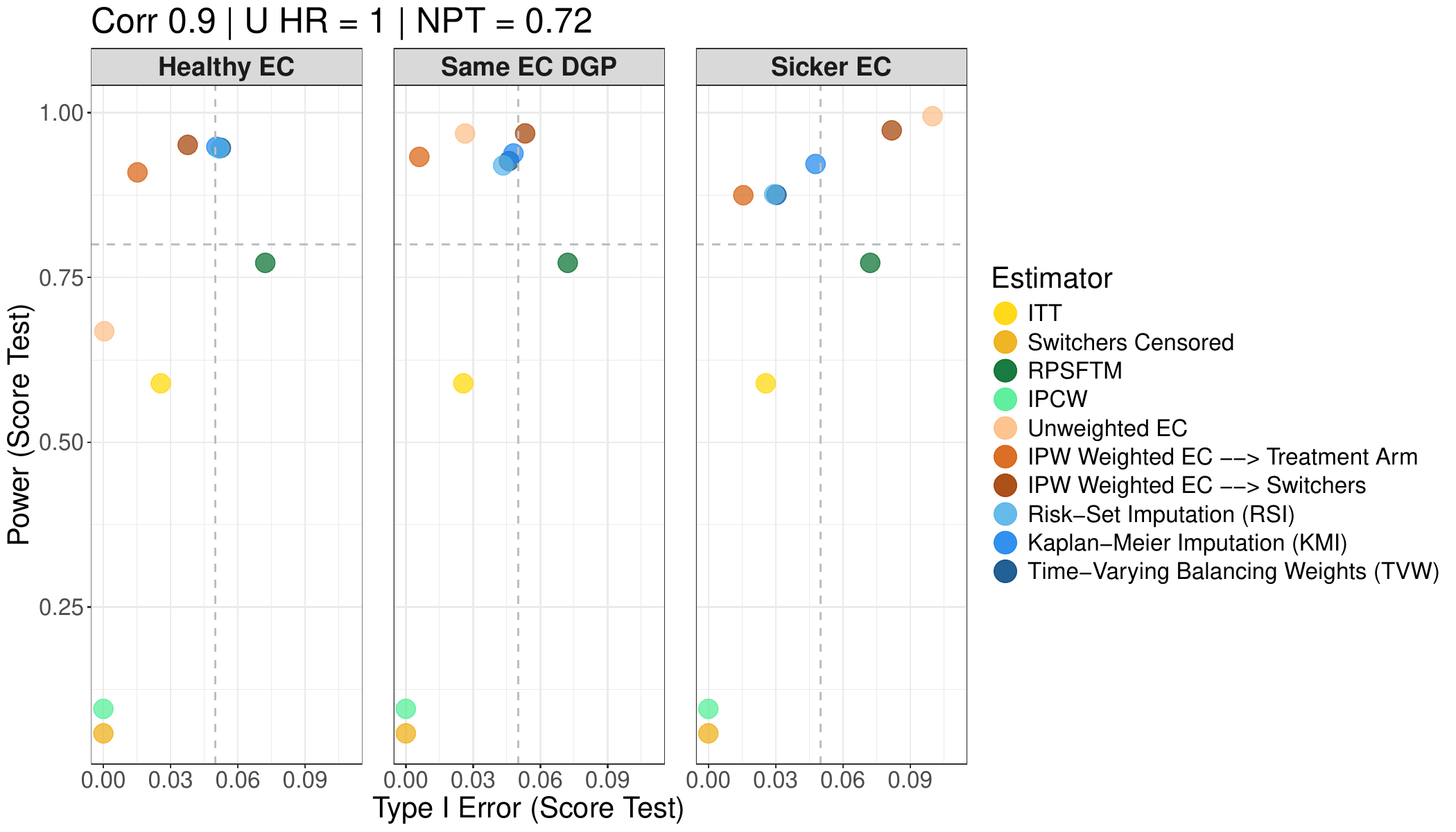}
        \caption{Power and Type I error estimates}
        \label{fig:corr0.9-u1-npt72-b}
    \end{subfigure}
    
    \caption{Simulation results for $\rho = 0.9$, $\exp(\beta_u) = 1$ and constant NPT. Individual panels represent covariate distribution of ECs relative to internal controls (e.g., healthier/sicker/same DGP). The first four estimates in each panel are identical since they do not use EC data. Refer to Section \ref{sec:method-summary} for a description of all methods used. Precise Type I error rates for estimators exceeding the visual truncation threshold of 0.1 include: Unweighted EC in the Sicker EC panel (0.995) and Weighted EC (treatment arm) in the Sicker EC panel (0.875).}
    \label{fig:corr0.9-u1-npt72}
\end{figure}

\begin{figure}[ht]
    \centering
    \begin{subfigure}[b]{0.49\linewidth}
        \centering
        \includegraphics[width=\linewidth]{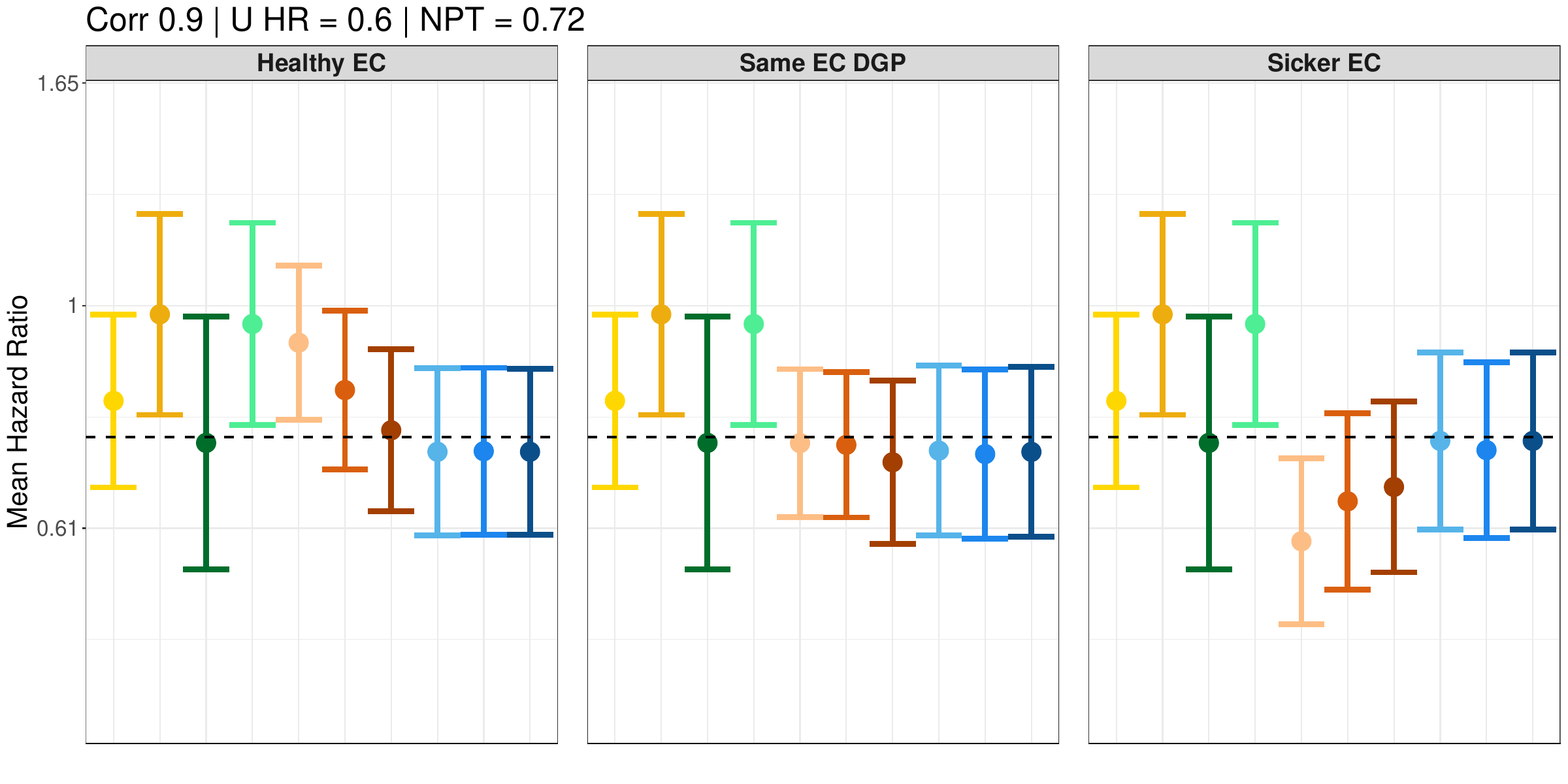}
        \caption{Hazard Ratio estimates}
        \label{fig:corr0.9-u0.6-npt72-a}
    \end{subfigure}
    \hfill
    \begin{subfigure}[b]{0.49\linewidth}
        \centering
        \includegraphics[width=\linewidth]{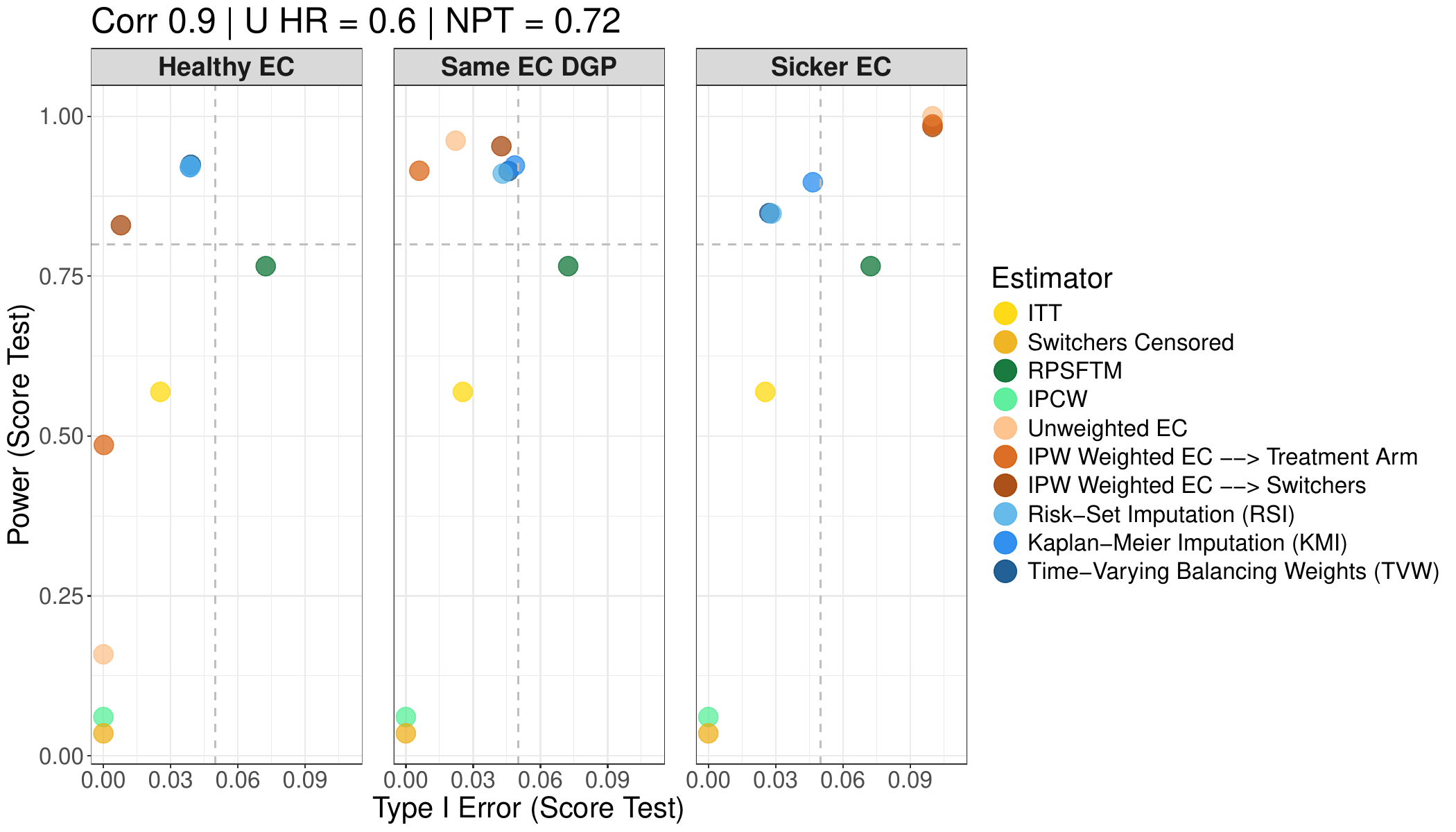}
        \caption{Power and Type I error estimates}
        \label{fig:corr0.9-u0.6-npt72-b}
    \end{subfigure}
    
    \caption{Simulation results for $\rho = 0.9$, $\exp(\beta_u) = 0.6$ and constant NPT. Individual panels represent covariate distribution of ECs relative to internal controls (e.g., healthier/sicker/same DGP). The first four estimates in each panel are identical since they do not use EC data. Refer to Section \ref{sec:method-summary} for a description of all methods used. Precise Type I error rates for estimators exceeding the visual truncation threshold of 0.1 include: Unweighted EC in the Sicker EC panel (0.704), Weighted EC (treatment arm) in the Sicker EC panel (0.2) and Weighted EC (switchers) in the Sicker EC panel (0.145).}
    \label{fig:corr0.9-u0.6-npt72}
\end{figure}

In Figures \ref{fig:corr0-u0.6-npt72} to \ref{fig:corr0.9-u0.6-npt72}, we present the remaining results for the simulation study (Section \ref{sec:simulation}) that were omitted from the main text. The same conclusions drawn from Section \ref{sec:sim-results} hold for these settings. 

\clearpage

\subsection{Non-constant switcher treatment effect} \label{sec:non-constant-npt}
The simulation study in \ref{sec:simulation} in the main text assumes the treatment effect for switchers is identical to the true treatment effect of the therapy. This assumption may be plausible in the case of treatment crossover, where control arm patients take the experimental therapy. However, in more general cases of treatment switching, such as to a non-protocol therapy (NPT), the treatment effect could differ.

We repeated the simulation study in Section \ref{sec:simulation}, changing the treatment effect for switchers to be 0.52 instead of 0.72 (as a hazard ratio). The results are presented in Figures TODO to TODO. The results are largely identical to those in the main text, with the exception that the RPSFTM method is biased due to the constant treatment effect assumption being violated.

\begin{figure}[ht]
    \centering
    \begin{subfigure}[b]{0.49\linewidth}
        \centering
        \includegraphics[width=\linewidth]{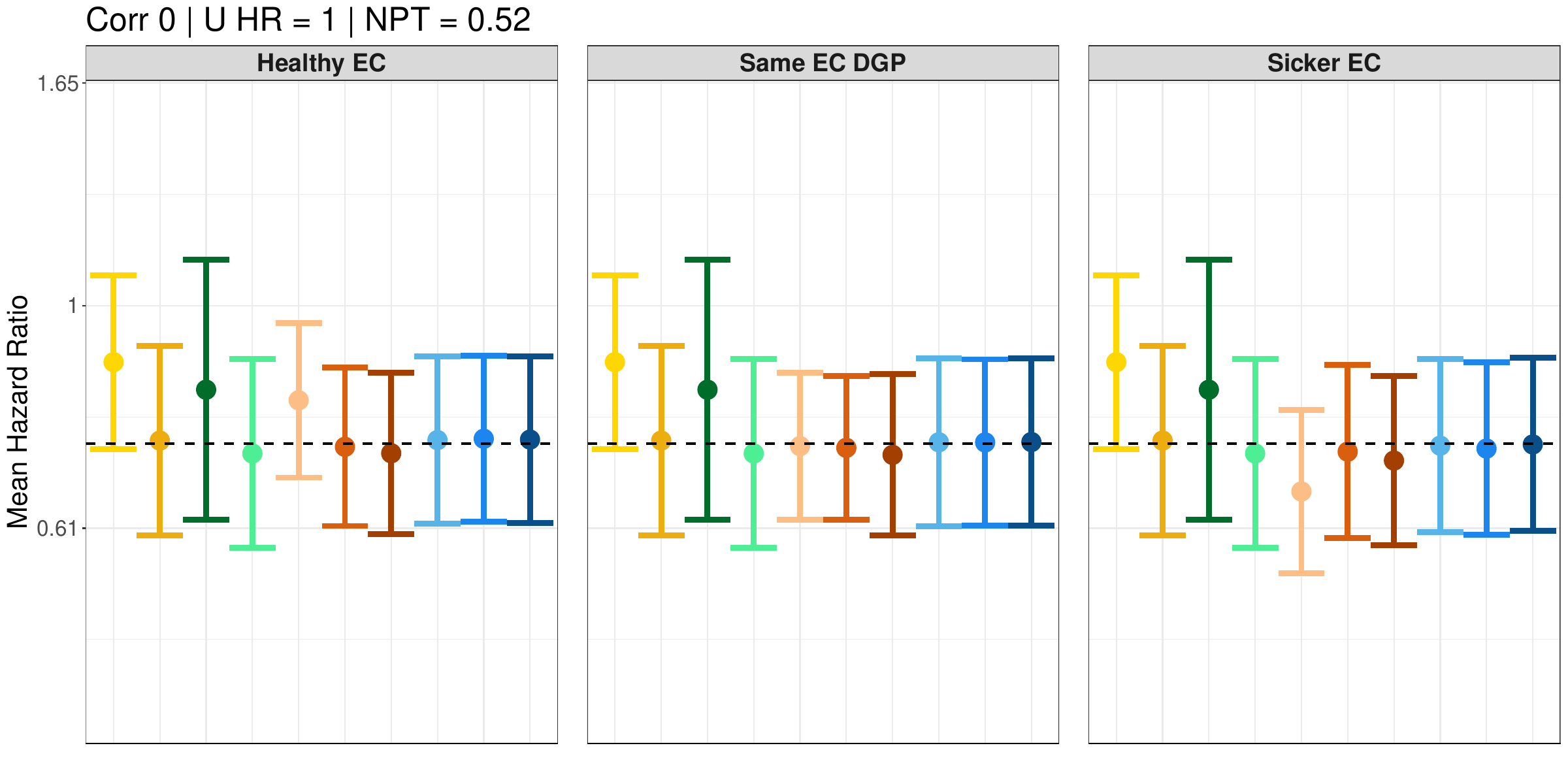}
        \caption{Hazard Ratio estimates}
        \label{fig:corr0-u1-npt52-a}
    \end{subfigure}
    \hfill
    \begin{subfigure}[b]{0.49\linewidth}
        \centering
        \includegraphics[width=\linewidth]{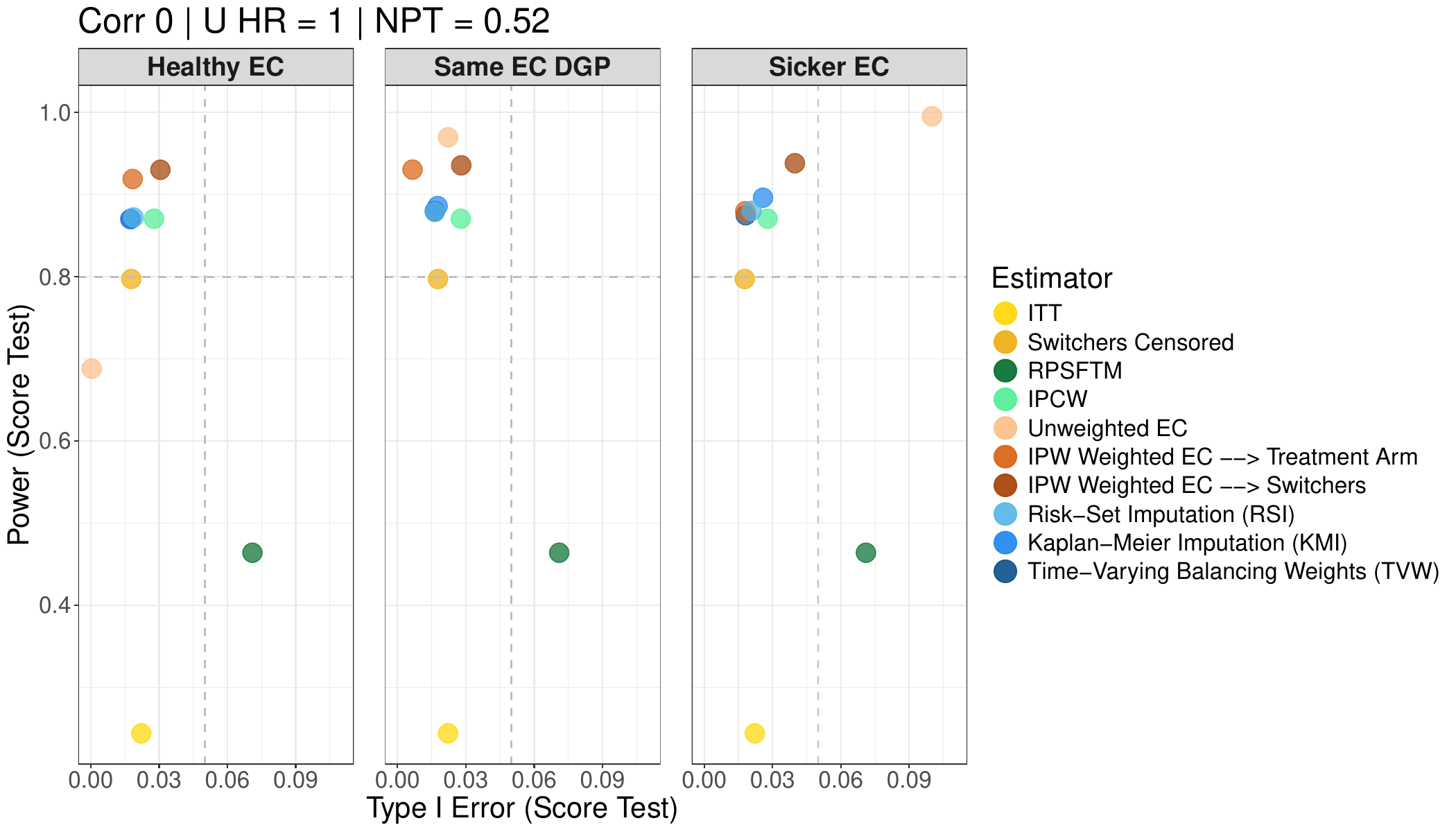}
        \caption{Power and Type I error estimates}
        \label{fig:corr0-u1-npt52-b}
    \end{subfigure}
    
    \caption{Simulation results for $\rho = 0$, $\exp(\beta_u) = 1$ and non-constant NPT. Individual panels represent covariate distribution of ECs relative to internal controls (e.g., healthier/sicker/same DGP). The first four estimates in each panel are identical since they do not use EC data. Refer to Section \ref{sec:method-summary} for a description of all methods used. Precise Type I error rates for estimators exceeding the visual truncation threshold of 0.1 include: Unweighted EC in the Sicker EC panel (0.212).}
    \label{fig:corr0-u1-npt52}
\end{figure}

\begin{figure}[ht]
    \centering
    \begin{subfigure}[b]{0.49\linewidth}
        \centering
        \includegraphics[width=\linewidth]{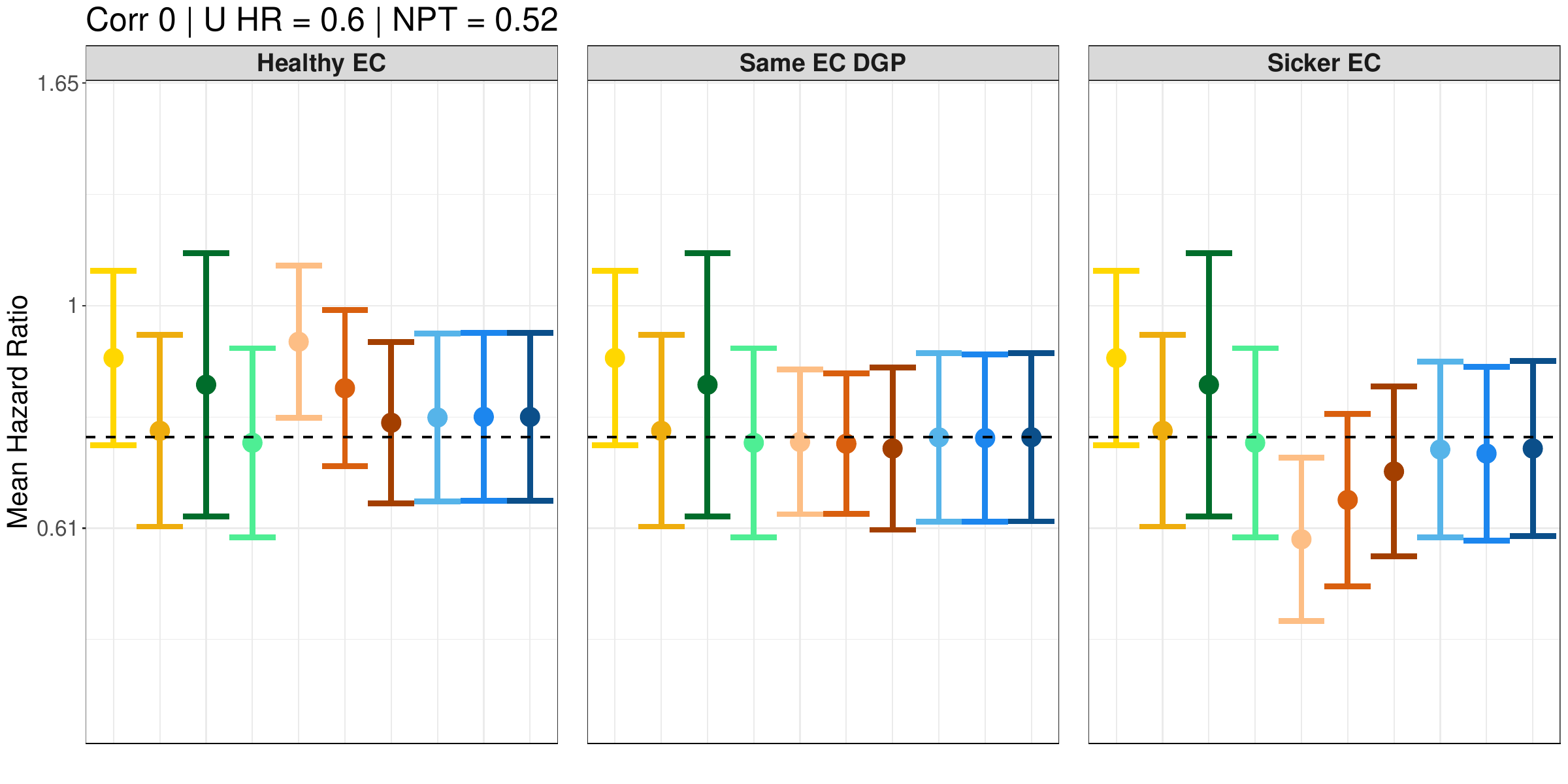}
        \caption{Hazard Ratio estimates}
        \label{fig:corr0-u0.6-npt52-a}
    \end{subfigure}
    \hfill
    \begin{subfigure}[b]{0.49\linewidth}
        \centering
        \includegraphics[width=\linewidth]{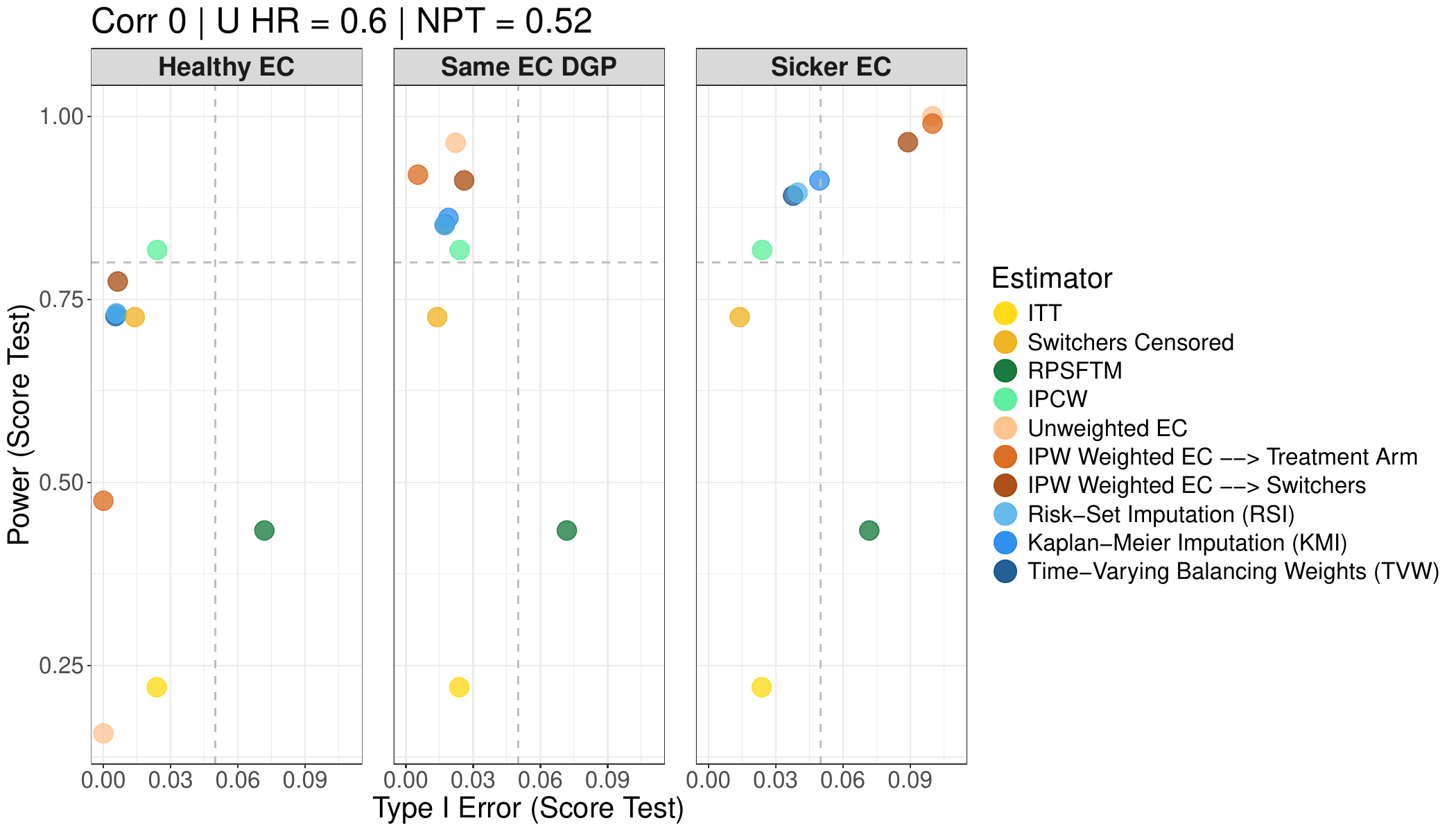}
        \caption{Power and Type I error estimates}
        \label{fig:corr0-u0.6-npt52-b}
    \end{subfigure}
    
    \caption{Simulation results for $\rho = 0$, $\exp(\beta_u) = 0.6$ and non-constant NPT. Individual panels represent covariate distribution of ECs relative to internal controls (e.g., healthier/sicker/same DGP). The first four estimates in each panel are identical since they do not use EC data. Refer to Section \ref{sec:method-summary} for a description of all methods used. Precise Type I error rates for estimators exceeding the visual truncation threshold of 0.1 include: Unweighted EC in the Sicker EC panel (0.698) and Weighted EC (treatment arm) in the Sicker EC panel (0.194).}
    \label{fig:corr0-u0.6-npt52}
\end{figure}

\begin{figure}[ht]
    \centering
    \begin{subfigure}[b]{0.49\linewidth}
        \centering
        \includegraphics[width=\linewidth]{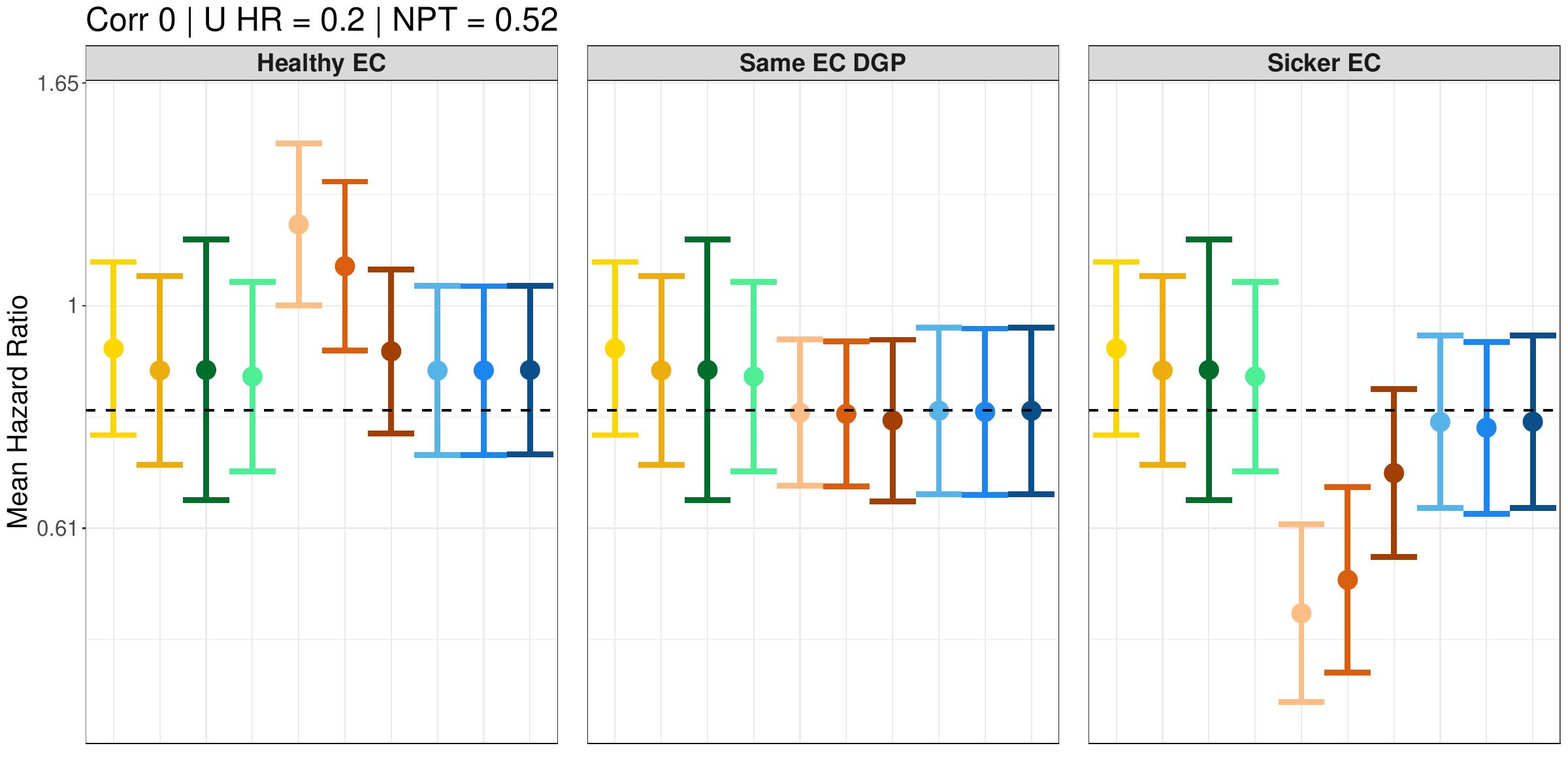}
        \caption{Hazard Ratio estimates}
        \label{fig:corr0-u0.2-npt52-a}
    \end{subfigure}
    \hfill
    \begin{subfigure}[b]{0.49\linewidth}
        \centering
        \includegraphics[width=\linewidth]{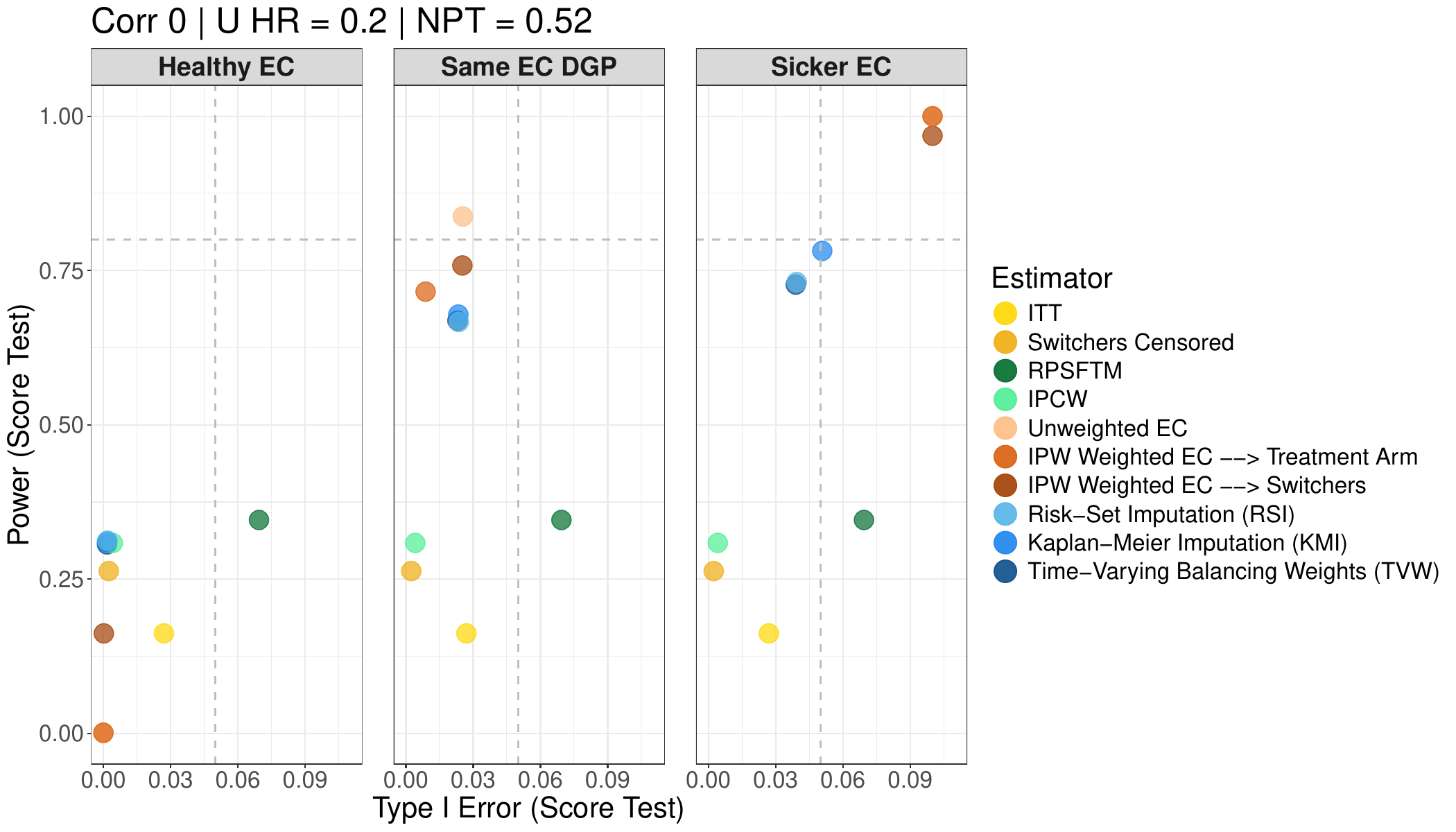}
        \caption{Power and Type I error estimates}
        \label{fig:corr0-u0.2-npt52-b}
    \end{subfigure}
    
    \caption{Simulation results for $\rho = 0$, $\exp(\beta_u) = 0.2$ and non-constant NPT. Individual panels represent covariate distribution of ECs relative to internal controls (e.g., healthier/sicker/same DGP). The first four estimates in each panel are identical since they do not use EC data. Refer to Section \ref{sec:method-summary} for a description of all methods used. Precise Type I error rates for estimators exceeding the visual truncation threshold of 0.1 include: Unweighted EC in the Sicker EC panel (0.997), Weighted EC (treatment arm) in the Sicker EC panel (0.918) and Weighted EC (switchers) in the Sicker EC panel (0.233).}
    \label{fig:corr0-u0.2-npt52}
\end{figure}

\begin{figure}[ht]
    \centering
    \begin{subfigure}[b]{0.49\linewidth}
        \centering
        \includegraphics[width=\linewidth]{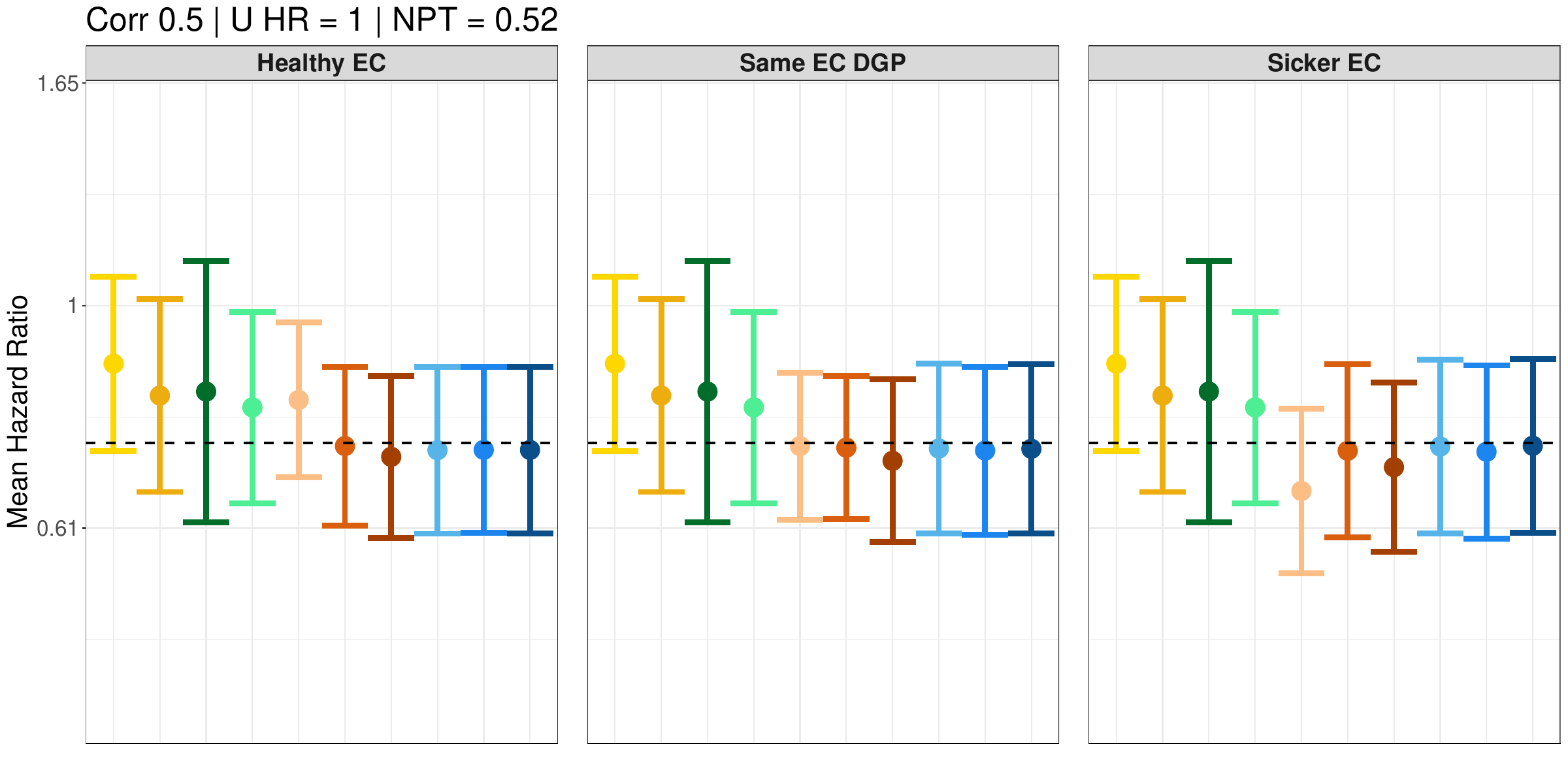}
        \caption{Hazard Ratio estimates}
        \label{fig:corr0.5-u1-npt52-a}
    \end{subfigure}
    \hfill
    \begin{subfigure}[b]{0.49\linewidth}
        \centering
        \includegraphics[width=\linewidth]{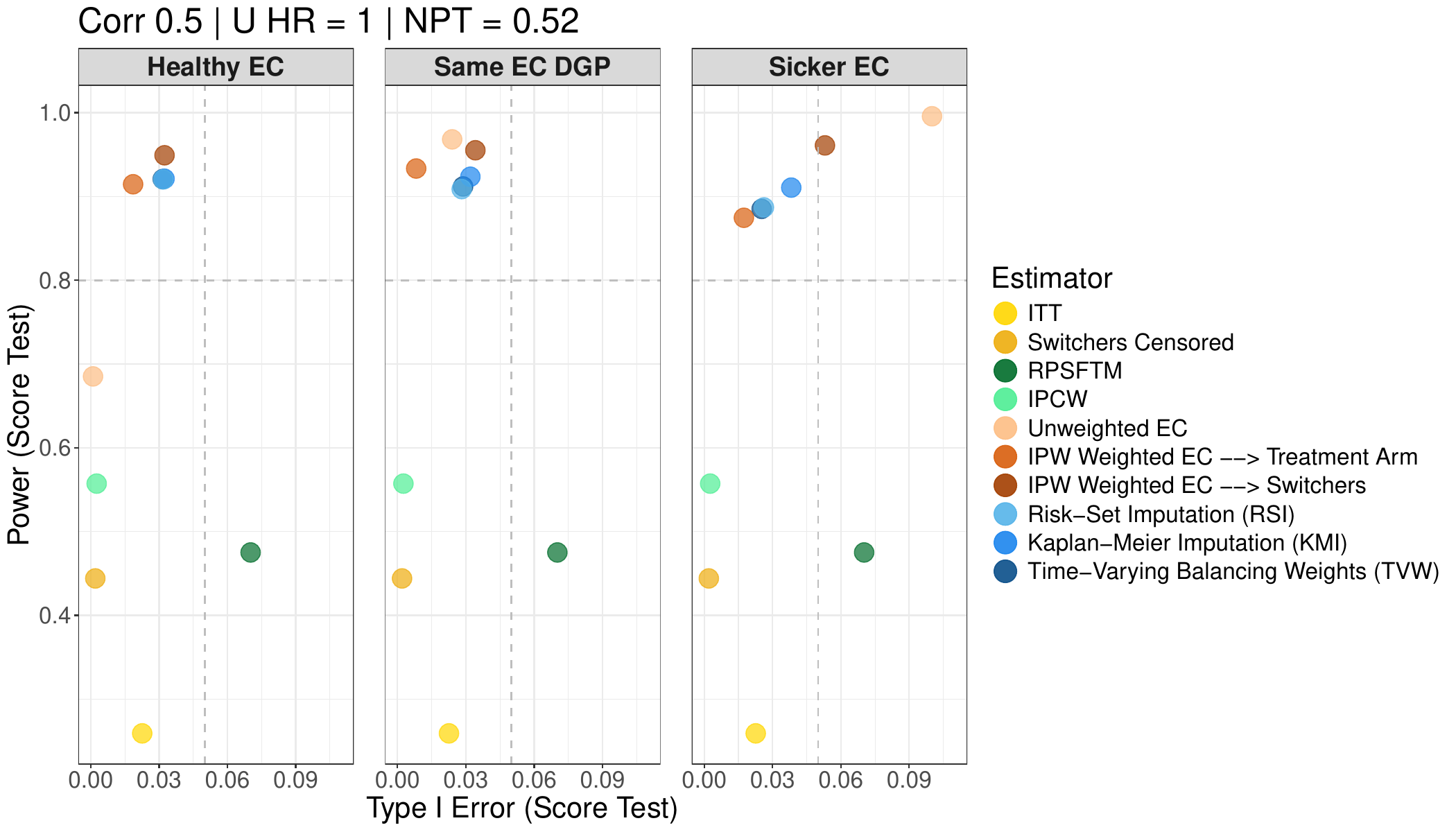}
        \caption{Power and Type I error estimates}
        \label{fig:corr0.5-u1-npt52-b}
    \end{subfigure}
    
    \caption{Simulation results for $\rho = 0.5$, $\exp(\beta_u) = 1$ and non-constant NPT. Individual panels represent covariate distribution of ECs relative to internal controls (e.g., healthier/sicker/same DGP). The first four estimates in each panel are identical since they do not use EC data. Refer to Section \ref{sec:method-summary} for a description of all methods used. Precise Type I error rates for estimators exceeding the visual truncation threshold of 0.1 include: Unweighted EC in the Sicker EC panel (0.216).}
    \label{fig:corr0.5-u1-npt52}
\end{figure}

\begin{figure}[ht]
    \centering
    \begin{subfigure}[b]{0.49\linewidth}
        \centering
        \includegraphics[width=\linewidth]{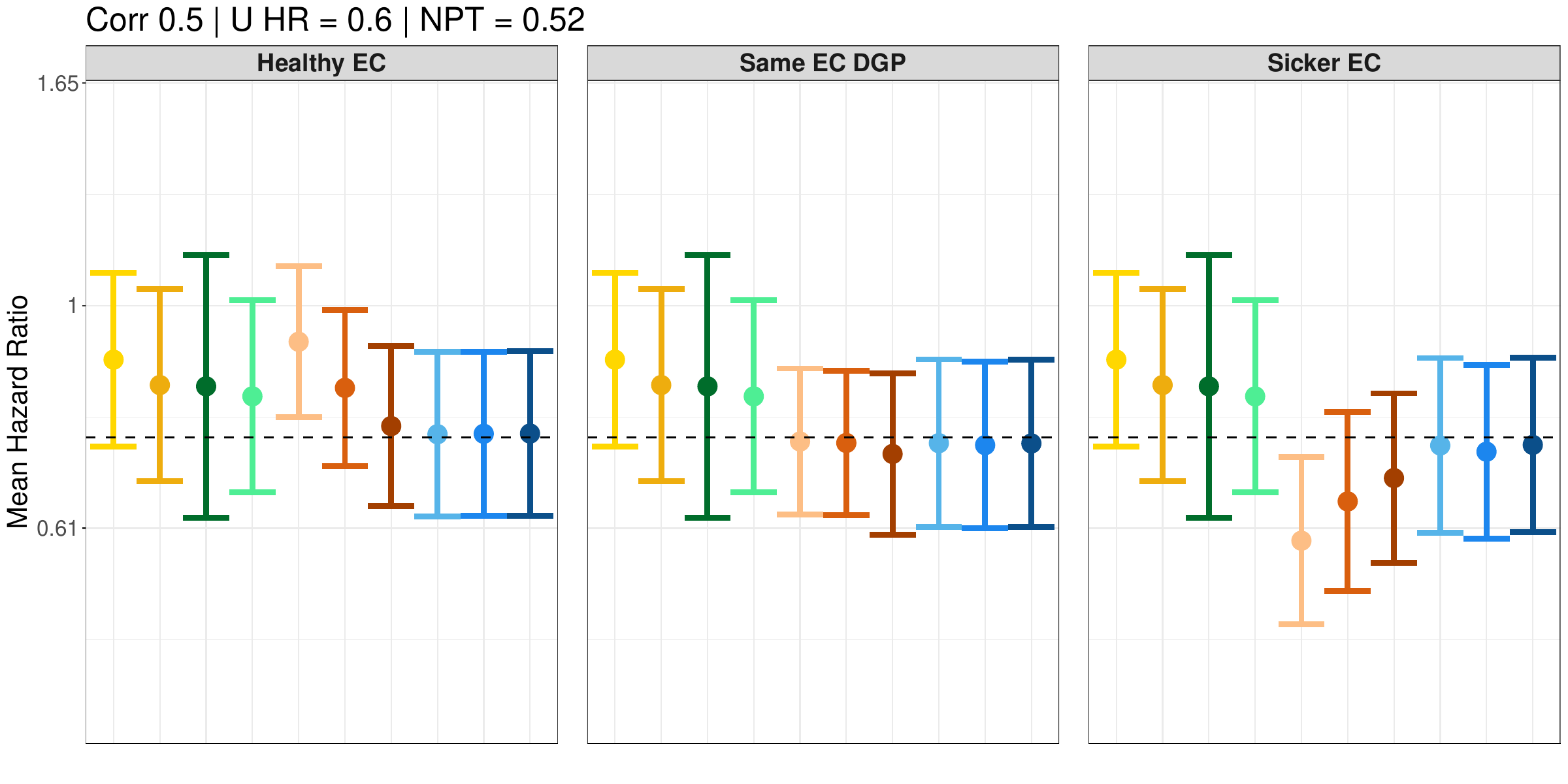}
        \caption{Hazard Ratio estimates}
        \label{fig:corr0.5-u0.6-npt52-a}
    \end{subfigure}
    \hfill
    \begin{subfigure}[b]{0.49\linewidth}
        \centering
        \includegraphics[width=\linewidth]{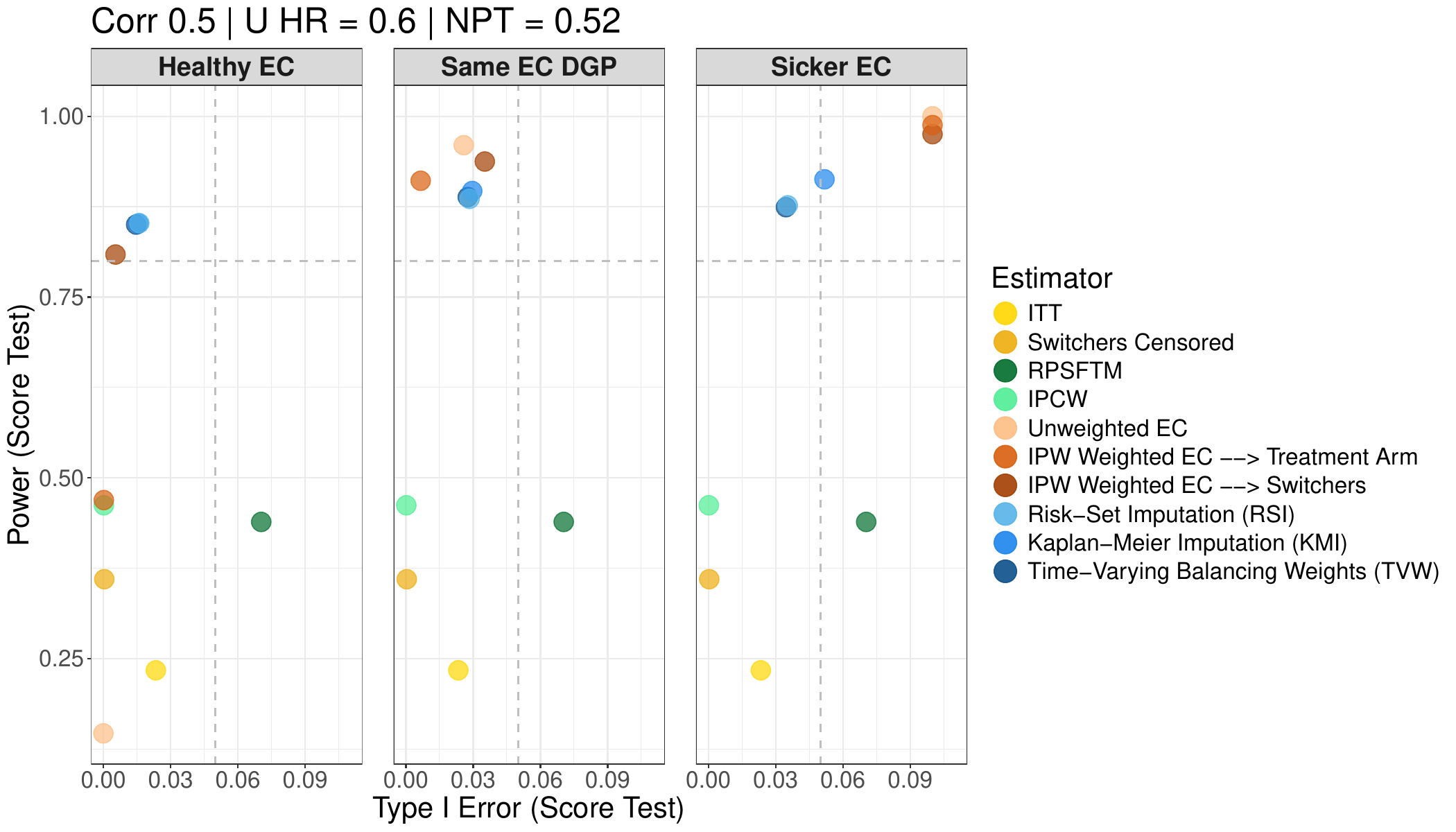}
        \caption{Power and Type I error estimates}
        \label{fig:corr0.5-u0.6-npt52-b}
    \end{subfigure}
    
    \caption{Simulation results for $\rho = 0.5$, $\exp(\beta_u) = 0.6$ and non-constant NPT. Individual panels represent covariate distribution of ECs relative to internal controls (e.g., healthier/sicker/same DGP). The first four estimates in each panel are identical since they do not use EC data. Refer to Section \ref{sec:method-summary} for a description of all methods used. Precise Type I error rates for estimators exceeding the visual truncation threshold of 0.1 include: Unweighted EC in the Sicker EC panel (0.692), Weighted EC (treatment arm) in the Sicker EC panel (0.205) and Weighted EC (switchers) in the Sicker EC panel (0.107).}
    \label{fig:corr0.5-u0.6-npt52}
\end{figure}

\begin{figure}[ht]
    \centering
    \begin{subfigure}[b]{0.49\linewidth}
        \centering
        \includegraphics[width=\linewidth]{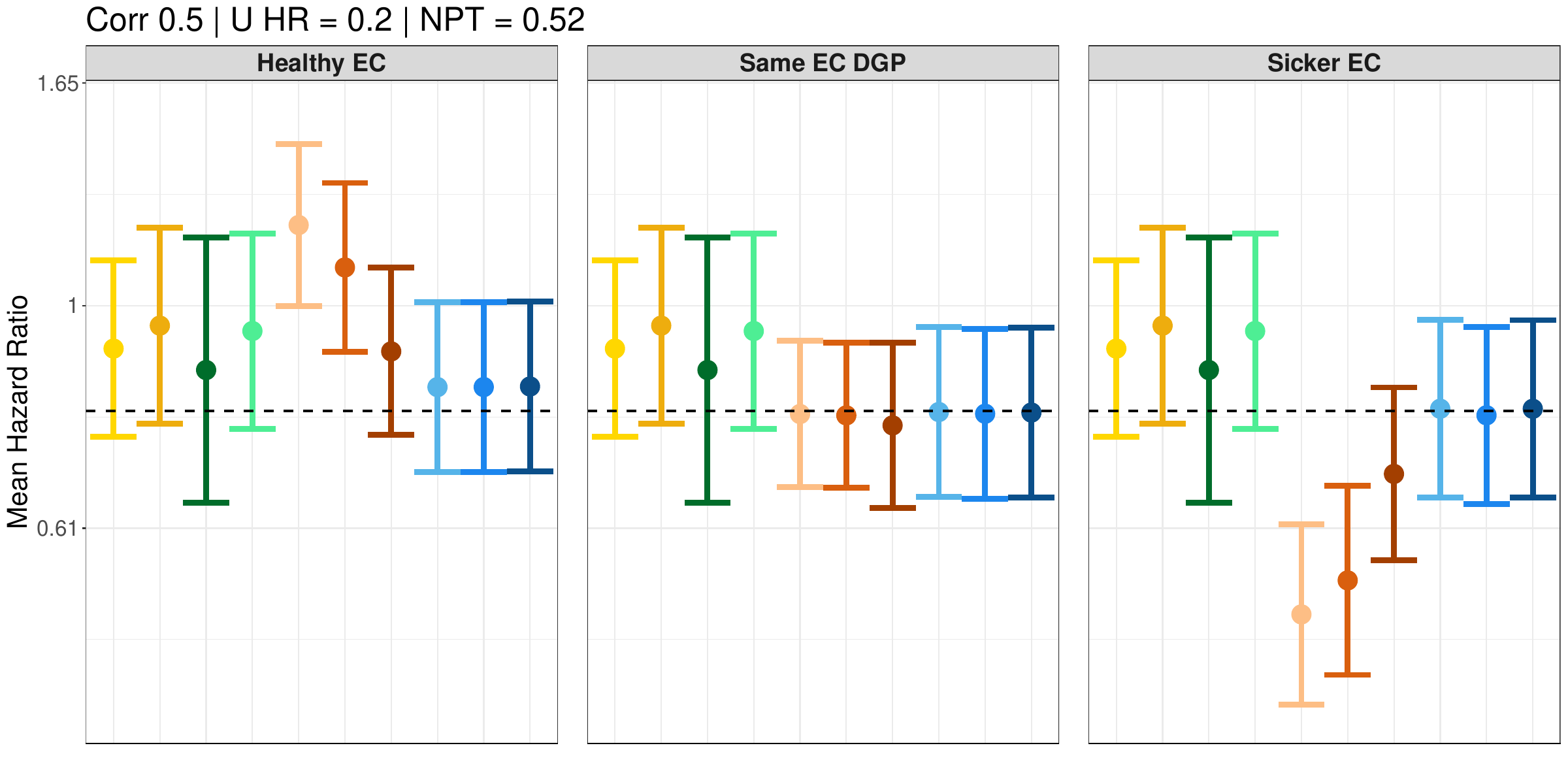}
        \caption{Hazard Ratio estimates}
        \label{fig:corr0.5-u0.2-npt52-a}
    \end{subfigure}
    \hfill
    \begin{subfigure}[b]{0.49\linewidth}
        \centering
        \includegraphics[width=\linewidth]{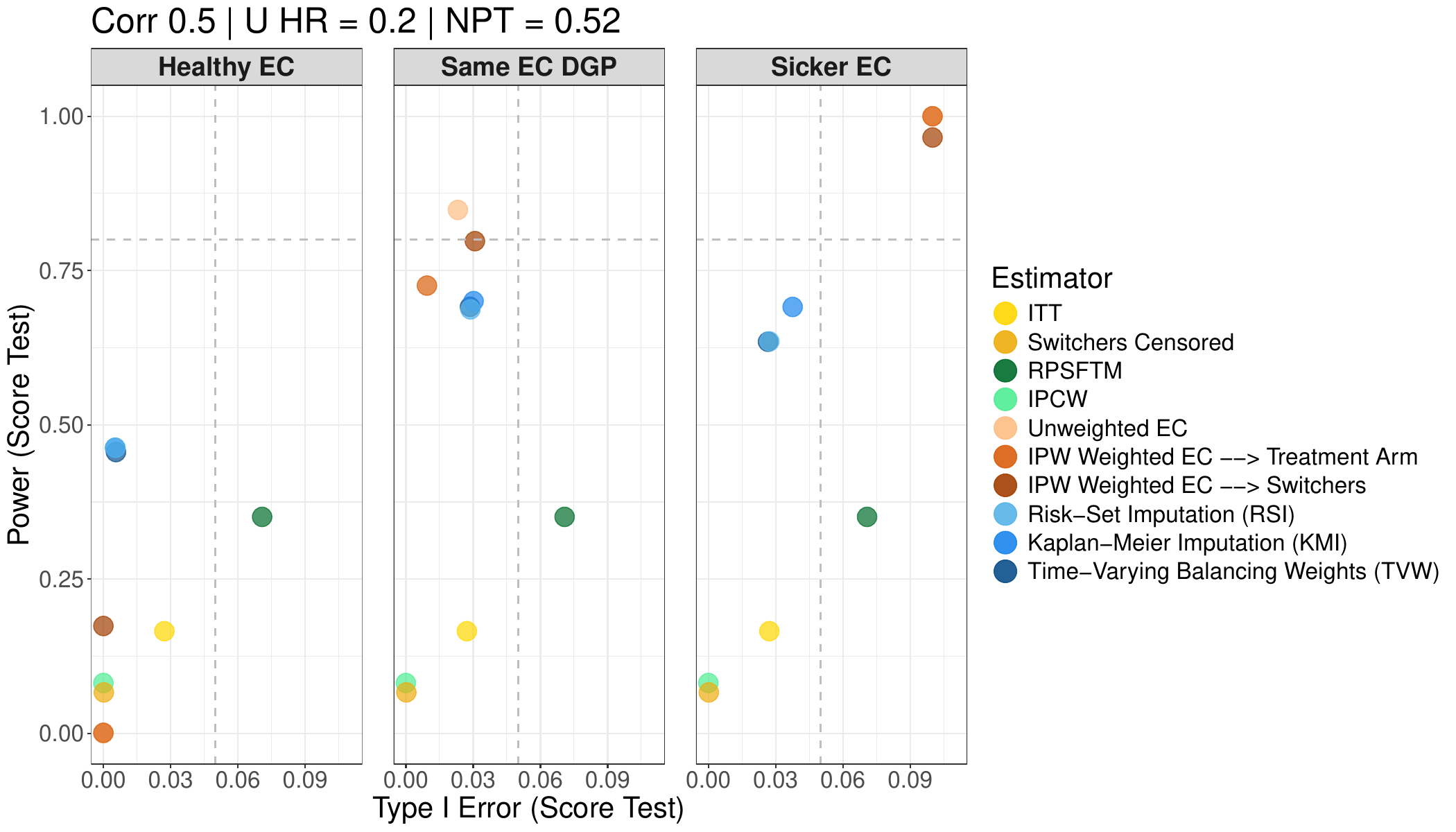}
        \caption{Power and Type I error estimates}
        \label{fig:corr0.5-u0.2-npt52-b}
    \end{subfigure}
    
    \caption{Simulation results for $\rho = 0.5$, $\exp(\beta_u) = 0.2$ and non-constant NPT. Individual panels represent covariate distribution of ECs relative to internal controls (e.g., healthier/sicker/same DGP). The first four estimates in each panel are identical since they do not use EC data. Refer to Section \ref{sec:method-summary} for a description of all methods used. Precise Type I error rates for estimators exceeding the visual truncation threshold of 0.1 include: Unweighted EC in the Sicker EC panel (0.995), Weighted EC (treatment arm) in the Sicker EC panel (0.911) and Weighted EC (switchers) in the Sicker EC panel (0.24).}
    \label{fig:corr0.5-u0.2-npt52}
\end{figure}

\begin{figure}[ht]
    \centering
    \begin{subfigure}[b]{0.49\linewidth}
        \centering
        \includegraphics[width=\linewidth]{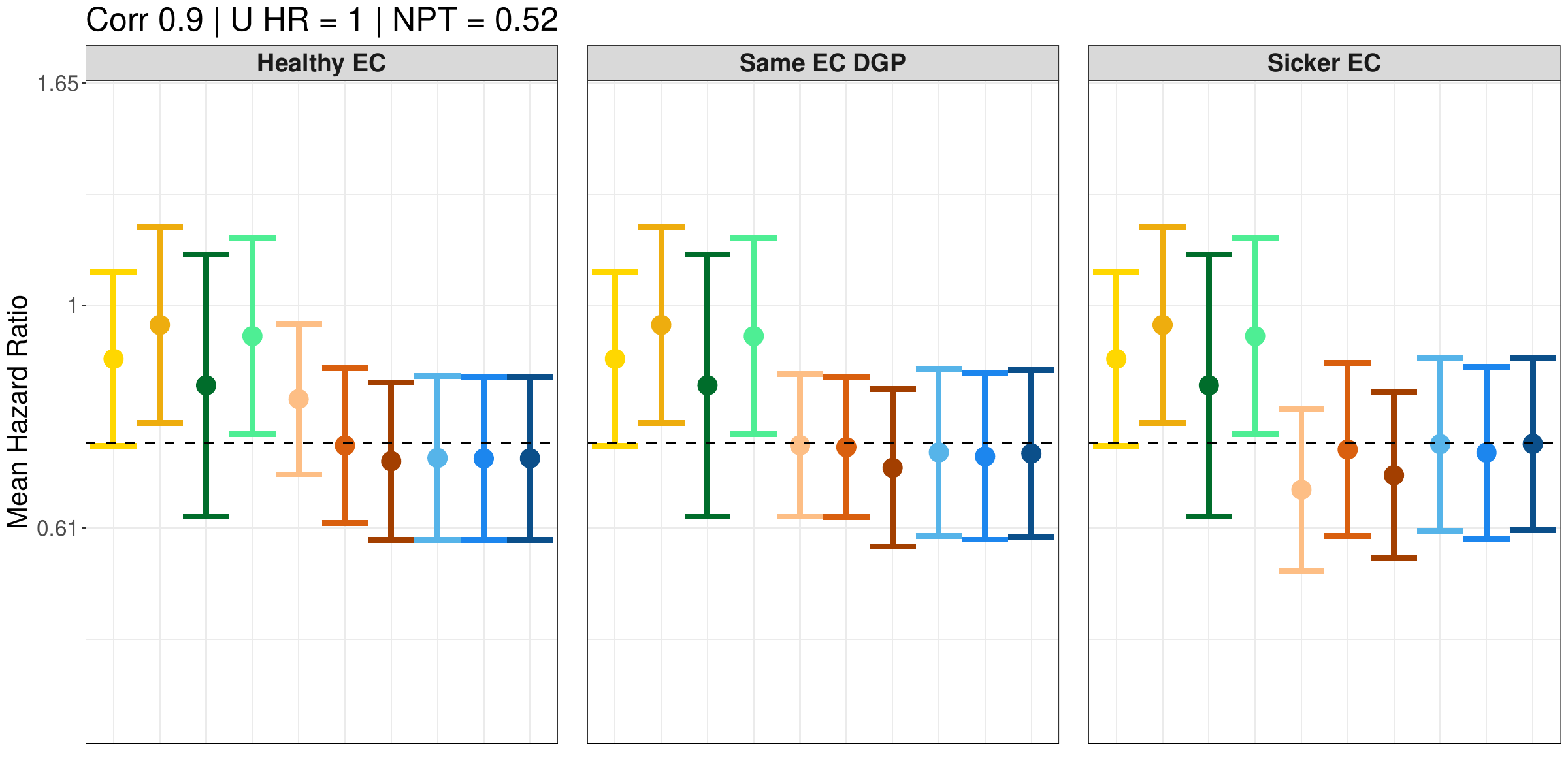}
        \caption{Hazard Ratio estimates}
        \label{fig:corr0.9-u1-npt52-a}
    \end{subfigure}
    \hfill
    \begin{subfigure}[b]{0.49\linewidth}
        \centering
        \includegraphics[width=\linewidth]{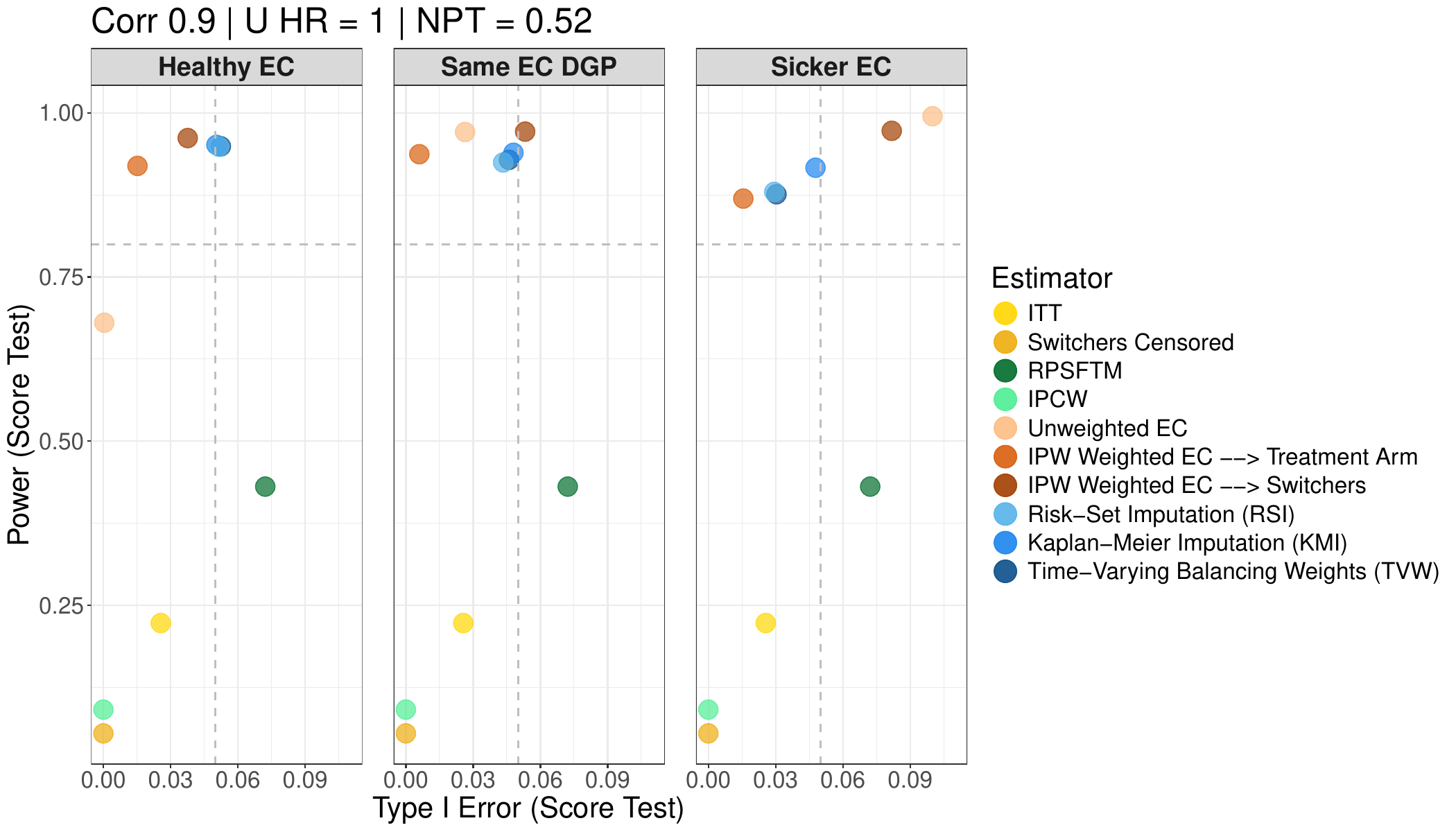}
        \caption{Power and Type I error estimates}
        \label{fig:corr0.9-u1-npt52-b}
    \end{subfigure}
    
    \caption{Simulation results for $\rho = 0.9$, $\exp(\beta_u) = 1$ and non-constant NPT. Individual panels represent covariate distribution of ECs relative to internal controls (e.g., healthier/sicker/same DGP). The first four estimates in each panel are identical since they do not use EC data. Refer to Section \ref{sec:method-summary} for a description of all methods used. Precise Type I error rates for estimators exceeding the visual truncation threshold of 0.1 include: Unweighted EC in the Sicker EC panel (0.215).}
    \label{fig:corr0.9-u1-npt52}
\end{figure}

\begin{figure}[ht]
    \centering
    \begin{subfigure}[b]{0.49\linewidth}
        \centering
        \includegraphics[width=\linewidth]{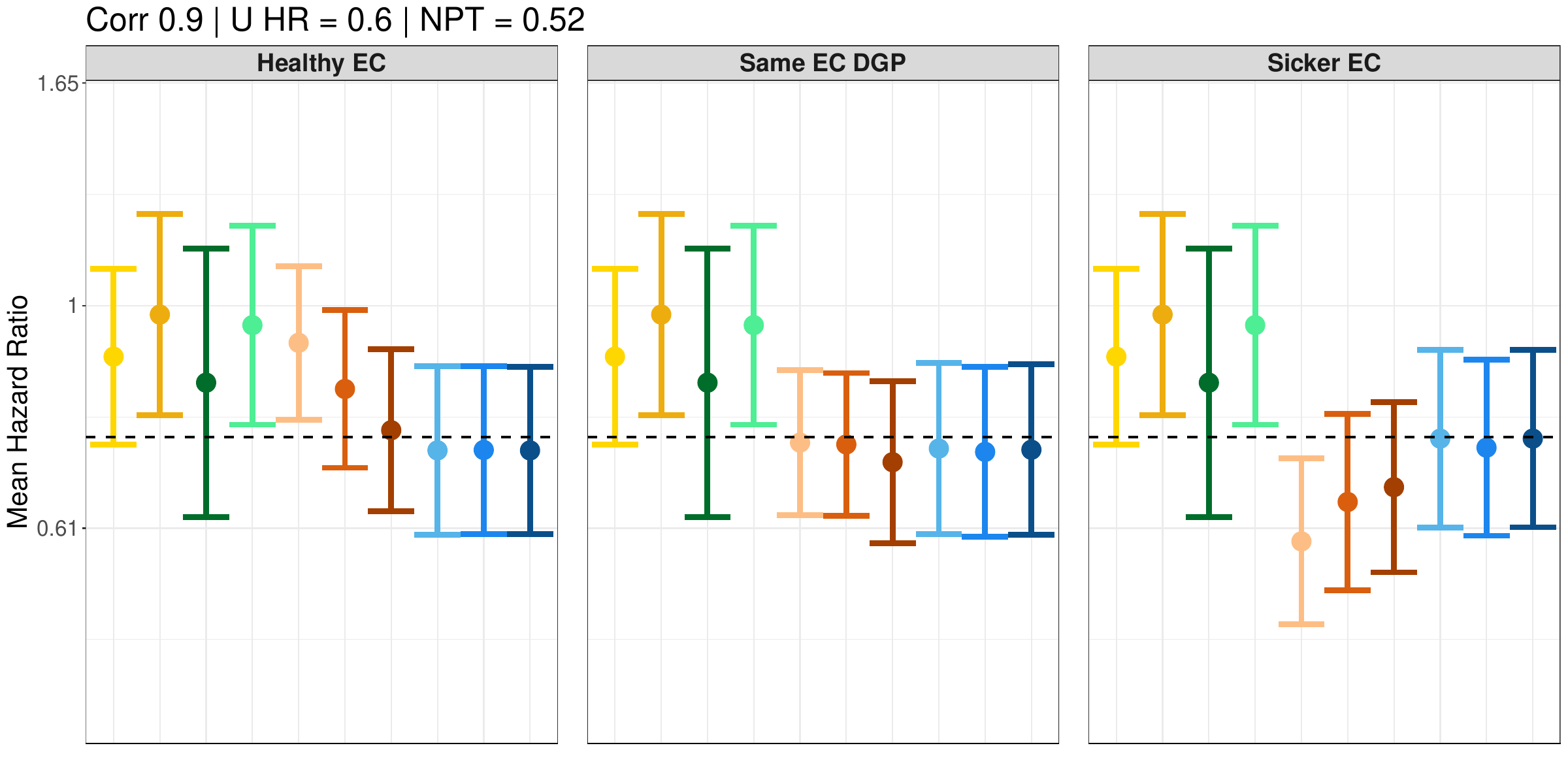}
        \caption{Hazard Ratio estimates}
        \label{fig:corr0.9-u0.6-npt52-a}
    \end{subfigure}
    \hfill
    \begin{subfigure}[b]{0.49\linewidth}
        \centering
        \includegraphics[width=\linewidth]{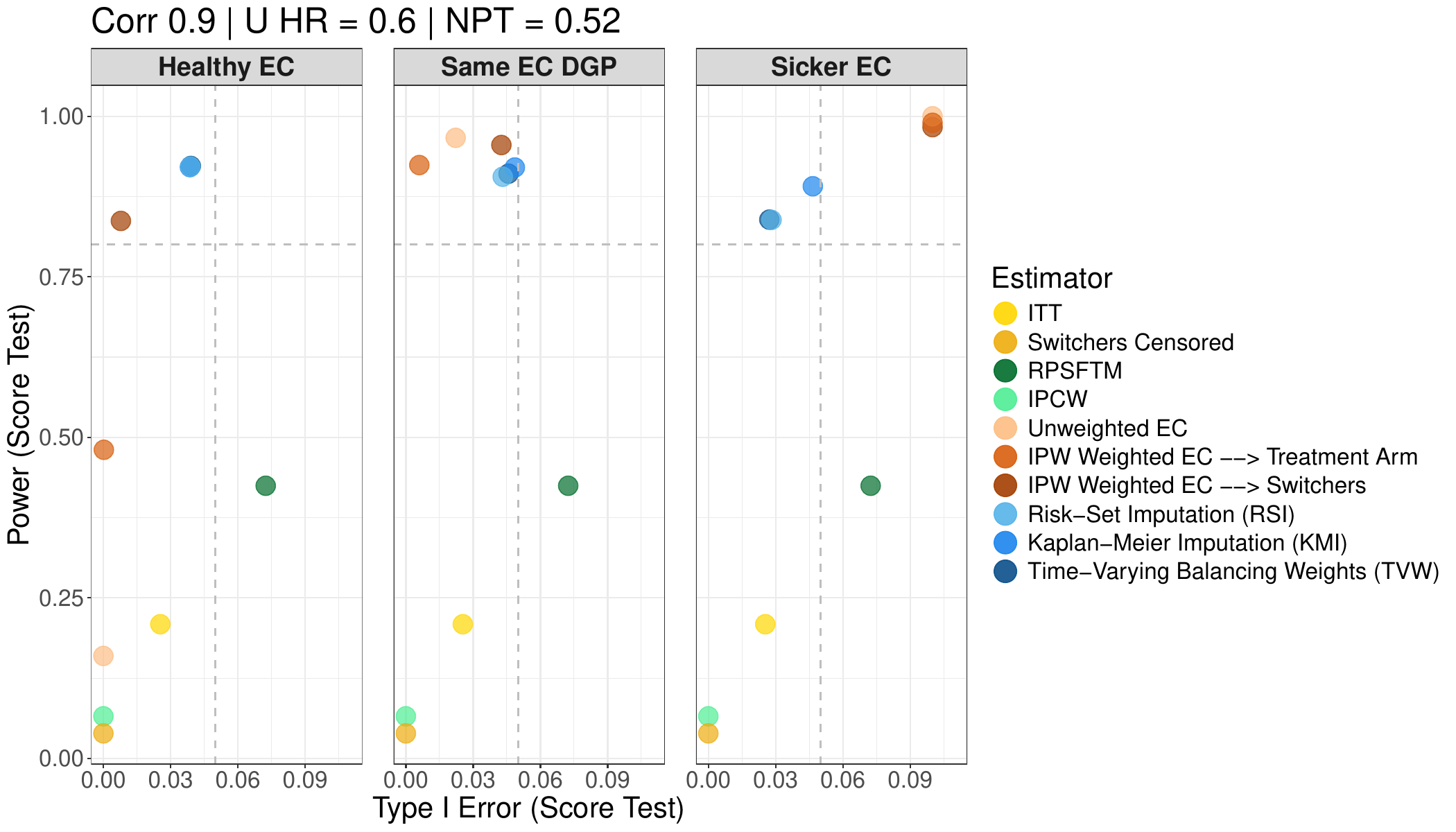}
        \caption{Power and Type I error estimates}
        \label{fig:corr0.9-u0.6-npt52-b}
    \end{subfigure}
    
    \caption{Simulation results for $\rho = 0.9$, $\exp(\beta_u) = 0.6$ and non-constant NPT. Individual panels represent covariate distribution of ECs relative to internal controls (e.g., healthier/sicker/same DGP). The first four estimates in each panel are identical since they do not use EC data. Refer to Section \ref{sec:method-summary} for a description of all methods used. Precise Type I error rates for estimators exceeding the visual truncation threshold of 0.1 include: Unweighted EC in the Sicker EC panel (0.704), Weighted EC (treatment arm) in the Sicker EC panel (0.2) and Weighted EC (switchers) in the Sicker EC panel (0.145).}
    \label{fig:corr0.9-u0.6-npt52}
\end{figure}

\begin{figure}[ht]
    \centering
    \begin{subfigure}[b]{0.49\linewidth}
        \centering
        \includegraphics[width=\linewidth]{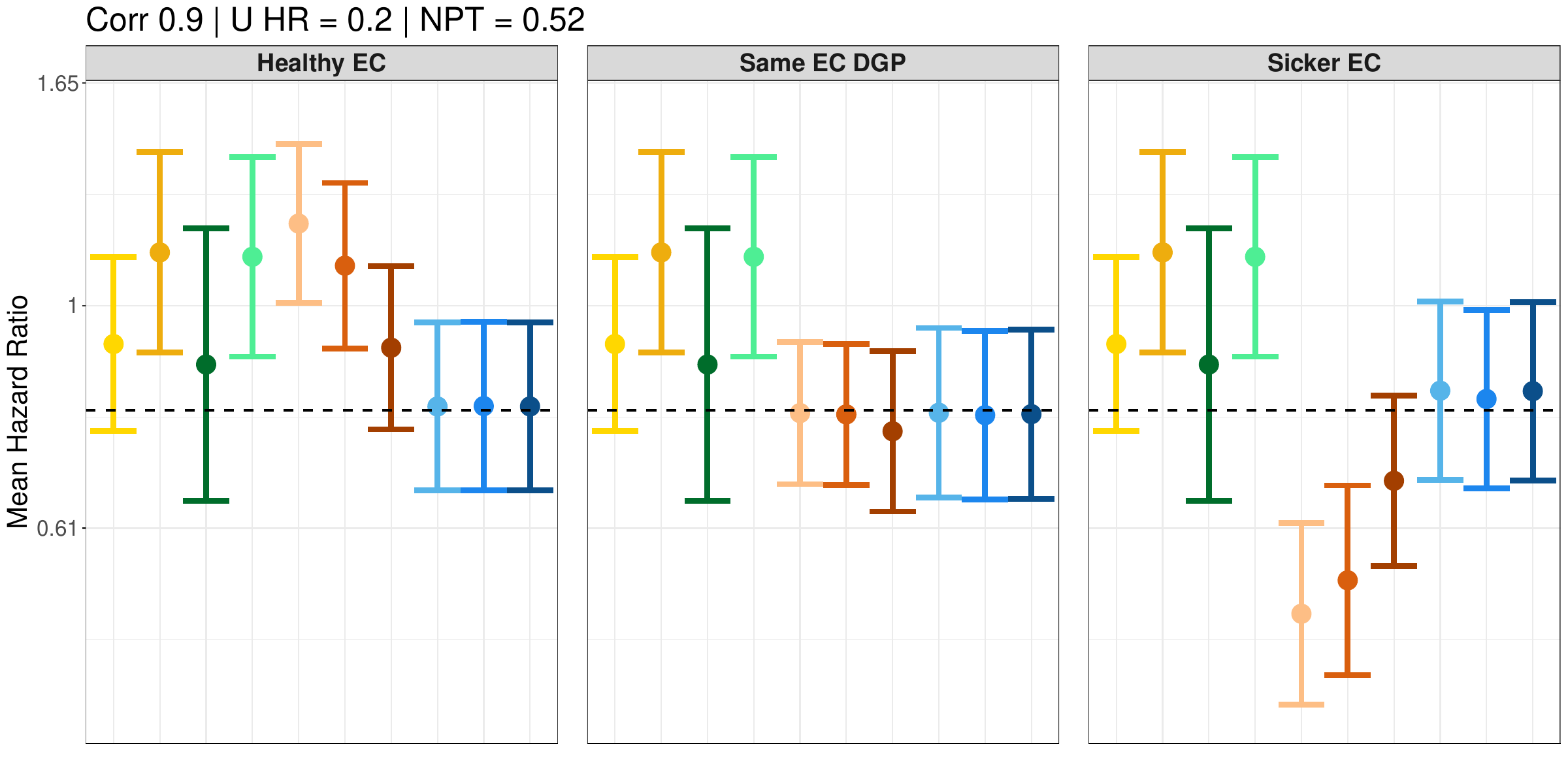}
        \caption{Hazard Ratio estimates}
        \label{fig:corr0.9-u0.2-npt52-a}
    \end{subfigure}
    \hfill
    \begin{subfigure}[b]{0.49\linewidth}
        \centering
        \includegraphics[width=\linewidth]{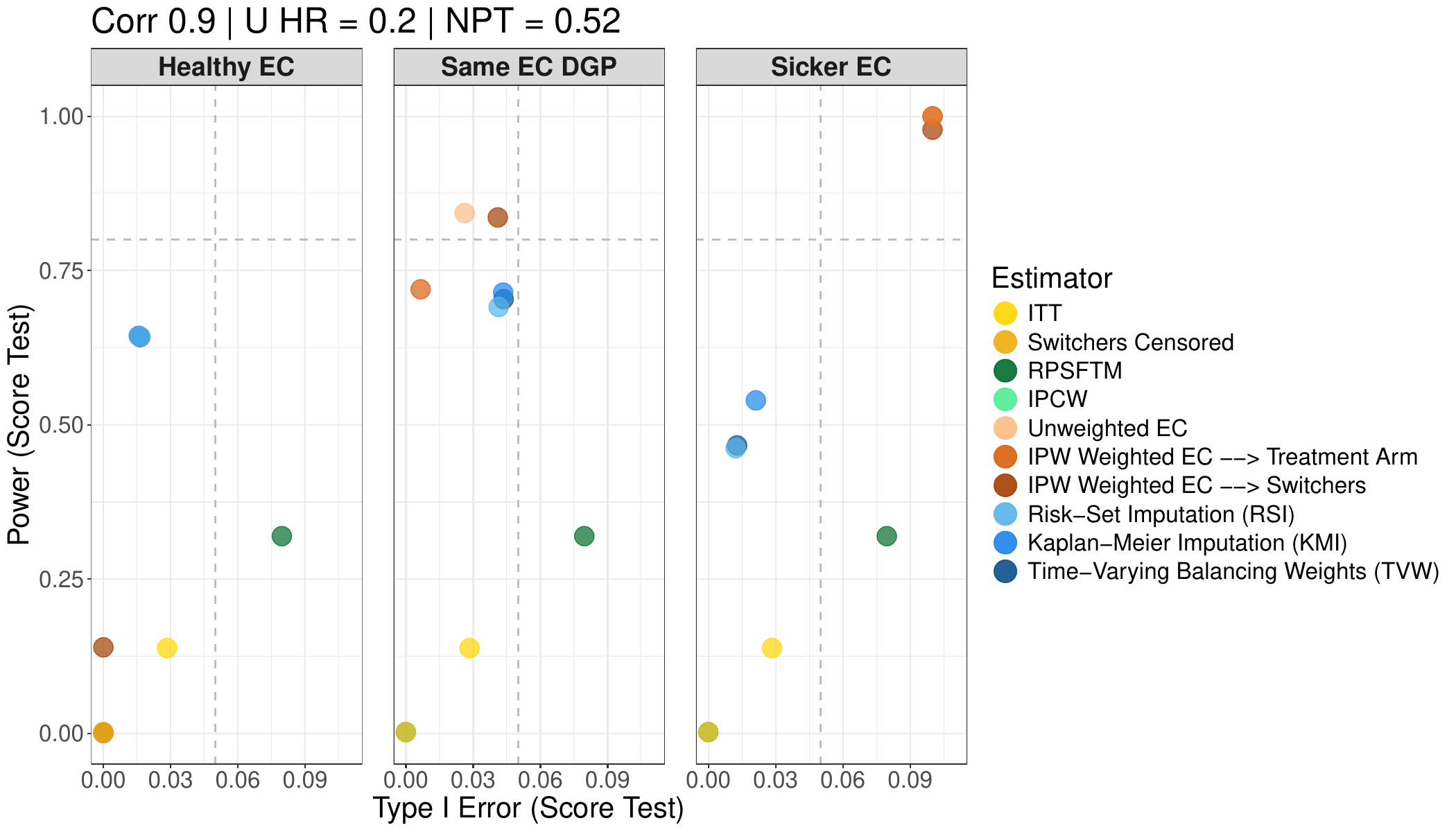}
        \caption{Power and Type I error estimates}
        \label{fig:corr0.9-u0.2-npt52-b}
    \end{subfigure}
    
    \caption{Simulation results for $\rho = 0.9$, $\exp(\beta_u) = 0.2$ and non-constant NPT. Individual panels represent covariate distribution of ECs relative to internal controls (e.g., healthier/sicker/same DGP). The first four estimates in each panel are identical since they do not use EC data. Refer to Section \ref{sec:method-summary} for a description of all methods used. Precise Type I error rates for estimators exceeding the visual truncation threshold of 0.1 include: Unweighted EC in the Sicker EC panel (0.997), Weighted EC (treatment arm) in the Sicker EC panel (0.907) and Weighted EC (switchers) in the Sicker EC panel (0.283).}
    \label{fig:corr0.9-u0.2-npt52}
\end{figure}

\clearpage

\section{Simulation EC Covariate DGPs} \label{sec:app-ec-dgp}
As discussed in Section \ref{sec:factors-vary-sim}, we vary several factors in our simulation to introduce bias. One factor is the covariate distribution of the EC population. In particular, we consider scenarios where ECs follow the same distribution or are healthier or sicker than the internal trial control population. The covariate distributions of these population are summarized in Table \ref{tab:simulation_covariates_supp}. The hazard ratio of each covariate on PD and OS are identical in the internal trial and EC cohorts.

\begin{table}[ht]
\centering
\scriptsize
\setlength{\tabcolsep}{3pt}

\begin{tabular}{lllllc}
\toprule
\textbf{Variable} & \textbf{Description} & \textbf{Trial DGP} & \textbf{Healthier EC} & \textbf{Sicker EC} & \textbf{\begin{tabular}{@{}c@{}}Hazard Ratio\\(PD \& OS)\end{tabular}} \\
\midrule
$X_1$ & Age ($\geq65$ vs. $< 65$) & $\text{Bern}(0.5)$ & $\text{Bern}(0.25)$ & $\text{Bern}(0.75)$ & 0.90 \\
$X_2$ & Sex (Female vs. Male) & $\text{Bern}(0.5)$ & $\text{Bern}(0.25)$ & $\text{Bern}(0.75)$ & 0.80 \\
$X_3$ & Biomarker (Median vs. Low) & $\text{Cat}(1/3, 1/3, 1/3)$ & $\text{Cat}(1/6, 2/6, 3/6)$ & $\text{Cat}(3/6, 2/6, 1/6)$ & 0.80 \\
      & Biomarker (High vs. Low) &  & & & 0.70 \\
$X_4$ & Tumor Burden (SLD) & $\text{LogNormal}(\ln(80), 0.66)$ & $\text{LogNormal}(\ln(60), 0.66)$ & $\text{LogNormal}(\ln(100), 0.66)$ & 1.005 \\
$U$   & Unmeasured Confounder & $\text{Bern}(0.5)$ & $\text{Bern}(0.25)$ & $\text{Bern}(0.75)$ & Varying \\
\midrule
$Z$ & Binary Treatment & 1:1 Randomization & N/A & N/A & 0.72 (OS) \\
          &                          &                   &    &    & 0.64 (PD) \\
\bottomrule
\end{tabular}
\caption{Summary of baseline covariate distributions for internal control and EC populations used in the simulation study.}
\label{tab:simulation_covariates_supp}
\end{table}

\section{Deomgraphics for IMpower130 and OAK Studies}

\begin{table}[ht]
\centering
\begin{tabular}{lcc}
  \toprule
  \textbf{Covariate} & \textbf{OAK Trial (EC)} & \textbf{IMpower130 Trial Switchers} \\ 
  & \textbf{Proportion (SE)} & \textbf{Proportion (SE)} \\ 
  \midrule
  Sex & 0.355 (0.021) & 0.484 (0.052) \\ 
  PD-L1 Expression (Level 2) & 0.783 (0.018) & 0.677 (0.048) \\ 
  Liver Metastasis & 0.800 (0.017) & 0.892 (0.032) \\ 
  Tobacco History & 0.193 (0.017) & 0.194 (0.041) \\ 
  \bottomrule
\end{tabular}
\caption{Baseline characteristics of external controls versus trial switchers prior to adjustment and methods application.}
\label{tab:baseline_covariates}
\end{table}




\end{document}